%% file: mass.tex
\def \farcs{\hbox{$.\!\!^{\prime\prime}$}}

\def \msun{\hbox{${\rm M}_\odot$}}
\def\arcsec{\hbox{$^{\hbox{\rlap{\hbox{\lower4pt\hbox{$\,\prime\prime$}}
          }\hbox{$\frown$}}}$}}

\def\arcmin{\hbox{$^{\hbox{\rlap{\hbox{\lower4pt\hbox{$\;\prime$}}
          }\hbox{$\frown$}}}$}}

\documentclass[natbib,amsmath,amssymb]{svjour3}  

\smartqed  
\usepackage[dvips]{graphicx}
\usepackage{amsbsy}
\usepackage{amssymb}
\usepackage{amsmath}
\renewcommand{\d}{\mathrm{d}}

\input{epsf.def}

\begin{document}

\title{Masses of galaxy clusters from gravitational lensing
}

\titlerunning{Cluster lensing masses}        

\author{Henk Hoekstra         \and
        Matthias Bartelmann   \and
        H{\aa}kon Dahle \and
        Holger Israel \and
        Marceau Limousin \and
        Massimo Meneghetti
}


\institute{H. Hoekstra \at
              Leiden Observatory, Leiden University, Leiden, The Netherlands    
\and M. Bartelmann \at
     Zentrum f{\"u}r Astronomie, Institut f{\"u}r Theoretische
Astrophysik, Heidelberg, Germany
\and H. Dahle \at
Institute of Theoretical Astrophysics, University of Oslo, Oslo, Norway
\and H. Israel \at
Argelander-Institut f{\"u}r Astronomie, Bonn University, Bonn, Germany
\and M. Limousin \at
Laboratoire d’Astrophysique de Marseille, Aix Marseille Universit{\'e}, CNRS, Marseille, France
\and M. Meneghetti \at
INAF - Osservatorio Astronomico di Bologna, Bologna, Italy \&\\
INFN, Sezione di Bologna, Bologna, Italy
}

\date{Received: date / Accepted: date}

\maketitle

\begin{abstract}
Despite consistent progress in numerical simulations, the observable
properties of galaxy clusters are difficult to predict ab initio.  It
is therefore important to compare both theoretical and observational
results to a direct measure of the cluster mass. This can be done by
measuring the gravitational lensing effects caused by the bending of
light by the cluster mass distribution. In this review we discuss how
this phenomenon can be used to determine cluster masses and study the
mass distribution itself. As sample sizes increase, the accuracy of
the weak lensing mass estimates needs to improve accordingly. We discuss
the main practical aspects of these measurements. We review a number
of applications and highlight some recent results.

\end{abstract}

\section{Introduction}
\label{intro}

Inhomogeneities in the matter distribution perturb the paths of
photons that are emitted by distant galaxies. The result is equivalent
as if we are viewing these sources through a piece of glass with a
spatially varying index of refraction: the images appear slightly
distorted and magnified. Both effects can be measured in principle,
and can be used to determine (projected) masses, because the amplitude
of the distortion provides a direct measure of the gravitational tidal
field, independent of the nature of the dark matter or the dynamical
state of the system of interest. This is particularly useful for
clusters of galaxies, which are dynamically young and often show signs
of merging.

If the deflection of the light rays is large enough, multiple images
of the same source can be observed: these strong lensing events
provide precise constraints on the mass on scales enclosed by these
images. This phenomenon was already discussed by \cite{Zwicky37}, but
the first observation was made more than 40 years later with the
discovery of a gravitationally lensing QSO \citep{Walsh79}. The
discovery of strong lensing by a cluster of galaxies was reported
by \cite{Soucail87}, and is now seen routinely, in particular in deep
observations with the {\it Hubble Space Telescope}. The study of strong
lensing systems has a number of applications, which are discussed in
\cite{Bartelmann12,Meneghetti12}. In this review we limit the discussion to 
the use of strong lensing to determine masses of galaxy
clusters. Unfortunately, the effect is confined to the inner regions,
and comparison with other proxies for mass requires extrapolation to
larger radii.

The lensing effect itself, however, is not limited to small scales. At
large radii the tidal field causes a subtle change in the shapes of
galaxies, resulting in a coherent alignment of the sources that can be
measured statistically. As the signal is proportional to the mass of
the system, galaxy clusters were the first objects to be studied using
this weak gravitational lensing technique, with the first detection of
the signal presented by \cite{Tyson90}.  The '90s saw a rapid
development of the field, and a shift from studying clusters of
galaxies to the statistical properties of the large-scale matter
distribution: the cosmic shear \citep[see e.g.,][for a
review]{Hoekstra08}. The latter application was made possible thanks
to the advent of mosaic cameras on large telescopes, but also led to
improved methods to measure the shapes of galaxies. In addition, the
desire to study the lensing signal as a function of redshift has
improved our knowledge of the source redshift distribution, which is
needed to convert the lensing signal into an actual mass. This
progress over the past 20 years has also benefited the work on
clusters of galaxies, which now can be studied out to larger radii,
and with improved accuracy.

Here we provide a brief introduction how the observable consequences
of gravitational lensing can be used to determine masses of galaxy
clusters, with a focus on a number of practical aspects. We contrast
the advantages (and disadvantages) of both strong and weak lensing and
discuss some recent results. The principles of gravitational lensing
are briefly reviewed in \S2, but readers interested in lensing results
can skip this, as well as \S3 where we discuss some of the technical
issues arising in the measurement and interpratation of the weak
lensing signal. Some highlights of mass reconstructions are reviewed
in \S4 and a discussion of the properties of cluster halos, such as
density profiles and halo shapes, can be found in \S5. Cluster mass
estimates are discussed in \S6 and some measurements of the halos of
cluster galaxies are presented in \S7. We conclude and present an
outlook in \S8.

\section{Gravitational lensing}
\label{sec:gravlens}

We start with a basic introduction of gravitational lensing, with a
particular focus on applications that pertain to clusters of
galaxies. We refer the interested reader to more thorough
introductions into the subject by \cite{Bartelmann01},
\cite{Schneider06} or \cite{Bartelmann10}.

As mentioned in the introduction, inhomogeneities along the
line-of-sight deflect photons that are emitted by distant galaxies
(the sources). As a result a source with a true position
$\vec{\beta}$ is actually observed at positions
$\vec{x}$ that satisfy the lens equation

\begin{equation}
\vec{\beta}=\vec{x}-\vec{\alpha}(\vec{x}),
\end{equation}

\noindent where $\vec\alpha$ is the deflection angle 
\citep[see Fig.~11 in][for a diagram]{Bartelmann01}. Note that 
the lens equation can have multiple solutions for $\vec{x}$, which
means that several images of the same source are observed. The case
when multiple images are produced is commonly referred to as ``strong
gravitational lensing''.  As discussed in \S\ref{sec:stronglens}, this
phenomenon can be used to provide very precise constraints on the mass
distribution in cluster cores (also see \cite{Meneghetti12} in this
volume).

An important feature of gravitational lensing is that it conserves the
surface brightness: for a source with a true surface brightness
distribution $f^{\rm s}({\vec x})$ the observed surface brightness
distribution is given by

\begin{equation}
f^{\rm obs}(\vec{x})=f^{\rm
s}[{\vec\beta}(\vec{x})].
\end{equation}

\noindent Away from the lens, where the deflection angle as well
as its spatial variation are typically small compared to the extent
of the source, this mapping can be linearized. This case is commonly
referred to as ``weak gravitational lensing'', and the resulting
remapping of the surface brightness distribution can be written as

\begin{equation}
f^{\rm obs}(x_i)=f^{\rm s}({\cal{A}}_{ij}x_{j}),
\end{equation}

\noindent where ${\bf x}$ is the position on the sky and $\cal{A}$ is
the distortion matrix (i.e., the Jacobian of the transformation),
which is specified by the projected surface density of the lens and
the redshifts of the lens and the source. It is convenient to
introduce the deflection potential $\Psi$

\begin{equation}
\Psi({\bf x})=\frac{1}{\pi}\int {\mathrm d}^2{\bf x'} \kappa({\bf x'})\ln|{\bf x-x'}|,
\end{equation}\label{eq:psi}

\noindent where the convergence $\kappa$ is the ratio of the projected
surface density $\Sigma({\bf x})$ and the critical surface density
$\Sigma_{\rm crit}$:

\begin{equation}
\kappa({\bf x})=\frac{\Sigma({\bf x})}{\Sigma_{\rm crit}},
\end{equation}

\noindent with $\Sigma_{\rm crit}$ defined as

\begin{equation}
\Sigma_{\rm crit}=\frac{c^2}{4\pi G}\frac{D_{\rm s}}{D_{\rm l} D_{\rm ls}}.
\label{eq:sigmacrit}
\end{equation}

\noindent Here $D_{\rm s}$, $D_{\rm l}$, and $D_{\rm ls}$ correspond
to the angular diameter distances between the observer and the source,
observer and the lens, and the lens and the source. Hence, the lensing
signal depends on the redshifts of both the lenses and the sources.
The implications of this are discussed in more detail in
Section~\ref{sec:redshift}.

The distortion matrix ${\cal A}$ can be written in terms of the second
derivatives of the deflection potential $\Psi$

\begin{equation}
{\cal A}=\delta_{ij}-\frac{\partial^2 \Psi}{{
\partial x_i\partial x_j}}=
\left(
    \begin{array}{cc}
        1-\kappa-\gamma_1 & -\gamma_2 \\
        -\gamma_2        & 1-\kappa+\gamma_1 \\
    \end{array}
\right)=(1-\kappa)
\left(
    \begin{array}{cc}
        1-g_1 & -g_2 \\
        -g_2        & 1+g_1 \\
    \end{array}
\right),
\label{eq:distort}
\end{equation}

\noindent where we used that $2\kappa=\nabla^2\Psi$, introduced
the complex shear ${\bf \gamma}\equiv\gamma_1+i\gamma_2$, and defined
the reduced shear $g_i=\gamma_i/(1-\kappa)$. The shear is related to
the deflection potential through

\begin{equation}
\gamma_{1}=\frac{1}{2}\left(\frac{\partial^2 \Psi}{\partial x_1^2}
-\frac{\partial^2 \Psi}{\partial x_2^2}\right)\hspace{1em}{\rm and}\hspace{1em}
\gamma_2=\frac{\partial^2\Psi}{\partial x_1\partial x_2},
\label{eq:shear}
\end{equation}

\noindent The effect of the remapping by ${\cal A}$ is to transform a circular 
source into an ellipse with axis ratio $\sim (1-|g|)/(1+|g|)$ and
position angle $\alpha=\arctan(g_2/g_1)/2$; the reduced shear
describes the anisotropic distortion of a source. If $\kappa\ll 1$
(i.e., the weak lensing regime), the observable $g_i\sim \gamma_i$.
In addition, the source is magnified by a factor

\begin{equation}
\mu=\frac{1}{\det \cal{A}}=\frac{1}{(1-\kappa)^2-|\gamma|^2}=
\frac{1}{(1-\kappa)^2(1-|g|^2)},
\end{equation}

\noindent boosting the flux by the same amount. To first order, the 
magnification depends on the convergence only; i.e. $\kappa$ describes
the isotropic distortion of a source (contraction or dilation). Both
the shearing and magnification of sources are observable effects,
although both are quite different in terms of techniques and
systematics.

Note that the shear and convergence are related to one another through
the deflection potential. As a result it is possible to express the
shear as a convolution of the convergence with a kernel $\chi({\bf
x})$. This can be seen by computing the shear following
Eqn.~\ref{eq:shear}, starting from Eqn~\ref{eq:psi}:

\begin{equation}
\gamma({\bf x})=\frac{1}{\pi}\int {\mathrm d}^2{\bf x'}\chi({\bf x-x'})
\kappa({\bf x'}),
\end{equation}

\noindent where the convolution kernel $\chi({\bf x})$ is given by

\begin{equation}
\chi({\bf x})=\frac{x_2^2-x_1^2-2ix_1 x_2}{|{\bf x}|^4}.
\end{equation}

As first shown by \cite{KS93}, this expression can be inverted,
i.e. it is possible to express the surface density in terms of the
observable shear:

\begin{equation}
\kappa({\bf x})-\kappa_0=\frac{1}{\pi}\int {\mathrm d}^2{\bf x'}
\chi^*({\bf x-x'})\gamma({\bf x'}),\label{eq:ks}
\end{equation}

\noindent where the constant $\kappa_0$ shows that the surface density
can only be recovered up to a constant, and reflects the fact that a
constant $\kappa$ does not cause a shear. In fact one can show that a
transformation $\kappa'=\lambda\kappa+(1-\lambda)$ leaves the reduced
shear unchanged. This is known as the mass-sheet
degeneracy \citep{Gorenstein88}. The possibility to reconstruct the
surface mass density is an important application of weak gravitational
lensing, which is discussed in more detail in
\S\ref{sec:massrec}. Note that $\kappa({\bf x})$ is a real
function, but that the reconstruction in principle can have an
imaginary part as well. This can be used to examine fidelity of the
mass map and may help identify issues with the analysis of the data
(see \S\ref{sec:shapemeasurement}).

\subsection{Strong gravitational lensing}
\label{sec:stronglens}

Strong gravitational lensing and its applications to galaxy clusters
are discussed in detail in a dedicated contribution to this
volume \citep{Meneghetti12}. Here we give a brief introduction and
refer the interested reader to the separate review for further
information.

Several different meaningful definitions of strong lensing are
possible and are being used. In the context of galaxy clusters, we define
gravitational lenses as strong if they are capable of producing
multiple images from single sources. For this to happen, the lens
mapping must become locally multiply defined, which means that the
Jacobian matrix $\mathcal{A}$ of the lens mapping must be singular
somewhere:

\begin{equation}
  \det\mathcal{A}(\vec x_\mathrm{c}) = 0\;.
\label{eq:2.1-1}
\end{equation}

\noindent The set of angular positions $\vec{x}_\mathrm{c}$ on the sky where
Eqn.~\ref{eq:2.1-1} is satisfied can be shown to form closed curves,
which are called critical curves. Their images $\vec\beta_\mathrm{c}$
under the lens mapping are called caustics. A sufficient, but not
necessary condition for a gravitational lens to be strong, or
synonimously to become supercritical, is that its surface-mass
density $\Sigma$ exceeds the critical value $\Sigma_\mathrm{crit}$ at
least somewhere within the lens. Where this happens, the convergence
exceeds unity, i.e. $\kappa>1$.

Sources near caustics are strongly distorted. Multiple images form in
pairs as sources cross caustics inward. Since a source very far away
from the lens must have a single image, the number of images produced
by gravitational lenses must thus be odd, unless the lens has a
singular lensing potential.  Caustic points with well-defined tangents
are called fold points, others (``tips'') are called cusp points. If a
source approaches a fold point or a fold section from within a
caustic, two of its images approach from either side of the
corresponding section of the critical curve, merge once the source
crosses the caustic outward and disappear as the source moves on. For
sources approaching cusp points from the inside out, three of their
images approach and merge to form a single, strongly distorted and
highly magnified image. The largest gravitational arcs, which are the
perhaps the most impressive manifestations of strong lensing by galaxy
clusters, are often formed from sources near cusp points.

Typically, two caustics or critical curves may be formed in a strong
lens, one for each of the two eigenvalues of the Jacobian matrix of
the lens mapping. They are distinguished by the preferential
orientation of the distortions experienced by sources near these
caustics. If two caustics are formed, typically the inner one is
mapped to the outer critical curve, where images appear tangentially
distorted relative to the lens' center. These curves are called the
tangential caustics or critical curves. Sources near the other
caustic, usually mapped to the inner critical curve, are imaged in a
radially distorted fashion. The respective caustics and critical
curves are hence called radial. Radially distorted images, commonly
called radial arcs, are often hard to recognise since they occur very
close to cluster centers where they are often outshone by bright,
central cluster galaxies. Where they can be found, they provide
important constraints on the local slope of the density profile.

The importance of strong cluster lensing and of the critical curves is
most easily illustrated with axially-symmetric lens models, even
though they certainly are not the most realistic. Such lenses can be
characterised by the dimensionless mass $m(r)$

\begin{equation}
  m(r) = 2\int_0^r r'\d r'\,\kappa(r'),
\label{eq:2.1-2}
\end{equation} 

\noindent where the radius $r=|\vec x|$. In axially-symmetric lenses, 
radial critical curves occur where

\begin{equation}
  \frac{\d}{\d r}\frac{m(r)}{r} = 1\;,
\label{eq:2.1-3}
\end{equation}

\noindent while tangential critical curves require

\begin{equation}
  m(r) = r^2.
\label{eq:2.1-4}
\end{equation}

The second relation immediately provides an estimate for the cluster
mass enclosed by a tangential critical curve at radius $r_{\rm
t}$. Multiplying with $\pi D_\mathrm{l}^2\Sigma_\mathrm{crit}$ converts
the dimensionless mass $m(r_{\rm t})$ on the left-hand side to the projected
mass enclosed by the physical radius $D_\mathrm{l}r_\mathrm{t}$ and
gives

\begin{equation}
  M(\theta_\mathrm{t}) = \pi D_\mathrm{l}^2r_\mathrm{t}^2\Sigma_\mathrm{crit} =
  \approx
  4.4\times10^{14}\,M_\odot\,\left(\frac{r_\mathrm{t}}{30\,\mbox{arcsec}}\right)^2\left(\frac{D_\mathrm{l}D_\mathrm{s}}{D_\mathrm{ls}\,\mathrm{Gpc}}\right)\;.
\label{eq:2.1-5}
\end{equation}

This simple relation is the basis for cluster mass estimates based on
strong gravitational lensing. The location of the critical curve can
be determined relatively well (especially compared to weak lensing
measurements). As a consequence the estimate of the projected mass is
precise, but not necessarily accurate as a caveat concerning estimates
obtained from (Eqn.~\ref{eq:2.1-5}) should be
mentioned. Equations~(\ref{eq:2.1-2}) and~(\ref{eq:2.1-4}) together
imply that the mean convergence within a tangential critical circle is
unity for axially-symmetric lenses. Under very general conditions, it
can be shown that realistic tangential critical curves enclose a lower
mean convergence. In simulated, asymmetric clusters, tangential
critical curves were found to enclose a mean convergence of
$\approx0.82$, while circles around cluster centers traced by
tangential arcs enclose a mean convergence of only
$\approx0.65$ \citep{Bartelmann95}. This systematic bias in cluster
mass estimates based on axially-symmetric models is due to the fact
that at fixed mass, the stronger gravitational tidal field of an
asymmetric mass distribution increases the extent of the critical curves.

More accurate mass estimates require detailed modelling of the
observable effects of strong lensing, either arcs or multiple
images. On galaxy cluster scales, this is commonly done via the
so-called {\em parametric} approach \citep[e.g.][]{Kneib96,Broadhurst05}. 
It consists of building up a model for the lensing cluster by combinations of
different potentials,
\begin{equation}
\psi=\sum_i\psi_i(\vec{p}) \;.
\end{equation}
Each potential is characterized by a set of parameters, defining for
instance the radial scaling of $\psi_i$, its ellipticity and
orientation, and its position. All these parameters are encoded in the
general vector $\vec{p}$ in the above formula. The parameters are
varied trying to fit the observed strong lensing features. To quantify
the goodness of the fit one defines $\chi^2$ variables, either in the
lens and/or in the source plane, to compare the observed and the
predicted properties of the strong lensing features. The best fit
model is obtained by minimizing the $\chi^2$. For example, if a set of
$n_i$ multiple images at the positions $\vec{x}_{\rm obs}^{ij}$ is
used to build the lens model, one can define the $\chi^2_i$ on the
lens plane
\begin{equation}
	\chi^2_i=\sum_{j=1}^{n_i}\frac{[\vec x_{\rm obs}^{ij}-\vec x^{ij}(\vec{p})]^2}{\sigma_{ij}^2} \;,
\end{equation}
where we have denoted with $\vec x^{ij}(\vec{p})$ the position of the
image $j$ of the multiple system $i$ predicted by the model for the
set of parameters $\vec p$, while $\sigma_{ij}$ is the associated
position error.  If $n$ sets of multiple images are available, the
contributions to the total $\chi^2$ by each set $i$ are summed:
\begin{equation}
	\chi^2=\sum_i \chi^2_i \;.
\end{equation} 
	 
Defining a $\chi^2$ on the source plane is slightly more
complicated. Let us consider the multiple image system $i$. For each
image $\vec x^{ij}_{\rm obs}$, the current lens model will predict a
source at $\vec y^{ij}(\vec p)$. If we impose that all multiple images
originate from the same source, which we assume to be located at
$\hat{\vec y}^{i}(\vec p)=1/n_i\sum_{j=1}^{n_i} \vec y^{ij}(\vec
p)$, we can define the $\chi^2$ on the source plane as
\begin{equation}
\chi^2_{S,i}=\sum_{j=1}^{n_i} \frac{[\vec y^{ij}(\vec p)-\hat{\vec{y}}^i(\vec p)]^2}{\sigma_{ij}^2\mu_{ij}^{-2}} \;.  
 \end{equation} 	 
Note that the position error on the source plane has been obtained by
correcting the position error on the lens plane by the magnification
factor $\mu_{ij}$.

In the construction of the model it is important to include the
contributions from all the relevant mass components of the lens. One
or more smooth large scale components, resembling the cluster dark
matter halo, are generally combined with smaller scale
perturbers. These are typically associated with the bright galaxies in
the cluster. Due to the complexity of the systems if each perturber
is optimized individually, the number of free parameters in the
model can rapidly saturate. To avoid this, a common practice is
to vary the strength of all perturbers simultaneously using
scaling relations \citep[e.g.][]{Jullo07}. As discussed
in \S\ref{sec:galhalo}, this can be used to also constrain the
properties of the halos of cluster galaxies.
	 
\cite{Meneghetti10} tested the robustness of parametric strong 
lensing modeling techniques using simulated observations of galaxy
clusters. They found that the mass is very well constrained around the
cluster critical lines, i.e. where the multiple image systems that are used to
optimize the model are located. The typical accuracy achieved on the
mass determination within the Einstein radius is $\sim 5\%$. 
However, the recovered mass profiles can deviate significantly
from the true profile away from the critical lines. Since these
typically extend to $\lesssim 100$ kpc, which is smaller than the
scale radius of a massive cluster-scale dark-matter halo, strong
lensing alone cannot robustly measure the concentration of the lenses,
unless it is combined with weak lensing data.

\paragraph{Flexion:}

The study of the transition from the strong lensing regime to the
small weak lensing distortions has recently been developed.  In this
regime the galaxies are no longer multiply imaged, but the variation
of the shear across the source itself cannot be ignored: the higher
order terms in the expansion of the distortion matrix become
important.  As a result the galaxies appear somewhat skewed and
"banana"-shaped \citep[e.g.,][]{Golberg02, Goldberg05, Irwin05,
Bacon06}. This can be measured and provides additional information,
compared to the shear alone \citep[e.g.,][]{Leonard07, Okura08,
Leonard10}.

Whereas the shear measures a density contrast which can be used to
determine masses, the measurement of the flexion signal is mostly
sensitive to the gradient of the density distribution. Furthermore, it
requires a substantial spatial variation of the shear
field. Consequently, the flexion signal is most suited for the study
of the density profiles and substructure in the inner regions of
galaxy clusters. It is therefore discussed in more detail
in \cite{Bartelmann12}.

\subsection{Weak gravitational lensing}
\label{sec:weaklens}

In the case of weak gravitational lensing the changes in the
observable properties of the galaxies are more subtle, and the
measurement of the lensing signal requires an ensemble of
sources. Compared to strong lensing studies, the statistical nature of
weak lensing limits the precision with which masses can be
determined. This shortcoming is, however, in part compensated by the
fact that any density fluctuation causes a distortion of the
gravitational tidal field. The resulting signal can therefore be
measured, provided the surface density of sources is high
enough. Furthermore, as is discussed in more detail
in \S\ref{sec:massrec}, it is possible to reconstruct an image of the
matter distribution.

When $\kappa$ and $\gamma$ are small, the effect of gravitational
lensing is to change the unlensed ellipticity of a galaxy,
$\epsilon_{\rm orig}$ (defined as $(a-b)/(a+b)$, where $a$ is the
semi-major axis, and $b$ the semi-minor axis) to an observed
value:

\begin{equation}
\epsilon^{\rm obs}=\frac{\epsilon^{\rm orig}+g}{1+g^*\epsilon^{\rm orig}}\approx\epsilon^{\rm orig}+\gamma,
\end{equation}

\noindent where the asterisk denotes the complex conjugate. Only if the shear 
is larger than $\epsilon^{\rm orig}$ does the observed ellipticity of
a single galaxy provide a useful estimate of the shear. A population
of intrinsically round sources would therefore be ideal. In practice
galaxies have an average intrinsic ellipticity of $\sim 0.25$ per
component \citep[e.g.,][]{Hoekstra00,Leauthaud07}, where the exact value 
depends somewhat on the galaxy type \citep[e.g.,][]{vanUitert12}.

Under the assumption that the original position angles of the galaxies
are distributed randomly \citep[see, however,
e.g.,][]{Hirata04,Mandelbaum06,Schneider10,Joachimi11}, the lensing shear can be
inferred by averaging over an ensemble of sources. The statistical
uncertainty in the measurement of the shear component $\gamma_i$ is
given by

\begin{equation}
\sigma_{\gamma,i}=\sqrt{\frac{\langle\epsilon_i^2\rangle}{N}}
 \approx\frac{0.25}{\sqrt{N}},
\end{equation}

\noindent where $N$ is the number of galaxies that is averaged to obtain 
the lensing signal. 

If we consider an isolated lens, the effect of weak lensing is a
systematic tangential alignment of the images of the background
galaxies with respect to the lens. The average tangential
distortion, defined as
\begin{equation}
\gamma_T=-(\gamma_1\cos 2\phi + \gamma_2 \sin 2\phi),
\end{equation}

\noindent can then be used to quantify the lensing signal, where $\phi$ 
is the azimuthal angle with respect to the lensing galaxy. This is a
convenient quantity, because for {\it any} mass distribution the
azimuthally averaged tangential shear $\langle\gamma_T\rangle$ as a
function of radius from the cluster center can be interpreted as a
mass contrast \citep{MiraldaEscude91}:

\begin{equation}
\langle\gamma_T\rangle(r)=\frac{\bar\Sigma(<r)-\bar\Sigma(r)}{\Sigma_{\rm crit}}=
\bar\kappa(<r)-\bar\kappa(r),
\end{equation}

\noindent where $\bar\Sigma(<r)$ is the mean surface density within
an aperture of radius $r$ and $\bar\Sigma(r)$ is the azimuthally
averaged surface density at distance $r$. In recent years it has
become common to present the lensing signal as
$\Delta\Sigma(r)=\Sigma_{\rm crit}\gamma_T(r)$, which is convenient
when redshifts for the sources and lenses are available (see
e.g. Figure~\ref{fig:Johnston07}), as it allows a direct
comparison of the amplitude of the lensing signal between
samples. Note, however, that the observed signal corresponds to the
reduced shear $g=\gamma/(1-\kappa)$.  This is not a concern when
considering fitting parametric models to the data, as the reduced
shear can be computed. Some commonly used examples are discussed
below. An important consideration for weak lensing studies, however,
is that not all clusters (if any) are well described by simple models.

A simple, but nonetheless useful, model to compare to the data is the
singular isothermal sphere (SIS) with $\rho(r)=\sigma^2/(2\pi G r^2)$,
where $\sigma$ is the line-of-sight velocity dispersion. The latter
can be compared to results from redshift surveys of cluster members,
which provide a dynamical mass estimate.  For this mass distribution
the tangential shear is given by

\begin{equation}
\gamma_T(r)=\kappa(r)=\frac{r_E}{2r},
\end{equation}

\noindent where $r_E$ is the Einstein radius, which can be expressed as

\begin{equation}
r_E=28\farcs 9\left( { \frac{\sigma}{1000~{\rm km/s}} }\right)^2\beta,
\end{equation}

\noindent where $\beta\equiv D_{ls}/D_s$ is the ratio of the angular
diameter distance between the lens and the source and the angular
diameter distance between the observer and the source, which quantifies
the dependence of the amplitude of the lensing signal on the source
redshift.

\begin{figure}[t!]
\begin{center}
\includegraphics[width=0.6\textwidth]{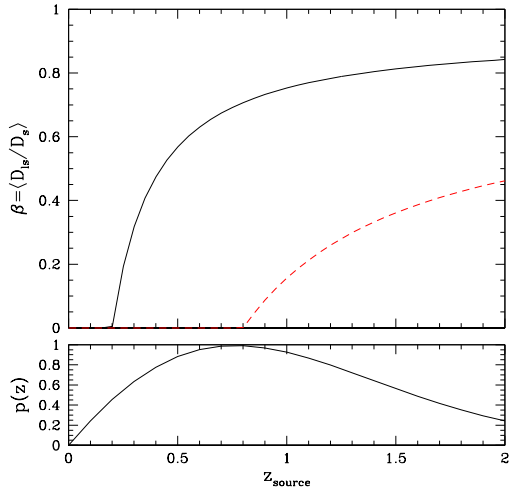}
\caption{\footnotesize The top panel shows the average value of 
$\beta=\langle D_{ls}/D_s\rangle$ as a function of source redshift for
a cluster at $z=0.2$ (black curve) and $z=0.8$ (red dashed curve).
Note that the asymptotic value is lower for clusters at higher
redshifts. The slope of the curve is a measure of the sensitivity to
errors or statistical variations in the source redshifts, which is
also larger for high redshift clusters. Furthermore, as is indicated
by the bottom panel, which shows an approximate redshift distribution
representative of deep optical observations, most of the background
sources are at redshifts where $\beta$ rises with a significant slope
for the $z=0.8$ cluster.}
\label{fig:betaz}
\end{center}
\end{figure}

For deep ground based observations the number density of sources is
$10-20$ arcmin$^{-2}$, which yields a statistical uncertainty of $\sim
1\farcs 5$ in the value for the Einstein radius, independent of the
cluster mass. The bottom panel of Figure~\ref{fig:betaz} shows an
approximate redshift distribution, representative of deep optical
observations, with a median redshift $\sim 0.9$. For a cluster at
intermediate redshift, one finds $\beta\sim 0.6$ when considering
typical ground based observations. This results in a signal-to-noise
ratio $\sim 12$ if $\sigma=1000$~km/s (which corresponds to a virial
mass of $M_{\rm vir}\sim 10^{15}M_\odot$). Hence, the masses of massive
clusters can be determined individually. To study individual lower
mass systems ($\sigma\sim 500$ km/s) it is necessary to increase the
source number density, which can only be done using deep {\it HST}
observations, because the faint galaxies are no longer resolved from
the ground.

Similarly, the study of high redshift objects (for which $\beta$ is
smaller) requires space based observations. This is demonstrated in
Figure~\ref{fig:betaz} where we plot the average value of
$\beta=\langle D_{ls}/D_s\rangle$ as a function of source redshift for
a cluster at $z=0.2$ (black curve) and $z=0.8$ (red dashed curve).  As
the amplitude of the lensing signal is proportional to $\beta$, it is
clear that the signal decreases as a function of lens
redshift. 

The slope $\partial\beta/\partial z$ quantifies the sensitivity to the
source redshift distribution, which is also larger at high redshifts.
This leads to additional noise in the mass measurements unless
redshift information for the individual sources is
obtained \citep[e.g.,][]{Hoekstra11b}. The importance of good
knowledge of the source redshift distribution is discussed further
in \S\ref{sec:redshift}.

\paragraph{NFW profile:} Numerical simulations have shown that halos over a 
wide range in mass can be described by a simple fitting
function \citep{NFW}. The density profile is characterized by two
parameters, typically the halo mass and the concentration $c$ (which
can be related to a characteristic scale). Observationally, the
concentration cannot be constrained well for individual clusters from
strong or weak lensing alone. As lensing is sensitive to the total
density profile, the contribution from baryons should be included,
although their impact should be most prominent in the cluster cores,
which are typically not included in the fit. To what extent the
profiles are changed depends on the baryon
physics \citep[e.g.,][]{Duffy10}, and the current situation is
unclear. The uncertainty in the mass estimate is enlarged further by
additional structures along the line-of-sight, as discussed
in \S\ref{sec:cosmicnoise}. Instead, it is possible to use the fact
that the parameters are found to be correlated in the simulations,
albeit with significant scatter. The NFW density profile itself is
given by:

\begin{equation}
\rho_{\rm NFW}(r)=\frac{M_{\rm vir}}{4\pi f(c)}\frac{1}{r(r+r_s)^2},
\label{eq:nfw}
\end{equation}

\noindent where $M_{\rm vir}$ is the mass enclosed within the virial 
radius $r_{\rm vir}$. The concentration parameter $c$ is defined as the 
ratio of $r_{\rm vir}$ and the scale radius $r_s$, and the function 
$f(c)=\ln(1+c)-c/(1+c)$.  Analytic expressions for the projected 
surface density and shear have been derived by \cite{Bartelmann96} and
\cite{Wright00}.

The virial mass is defined with respect to the mean density of the
Universe at the redshift of the cluster where the virial overdensity
is given by $\Delta_{\rm vir}\approx(18\pi^2+82\xi-39\xi^2)/\Omega(z)$, 
with $\xi=\Omega(z)-1$ \citep{Bryan98}. For a standard $\Lambda$CDM
cosmology $\Delta_{\rm vir}(z=0)=337$. Other definitions are also
common in the literature, with overdensities defined with respect to
the critical density $\rho_c$ at the cluster redshift:

\begin{equation}
M_\Delta=\frac{4\pi}{3}\Delta\rho_c(z) r_\Delta^3,
\end{equation}

\noindent where $r_\Delta$ corresponds to the radius where the mean 
density is $\Delta\times \rho_c(z)$, and $M_\Delta$ is the
corresponding enclosed mass. Note that $M_{200}$ is often referred to
as the virial mass.  Although similar in value, its definition is in
fact different. Given one has to choose a definition of the mass, it
is important that this is clearly defined when listing results.

\paragraph{Aperture mass:} An important advantage of  weak lensing is that 
the shear can be converted into an estimate of the projected mass within
an aperture, with relatively few assumptions about the actual mass
distribution. For instance the $\zeta_c$ estimator proposed
by \cite{Clowe98}:

\begin{equation}
\zeta_c(r_1)=2\int_{r_1}^{r_2}d\ln r\langle\gamma_t\rangle+
\frac{2r_{\rm max}^2}{r_{\rm max}^2-r_2^2} \int_{r_2}^{r_{\rm max}}
d\ln r \langle\gamma_t\rangle,
\end{equation}

\noindent can be expressed in terms of the mean dimensionless surface density 
interior to $r_1$ relative to the mean surface density in an annulus
from $r_2$ to $r_{\rm max}$:

\begin{equation}
\zeta_c(r_1)=\bar\kappa(r'<r_1)-\bar\kappa(r_2<r'<r_{\rm max}).
\end{equation}

This result demonstrates that we can determine the surface density
within an aperture up to a constant, as was the case for the
2-dimensional reconstruction discussed in \S\ref{sec:gravlens}.
Although the surface density in the outer annulus should be small for
results based on wide field imaging data, it cannot be ignored. One
way to estimate its value, is to use the results from a parametric model
that is fit to the data \citep{Hoekstra07}. This leads to a rather
weak model dependence ($<10\%$) in the estimate of the projected mass.

\begin{figure}[t!]
\begin{center}
\includegraphics[width=0.6\textwidth]{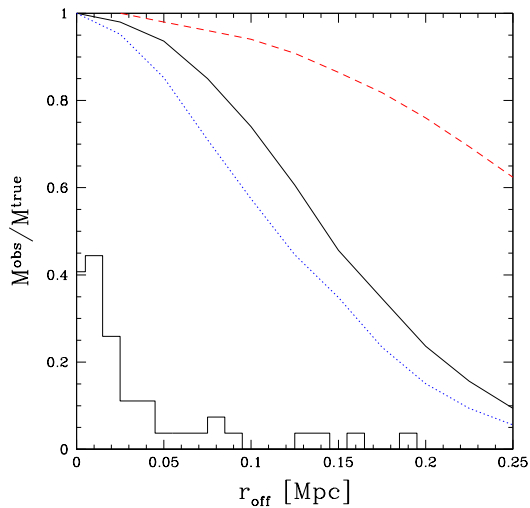}
\caption{\footnotesize From \cite{Hoekstra11a}: plot of the ratio
of the inferred lensing mass and true mass as a function of centroid
offset $r_{\rm off}$. The lensing mass is obtained by fitting an NFW model to the
shear at $200-750$~kpc (solid black curve) and $0.5-1.5$Mpc (dashed red
curve).  The blue dotted line corresponds to the bias if all data
within 750 kpc are used. The distribution of offsets observed for
massive clusters by \cite{Bildfell08} is indicated by the histogram.}
\label{biasoffset}
\end{center}
\end{figure}

\paragraph{Choice of cluster center:}
\label{sec:center}

The choice of the cluster center is important when deriving the
lensing mass. If the adopted center is offset from the correct value,
the contrast is lowered, resulting in a reduction in the observed
signal and consequently the lensing mass, as shown in
Figure~\ref{biasoffset}. The size of the bias depends on the range in
scales used to infer the lensing mass: if the lensing signal at
relatively large radii is used, the bias is diminished. Massive
clusters typically have well defined central galaxies, whose locations
coincide with the peak of the X-ray emission \citep{Bildfell08}. Note,
however, that in merging clusters of galaxies, such as the ``Bullet
Cluster'', the peak of the X-ray emission may not coincide with the main
cluster halos \citep{Clowe96}.

The reconstruction of the projected mass distribution
(see \S\ref{sec:massrec}) can in principle be used to select the
cluster center. However, even for the most massive clusters, the
signal-to-noise ratio is only of order 10, which means that one has to
be careful to use the peak position itself, as it will bias the mass
estimate high. It is possible to account for this when studying an
ensemble of clusters, but it is preferable to use external constraints
instead. Weak lensing cluster studies therefore typically adopt the
location of the brightest cluster galaxy or the peak in the X-ray
emission as the center.  The situation is more complicated when
considering large samples of clusters, for which high quality X-ray
observations may not be available or cannot be obtained because the
masses of the targets are low. When considering these large samples of
clusters the ``handpicking'' of the center is therefore not practical,
nor preferable. Instead the effects of miscentering can be taken into
account statistically \citep{Johnston07a}.

The presence of substructure in the central regions not only
complicates the choice of center, but it also leads to biases in the
mass estimates when parametric models are fit to the signal. This is
because the substructure also lowers the mean density contrast, and thus
the shear, compared to a unimodal density distribution
\citep[see e.g.,][]{Hoekstra02a}. 

\subsubsection{Cosmic noise}
\label{sec:cosmicnoise}

Clusters are integral parts of the cosmic web and are located at the
intersections of filaments. Therefore some of this large-scale
structure is physically associated with the cluster, which complicates
the interpretation of the lensing results, in particular when one
wants to compare to other mass indicators (see \S\ref{sec:mass}) or
determine density profiles
\citep{Meneghetti12}. The correlated large-scale structure has been studied 
using numerical simulations
\citep[e.g.][]{Metzler01, Marian10, Becker11, Bahe12}. 

As discussed in \cite{Becker11} and \cite{Rasia12}, the bias that is
introduced depends on how the mass is inferred from the lensing
signal. For instance, restricting the model fit to $\sim R_{200}$ can
reduce the bias, compared to fits to the signal at larger radii. Part
of this reduction arises from the fact that at large radii the NFW
model is not a good description of the signal, for instance due to the
presence of substructure and the fact that clusters themselves are
clustered. Another approach is to identify additional structures and
fit several halos simultaneously, although \cite{Hoekstra11b} show that
this of limited use, due to the inherent variation in the density
profiles.

For ensemble averaged studies the correlated structure can be
accounted for using a halo model approach 
\citep[e.g.][]{Johnston07b,Leauthaud10} (also
see \S\ref{sec:halomodel}). The results also depend on whether one
adopts a functional form for the relation between the mass and
concentration, or whether both are fit simultaneously. This clearly
complicates an easy comparison of results, especially when the
increasing sample sizes lead to smaller statistical
uncertainties. Using numerical simulations \cite{Becker11}, however,
show that the correlated structures are not a major source of
uncertainty and that the main source of scatter in scaling relations
between lensing mass and other indicators arises from the intrinsic
triaxiality of dark matter halos \citep[also see, e.g.,][]{Clowe04,
Corless07,Limousin12}.

Gravitational lensing, however, is sensitive to {\it all} structure
along the line-of-sight, and thus uncorrelated inhomogeneities along a
given line-of-sight also contribute to the lensing signal in that
direction. The large scale structure (LSS) introduces excess
correlations in the shapes of the galaxies, also known as `cosmic
shear', which can be used to infer cosmological parameters or study
theories of modified gravity \citep[see e.g.,][for a recent
review]{Hoekstra08}.  However, for weak lensing cluster studies this
excess variance acts as an additional source of noise, which was first
studied by \cite{Hoekstra01a}. It arises because we cannot distinguish
the cluster signal ($M_{\rm cl}$) from the LSS contributions ($M_{\rm
LSS}$).  Note that the latter contribution vanishes on average, and
thus does not bias the mass measurements.
 
The noise introduced by the large-scale structure can be expressed in
terms of the matter power spectrum and a scale dependent filter, similar
to what is done for cosmic shear \citep{Hoekstra01a}:

\begin{equation}
\langle M_{\rm LSS}^2\rangle(\theta)=2\pi\int{\rm d}l P_\kappa(l)g(l,\theta)^2,
\end{equation}

\noindent where $P_\kappa(l)$ is the projected convergence power spectrum 
and the expression for the filter function $g(l,\theta)$ depends on
the adopted statistic. A detailed discussion, including expressions for
$g(l,\theta)$, can be found in \cite{Hoekstra01a}.

\begin{figure}[t!]
\begin{center}
\hbox{%
\includegraphics[width=0.5\textwidth]{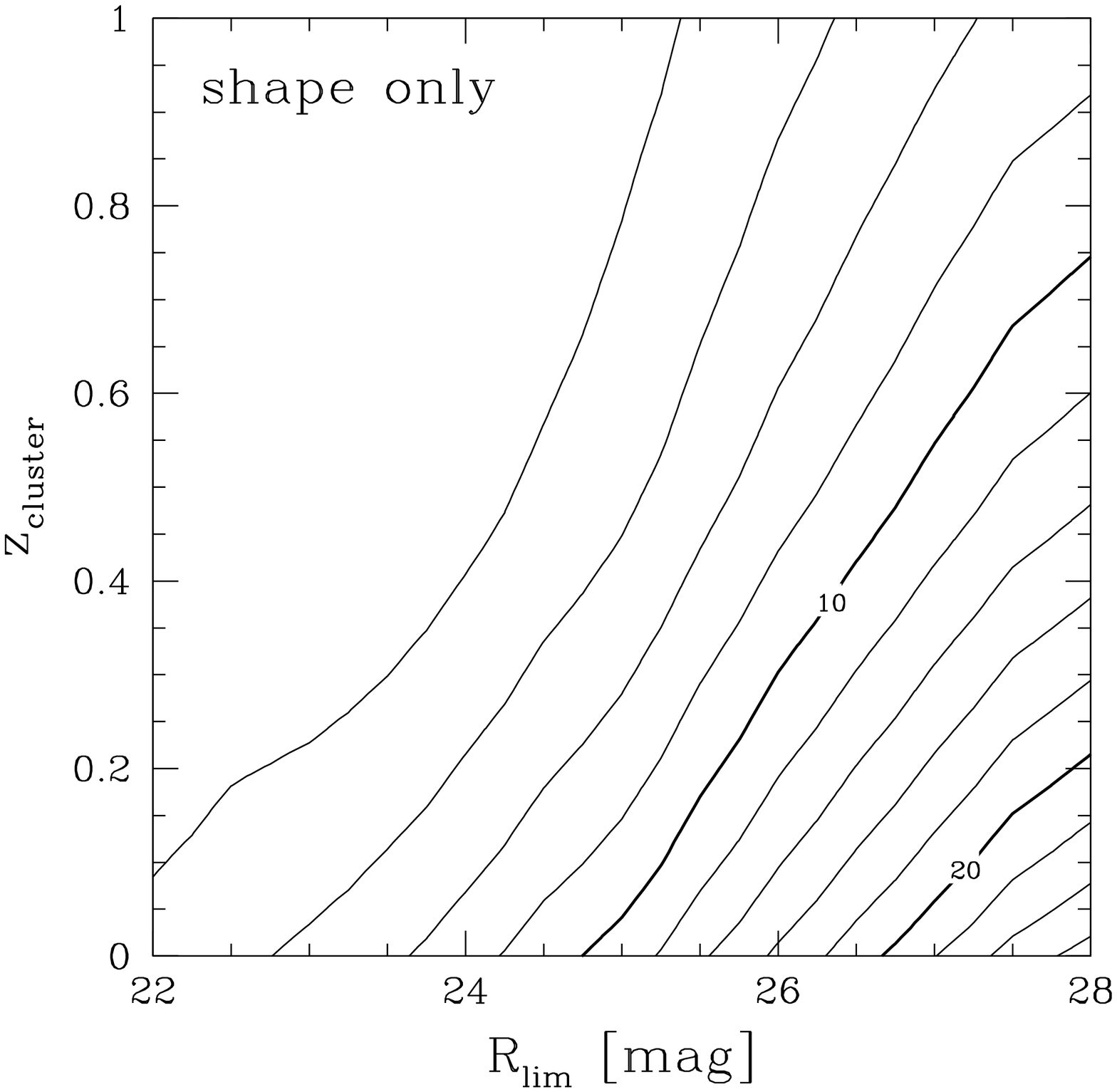}
\includegraphics[width=0.5\textwidth]{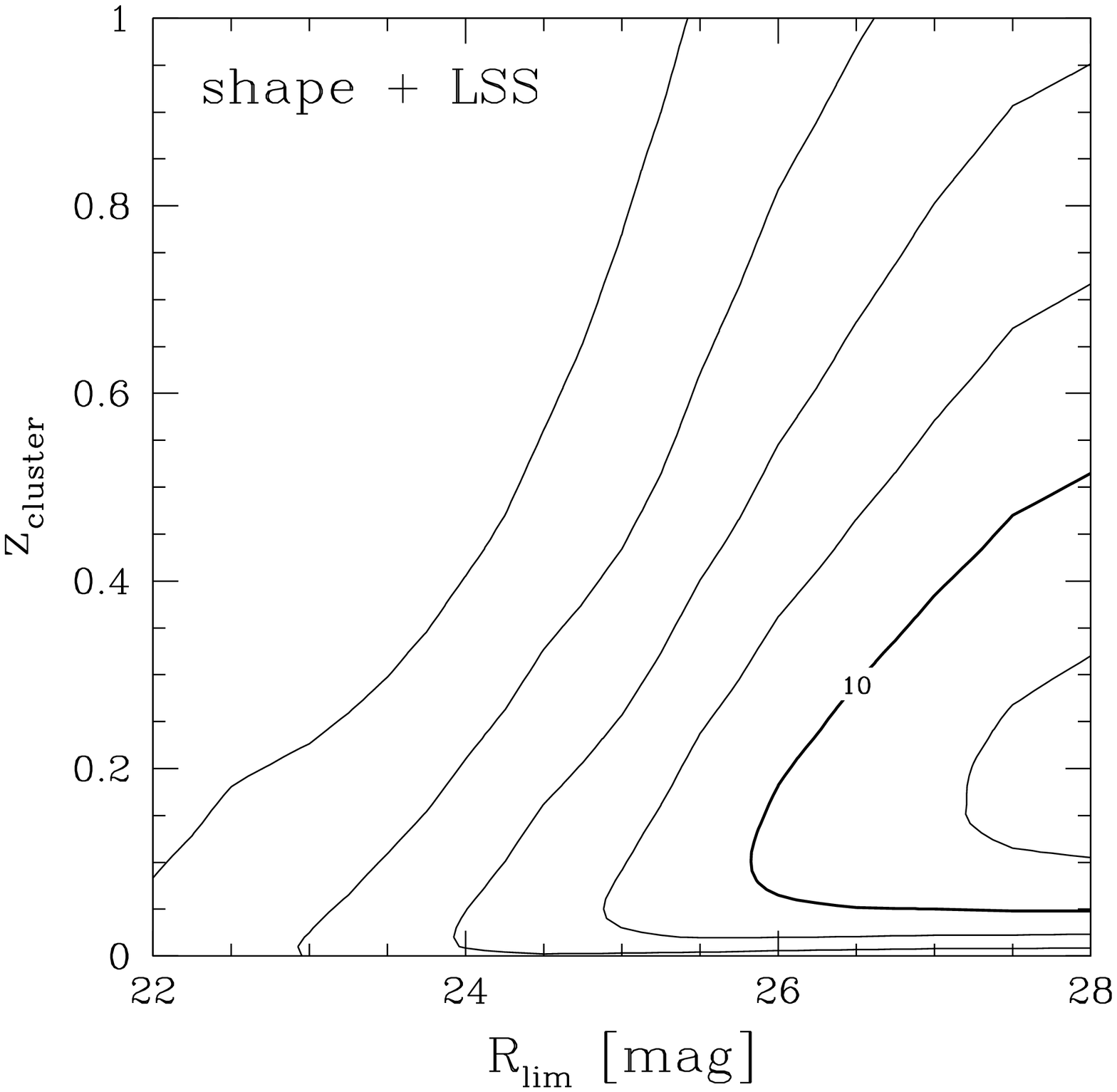}}
\caption{\footnotesize Adapted from \cite{Hoekstra01a}; {\it Left panel:}
contour plot of the signal-to-noise ratio of a cluster with a SIS
density profile and $\sigma=1000$km/s, as a function of limiting
magnitude in the $R$-band and cluster redshift, when only the shape
noise is considered. {\it Right panel:} The expected signal-to-noise
ratio when the effects of cosmic noise are included. At high redshifts
the uncertainty in the weak lensing mass remains dominated by the
intrinsic ellipticities of galaxies, but at low redshift cosmic noise
becomes an important source of uncertainty.}
\label{cosmicnoise}
\end{center}
\end{figure}

\cite{Hoekstra01a} found that the impact of cosmic noise is most important for 
clusters at low redshift (see the comparison presented in
Fig.~\ref{cosmicnoise}) and if one wants to determine the mass at
large radius. This work was extended in \cite{Hoekstra03a} who studied
how cosmic noise limits the precision of the determination of mass
density profiles from weak lensing. For instance, when an NFW model is
fit with mass and concentration as free parameters, the formal
uncertainties in both the mass and the concentration can be doubled,
compared to the simple statistical uncertainty from the noise of the
shapes of the sources.  The analytical results obtained in these
papers have been confirmed using numerical
simulations \citep{Hoekstra11b,Becker11}.

An interesting question is whether it is possible to reduce or even
remove the effects of cosmic noise. \cite{Dodelson04} proposed a
statistical approach which reduces the noise in mass reconstructions,
although this might also suppress real structures in individual
cases. The main contributors to the cosmic noise signal, however, are
clusters and groups of galaxies that can in principle be
identified. This was first investigated by \cite{dePutter05} who
showed that one would have to include halos down to very low masses
($<5\times 10^{13}\msun$). More importantly the variation in halo
profiles and shapes fundamentally limits the precision with which the
cosmic noise signal can be predicted. This was investigated in more
detail in \cite{Hoekstra11b} who showed that the precision of the mass
measurement can be improved only slightly in practice.

A number of studies have now demonstrated that uncorrelated
large-scale structure is a significant source of uncertainty for weak
lensing mass estimates. It is particularly important for studies that
aim to constrain the density profile, as the formal uncertainties on
the model parameters are increased substantially. It is therefore
important that weak lensing studies of clusters of galaxies take
cosmic noise into account.

\subsubsection{Cluster-mass cross-correlation function}
\label{sec:halomodel}

The limited number of background galaxies restricts the study of
individual clusters to virial masses that are a few times
$10^{14}\msun$ or larger. It is, however, possible to study lower mass
systems by averaging, or 'stacking' their signals: the resulting
signal is that of the mass-weighted ensemble. This approach is widely
used, and an important application of weak lensing is the study of the
properties of dark matter halos around galaxies
\citep[e.g.][]{Brainerd96, Hoekstra04, Mandelbaum05, vanUitert11}. 
It has also been employed in studies of galaxy groups discovered in
the CNOC2 field redshift survey \citep{Hoekstra01b,Parker05} and
COSMOS \citep{Leauthaud10}. In the case of a shallow, but wide survey,
stacking the signal around the large number of lenses can be used to
compensate for the low number density. This approach has been applied
with great success to the SDSS data to measure the ensemble averaged
lensing signal around galaxies \citep[e.g.,][]{Fischer00, Guzik02,
Mandelbaum05}, groups and
clusters \citep[e.g.,][]{Mandelbaum06,Sheldon09}.

\begin{figure*}
\begin{center}
 \includegraphics[width=\textwidth]{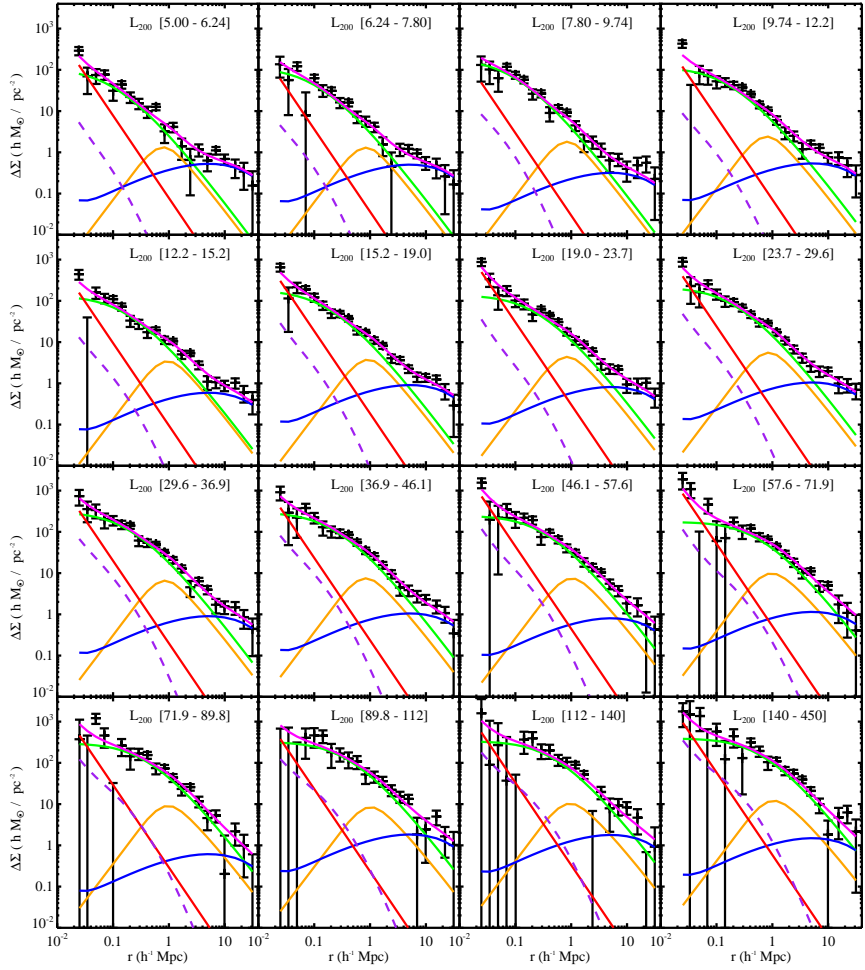}
\end{center}
\caption{Halo model fits to the $\Delta\Sigma$ profiles of SDSS MaxBCG clusters.
For each bin in total $i$-band luminosity, a model (magenta) is fitted,
consisting of these components: BCG (red), miscentering correction (orange),
NFW main halo (green), halo clustering (blue), and non-linear component 
(purple dashed). Figure from \citet{Johnston07b}.}
\label{fig:Johnston07}
\end{figure*}

By dividing the sample of clusters into subsets based on an observable
property, such as X-ray luminosity or optical richness, it is possible
to determine scaling relations between the observable and the mean
mass of the
sample \citep[e.g.,][]{Johnston07b,Rykoff08,Leauthaud10}. It is
important to note here that the interpretation of the result requires
knowledge of the intrinsic scatter between the mass and the quantity
of interest. Only for high masses the latter can be determined
observationally. For low masses one relies on numerical simulations to
provide guidance.

Figure~\ref{fig:Johnston07} shows the lensing signal as a function of
luminosity $L_{200}$ around clusters from the SDSS MaxBCG
sample \citep{Koester07}, which contains 13,823 $0.1<z<0.3$ clusters
identied in 7500 deg$^2$ of SDSS data. The lensing signal was
measured by \cite{Sheldon09} using a source catalogue of
$2.8\times\,10^{6}$ galaxies for which SDSS lensing-quality data and
photometric redshifts are available. A clear lensing signal is
detected in all luminosity bins.

The ensemble averaged signal corresponds to the cluster-mass
cross-correlation function. This reflects that only on small scales
the signal arising from the main cluster halo dominates, whereas at
large radii it is sensitive to the clustering of clusters, which in
fact provides additional information. It is possible to compute the
combined signal using a (semi-)analytical approach suggested
by \cite{Seljak00}: the halo-model. It uses the fact that individual
dark matter halos are well described by the NFW profile, and that we
can compute the halo mass function and the clustering of halos for a
given cosmology. Much ongoing work focusses on ways to improve the
shortcomings of the halo-model.

Here we briefly review the main components of the model that are used
to model the lensing signal around clusters of galaxies \citep[see
e.g.][for more details]{Johnston07b}. On small scales the lensing
signal is computed from the NFW model, which is convolved with a
probability density distribution of offsets from the adopted center.
As discussed in \S\ref{sec:center} this can be determined fairly
directly for massive clusters, but for large samples of (optically
selected) clusters a distribution of offsets needs to be adopted, for
instance based on numerical simulations. This first contribution
describes the cluster dark matter halo (and implicitly assumes that
the ICM follows a similar distribution; see \cite{Semboloni11} for a
study where this assumption is relaxed). The signal of the BCG is also
added, and can be approximated by a point mass.

On large scales the fact that clusters themselves are part of the
cosmic web and thus are clustered causes an additional signal. This
can be seen clearly in Figure~\ref{fig:Johnston07} for the lower mass
clusters (top-left corner): the lensing signal shows a clear excess
compared to the NFW signal (green curve). The signal from these
neighbouring halos is described by the two-halo term \citep{Seljak00}
and is indicated by the blue curves. This term requires the
calculation of the correlation function in linear perturbation theory
and a prescription for the linear bias as a function of mass and
redshift. Comparison of the predicted amplitude to the observed value
can be used to test models of structure formation or to constrain the
normalisation of the matter power spectrum \citep[see \S5.3
in][]{Johnston07b}.

\subsubsection{Magnification}
\label{sec:magnification}

The measurement of the magnification of galaxies provides a
complementary way to study the mass distribution. Especially in the
case of observations in multiple filters, the data used for the shear
analysis allow for a magnification measurement as well.  The actual
magnification cannot be measured for a single object because the
intrinsic flux is typically unknown. Instead, the signal can be
inferred from the change in the source number counts. Such a change
arises from the balance between two competing effects. On the one hand
the actual volume that is surveyed is reduced, because the solid angle
behind the cluster is enlarged. However, the fluxes of the sources in
this smaller volume are boosted, thus increasing the limiting
magnitude. As a consequence, the net change in source surface density
depends not only on the mass of the lens, but also on the steepness of
the intrinsic luminosity function of the sources.  If it is steep, the
increase in limiting magnitude wins over the reduction in solid angle,
and an excess of sources is observed. If the number counts instead are
shallow, a reduction in the source number density is observed.

A correct interpretation of the result only requires accurate
photometry and knowledge of the (unlensed) luminosity function.
Therefore the requirements on the PSF are much less stringent compared
to the shear-based approach. The main systematics in this case are the
uniformity of the photometry and completeness of the source sample (to
establish the slope of the luminosity function). However,
contamination of the source population by cluster members needs to be
avoided as well. This is a larger issue for magnification studies,
compared to measurements using shear. As discussed
in \cite{Hildebrandt09} Lyman Break Galaxies (LBGs) can provide such a
clean sample of high redshift sources, although their number density
is relatively low.

The uncertainty in the measurement of magnification is determined by
variations in the number density (i.e., a combination of Poisson noise
and the clustering of the sources). As a result, the signal-to-noise
ratio per source galaxy is typically lower, compared to a shear
measurement. It is difficult to measure reliable shapes for distant
sources using typical ground based data, because they cannot be
resolved. Their magnitudes, however, can still be measured, and thus
also the magnification signal. Hence, the signal-to-noise of the
magnification signal can be higher than the shear signal in the case
of high redshift objects. The low number density of LBGs prevents
precise measurements for individual systems, but it has recently been
used to constrain masses of a sample of $z\sim1$ clusters
by \cite{Hildebrandt11} for the first time using this approach.
The same technique was used by \cite{Ford12} to study galaxy groups.

If photometric redshifts are available, it should be possible to
increase the number density of sources that can be used, although the
performance of photometric redshift codes in the presence of a large
number of cluster members needs to be investigated in detail. Another
approach is to consider the number density of galaxies redder than the
cluster red-sequence as proposed by \cite{Broadhurst95}. This may be
better suited for ensembles of clusters as well, as the stronger
clustering of red galaxies increases the noise in the measurement. For
a sample of massive clusters
\cite{Umetsu11} obtained good agreement between the magnification signal and 
that inferred from the shear analysis.

\cite{Schmidt12} used {\it HST} imaging of the COSMOS field to measure the 
magnification signal by combining the magnitudes (and redshifts) with
estimates of the sizes of the source galaxies. The signal-to-noise
ratio is 40\% of what is achieved from shear measurement. Although
this approach requires excellent imaging quality data, it comes at no
additional cost.  Finally \cite{Bauer12} used a sample of variable
quasars from the SDSS. In this case the magnification signal is
determined by comparing the luminosity to the one expected based on
the correlation between the amplitude of variability and the intrinsic
luminosity. These recent results highlight the growing interest in
using magnification as an additional way to determine cluster masses.

\section{Measuring the weak lensing signal}

Having introduced the observables and discussed how these can be used
to map the matter distribution and determine cluster masses, we now
turn to the practical aspects of the measurements. Although cluster
lensing studies share many common issues with other applications, such
as galaxy-galaxy lensing and cosmic shear, there are also a number of
issues that are particularly important for cluster lensing. 

Starting with the measurement of the galaxy shapes, discussed in more
detail in \S\ref{sec:shapemeasurement}, the shear in the inner regions of
clusters can be substantial, to a level where the biases in the shape
measurement algorithms are not (yet) well understood. However, even a
perfect measurement of the shapes of galaxies does not imply an
unbiased mass estimate. This is because the interpretation of the
lensing signal requires estimates for the level of contamination by
cluster members and knowledge of the source redshift distribution,
both of which are detailed in \S\ref{sec:redshift}.

\subsection{Shape measurements}
\label{sec:shapemeasurement}

The alignment in the shapes of the sources induced by weak lensing is
subtle and is often dwarfed by the anisotropy of the PSF, which
introduces an additive bias. The PSF also circularizes the images,
leading to a reduction of the amplitude of the lensing signal, i.e. a
multiplicative bias. Although this bias can be reduced using space
based observations, the PSF circularization can never be ignored. The
resulting images are sampled onto the pixels of the detector, which
itself can introduce problems: charge transfer inefficiency is an
important systematic in space based observations, where the background
is low \citep[see][for a detailed discussion of how the various
systematic biases affect weak lensing studies]{Massey12}. Finally the
images of the faint galaxies that carry most of the signal are noisy,
which complicates matter further \citep[e.g..][]{Melchior12, Refregier12}.

These effects need to be corrected for if we want to determine
accurate cluster masses from weak gravitational lensing. This not only
requires an algorithm that can undo the bias caused by the PSF, but
also an accurate model for the PSF itself \citep{Hoekstra04}. The
model is typically determined from a sample of moderately bright stars
that are identified in the actual data. Depending on the correction
scheme, the images of these stars are either modeled or shape
parameters are determined.  To describe the spatial variation of the
PSF the model is interpolated, typically using a low order polynomial
function of the coordinates. Much of the recent cluster studies use
data from cameras that cover the focal plane with a mosaic of
detectors.  The gaps between the detectors are filled by dithering the
observations. This can lead to additional structure in the PSF, where
the impact depends on the observing strategy. Alternatively, one can
homogenize the PSF before the images are combined, which can also
improve the fidelity of photometric redshift
measurements \citep{Hildebrandt12}. Another approach is to develop
techniques that can operate on unstacked images \citep{Miller12}.

Fortunately, in comparison to cosmic shear studies, the uncertainties
in the PSF model are not an important source of systematic when
determining cluster masses. This is because the mass estimate involves
integrating over the tangential shear as a function of radius, and
local deviations are suppressed in the averaging. Consequently the
cluster mass measurements are also relatively insensitive to
limitations of the algorithms to correct for PSF anisotropy, although
the latter can be important when considering mass reconstructions.

Once the spatial variation of the PSF has been modelled, the next step
is to measure the galaxy shapes and correct them for the various
observational distortions. This is the most critical part of the
analysis, and improving the correction algorithms continues to be an
active area of research. It is therefore not surprising that a range
of algorithms has been developed to tackle this problem. An overview
of a number of these can be found in \cite{STEP1}, \cite{STEP2},
\cite{GREAT08} and \cite{GREAT10}.

\begin{figure}[t!]
\begin{center}
\includegraphics{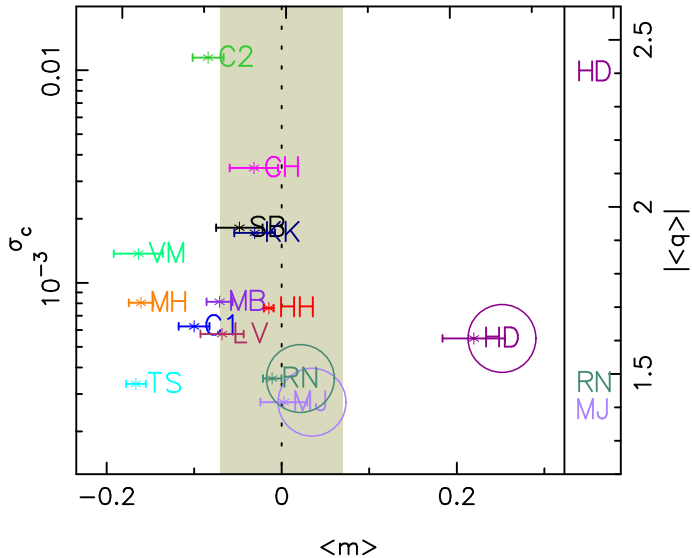}
\caption{\footnotesize Measurement of the calibration bias $m$ and
PSF residuals $\sigma_c$ from \cite{STEP1}. The ideal method
has $m=0$ and small $\sigma_c$. The shaded region indicates
a bias of less than 7\%.  Methods that were used for 
published cosmic shear results were found to have biases of
the order of a few percent. We refer the reader to \cite{STEP1}
for a detailed description of the symbols and methodology.}
\label{fig:step}
\end{center}
\end{figure}

Shape measurement methods can be roughly separated into two
classes. One approach is to measure the weighted moments of the
images, and correct these analytically for the effects of the PSF. The
KSB algorithm proposed by \cite{KSB} falls into this
category \citep[for modifications to the original algorithm
see][]{LK97,Hoekstra98}.  Despite a number of shortcomings, it remains
one of the most widely used methods. A nice feature is that the
corrections for PSF anisotropy and circularization are separate
operations. The algorithm is, however, limited by the assumptions made
about the PSF and galaxy profiles. 

A key assumption in the derivation is that the PSF can be described as
a convolution of a compact anisotropic kernel and a large isotropic
kernel. This may be a reasonable approximation for ground-based data,
but it is less so for diffraction-limited observations from space. An open
question is whether this can be circumvented by PSF homogenisation
with a kernel that ensures that the resulting PSF is
Gaussian \citep{Kuijken08}. More general algorithms have also been
proposed by \cite{Melchior11} and \cite{Bernstein10}, and work is
continuing to improve the algorithms.

The other class of methods attempts to fit the observed images with
sufficiently versatile models. The basis functions are often chosen
such that the (de)convolution can be performed analytically. For
instance \cite{Kuijken99} explored the possibility of modeling
galaxies as sums of Gaussians. Another basis set was proposed
by \cite{Refregier03}. This shapelet expansion has proven useful to
model the PSF of real data \citep{Kuijken08,Hildebrandt12}, but has
not been used widely for actual lensing measurements. Finally we
mention lensfit \citep{Miller07,Miller12}, which is a Bayesian
algorithm that fits galaxies with a bulge+disk model and has been used
recently to analyse data from the CFHT Legacy Survey \citep{Heymans12}.

\paragraph{Testing algorithm performance:} An important advantage of 
weak lensing is that the algorithms can be tested using realistic
simulations of the actual data \citep[see e.g.,][for
early tests]{Hoekstra98,Erben01,Hoekstra02b}.  To compare how well
methods perform, it is conventient to consider the bias in the shape
measurement as the combination of a multiplicative bias $m$ and an
additive bias $c$

\begin{equation}
\gamma_{\rm obs}=(1+m)\gamma_{\rm true}+c.
\end{equation}

\noindent This parametrization was used by the Shear TEsting Programme 
\citep[STEP;][]{STEP1}, which was the first collaborative effort of the
weak lensing community to test the performance of the various
algorithms. The first study presented in \cite{STEP1} involved the
blind analysis of simulated galaxies with relatively simple
morphologies. In \cite{STEP2} more complicated galaxies were examined.
The results of \cite{STEP1} are reproduced in
Figure~\ref{fig:step}. These studies showed that the dominant source of
bias is the correction for the size of the PSF, quantified by $m$,
although the most successful methods were able to achieve $1-2\%$
accuracy. 

Such biases may seem small, but the limitations become dominant for
cluster samples that exceed $\sim 100$ massive clusters. We also note
that these studies did not test the high shear regime. However, in the
case of aperture masses or when the model fits are restricted to
larger radii, this is not a serious problem.

\cite{STEP1} and \cite{STEP2} provided an important benchmark for
the performance of ground-based weak lensing studies, but also clearly
indicated that further progress is needed. For instance biases as a
function of object size and magnitude (or signal-to-noise ratio) have
been identied. To stimulate new ideas to solve the problem of shape
measurement for future projects, the Gravitational LEnsing Accuracy
Testing 2008 (GREAT08) Challenge \citep{GREAT08} aimed to involve
researchers outside the traditional weak lensing community. This early
challenge was followed by GREAT10, which improved the comparison
methodology. The results presented in \cite{GREAT10} showed a factor
of 3 improvement compared to the earlier studies, with the best
methods now achieving sub-percent biases on average. Further progress
is expected in coming years, in preparation for ambitious dark energy
projects such as {\it Euclid}\footnote{http://www.euclid-ec.org} \citep{Euclid}.

\subsection{The importance of redshift information}
\label{sec:redshift}

To relate the lensing signal to physical quantities, such as the mass,
we need to determine the value for the critical surface density
$\Sigma_{\rm crit}$ (Eqn.~\ref{eq:sigmacrit}), which requires knowledge of
the redshifts of the faint source galaxies. Furthermore, the sample of
galaxies used to compute the lensing signal may contain cluster
members which dilute the signal. In this section we review these
two important issues.

\subsubsection{Source redshifts}
\label{sec:zsource}

The faintest galaxies that are suitable for weak gravitational lensing
measurements are typically much fainter than galaxies targeted
efficiently in spectroscopic redshift measurements. Even if some
redshifts could be obtained, the resulting sample would be highly
incomplete at the magnitudes of interest. Hence, it is generally
unfeasible (because of the enormous cost in telescope time) to
determine the redshifts of a significant fraction of the weakly lensed
sources.

Fortunately, the redshifts do not need to be determined with high
precision, and photometric redshift estimates are sufficient. As
discussed in \S\ref{sec:weaklens} and shown in Figure~\ref{fig:betaz},
for a cluster at a known distance $D_{l}$, the relevant quantity is
$\beta \equiv D_{ls}/D_s$: the derived lens mass will scale linearly
with $\beta$, and any systematic error in the determination of this
parameter will translate directly into a corresponding bias in the
mass. Note that the inclusion of unlensed foreground galaxies is not a
problem {\it per se} as long as the foreground contribution in the
source galaxy catalogue can be modeled and included in the estimate of
$\langle \beta \rangle = \langle {\rm max} (0, D_{ls}/D_s) \rangle$.

Typically, the value of $\beta$ is determined by relating the source
galaxy catalogues to photometric redshift catalogues in the
literature, e.g. from the COSMOS survey \citep{Ilbert09} or the
Hubble Ultra Deep Field \citep{Coe06} or the CFHTLS Deep fields 
\citep{Ilbert06}, while taking into account the weights assigned to 
the background galaxies of different magnitudes when making the weak
shear measurement. Hence it is not necessary to determine photometric
redshifts of the galaxies in the actual cluster observations, although
the benefits of such ancillary information are discussed below.  In
cases where the imaging data for the lensing measurements go even
deeper than the photometric redshift data sets, extrapolation will be
required, typically using a parametrization of the galaxy redshift
distribution as a function of magnitude \citep[see
e.g.,][]{Schrabback10}. As such galaxies are often at high redshifts,
the lack of deep near-IR data provides another challenge, although
the situation keeps improving.

The required accuracy is naturally determined by the scientific goals
of a cluster study. If the goal is to determine lensing masses for a
small number of individual clusters, it is typically sufficient to
determine an effective mean value of $\beta$. Generally the source
galaxy population will follow a broad distribution in redshift.
However, it is possible to define an effective source galaxy redshift
such that $D_{ls}/D_{s} (z_{\rm eff}) = \langle \beta \rangle$ and
proceed as if all background galaxies are located at this redshift.
Note that it is important to correct this value for the contamination
by cluster galaxies (see \S\ref{sec:contam}). For statistical studies
based on larger samples of clusters (e.g., scaling relations), care
must be taken to avoid systematic effects caused by inadequate
redshift information for the source galaxy population.

The effective source redshift can be used for low redshift clusters
$(z<0.3)$, where the estimate for $\beta$ is robust and systematic
errors are negligible compared to the statistical errors. For high
redshift clusters $(z>0.6)$ the slope $\partial\beta/\partial z$ is
steeper (see Figure~\ref{fig:betaz}), which implies a larger
sensitivity to systematic errors in the source redshifts, but also a
larger sensitivity to the width of the redshift distribution. As
discussed by \cite{Seitz97} and \cite{Hoekstra00} this will introduce
an overestimate of the shear of order

\begin{equation}
1 + \left[ \frac{\langle \beta^2 \rangle}{\langle \beta \rangle^2} -1 \right] 
\kappa . 
\end{equation} 

\noindent This systematic error will be small for clusters at low redshift 
and is frequently ignored in the literature, but it can be relevant
for clusters at high redshift, in particular for lensing-based studies
of matter density profiles which use measurements at radii where the
convergence $\kappa$ is not $\ll 1$.

Although it is possible to interpret the results of large samples of
clusters with an average redshift distribution, the actual redshift
distribution of sources will vary. At large radii, where the signal is
averaged over many galaxies on relatively large scales, the source
redshift distribution is expected to be close to the average. At small
radii this is not a good representation, because the small number of
sources samples the average distribution only sparsely. This leads to
additional noise if photometric redshifts for the sources are not
available. At low redshift where variations in the redshift
distribution lead to only small variations in $\langle\beta\rangle$,
this is still a minor source of error. This is no longer the case at high
redshift, where the number density of background galaxies is low and
$\langle\beta\rangle$ depends strongly on the actual redshift
distribution. 

This was studied in \cite{Hoekstra11b} who showed that the lack of
redshift information for the individual sources leads to an additional
uncertainty in the shear estimate on scales $<4'$. As expected, the
scatter increases with redshift and becomes a relevant source of noise
for high redshift clusters $z>0.6$. Hence, studies of high redshift
clusters benefit greatly from photometric redshifts for the sources.
This not only reduces the noise due to variations from the average
redshift distribution, but also boosts the signal by removing
foreground (and cluster) galaxies.

\begin{figure}[t!]
\begin{center}
  \includegraphics[width=0.75\textwidth,bb = 1 212 590 630]{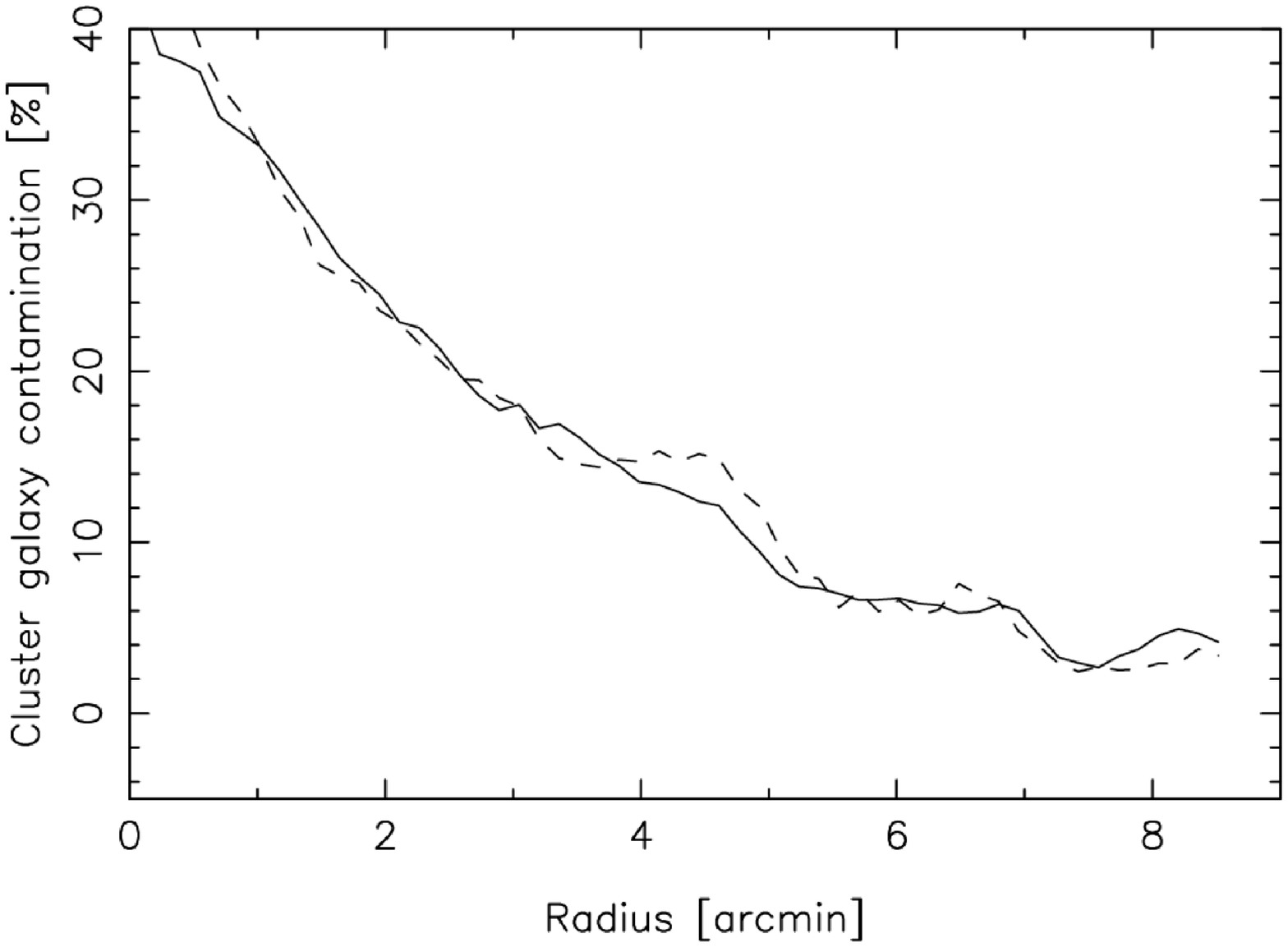}
\caption{\footnotesize Illustration of typical observed levels of cluster 
galaxy contamination in faint galaxy catalogs as a function of
distance from the cluster center.  The solid line represents an
average of six X-ray luminous ($L_{X, 0.1-2.4 {\rm keV}} \geq 6 \times
10^{44}\, {\rm ergs\, s}^{-1}$) clusters at an average redshift
$\langle z \rangle = 0.31$, while the dashed line represents an
average of five similarly X-ray luminous clusters at $\langle
z \rangle = 0.23$. Figure taken from \cite{Pedersen07}.}
\label{fig:background}  
\end{center}
\end{figure}

\subsubsection{Contamination by cluster galaxies}
\label{sec:contam}

Without redshift information for individual galaxies, one encounters
the problem of contamination by (unlensed) cluster galaxies which will
dilute the observed shear signal and cause an underestimate of the
cluster mass. Since the faint galaxies are clustered (though not
necessarily with the same radial distribution as their brighter
counterparts), the amount of such dilution will be a function of
radius. As shown in Figure~\ref{fig:background} the level of
contamination ranges from tens of percent in the innermost parts of
the cluster, while it is essentially negligible at large radii.  If
left unaccounted for, the resulting reduction in the lensing signal at
small radii may be misinterpreted as a flattening of the density
profile. Hence a careful treatment of the contamination by cluster
members is particularly important when studying the density 
profiles of clusters using weak lensing (see \S\ref{sec:profiles}).

The reliability of the separation of background and cluster galaxies
generally increases with the number of passbands available, with full
photometric redshift information for the sources performing best.
However, the limited availability of telescope time at major
observatories generally dictates the use of a minimum set of filters
sufficient to limit the bias caused by cluster galaxy contamination to
an acceptable level. In this section we discuss possible approaches in
increasing order of accuracy (and therefore cost).

\paragraph{Statistical correction based on galaxy overdensity as a function 
of radius:} State-of-the-art wide-field mosaic CCD cameras
allow us to efficiently image well beyond the virial radius of all but
the most nearby clusters. The galaxy number density measured in the
outskirts of the observed field can then be taken as the background
level, assumed to be unaffected by the cluster. A radial correction
for the cluster galaxy contamination can be derived from the measured
excess of counts at smaller radii (see Figure~\ref{fig:background}).  

This correction becomes unreliable in the innermost regions where the
observed background galaxy number density can be affected by
magnification (see \S\ref{sec:magnification}), or substructure in the
cluster core. In addition, there will be large statistical
uncertainties because of the low numbers of galaxies. This approach
can be used to estimate cluster masses \citep{Pedersen07,Hoekstra07},
provided the small scale lensing signal is avoided. We note, however,
that this correction is less suitable for studies of cluster density
profiles, which require reliable weak lensing measurements over a wide
range of radii.
 
\paragraph{Exclusion of cluster galaxies in a color-magnitude diagram:} 
The early-type cluster galaxies form a tight sequence in a
color-magnitude diagram, which can easily be recognized for massive,
low-z clusters.  For the color indices of most typical combinations of
two broadband optical filters, and for K-corrections given by the
spectral energy distributions of normal galaxies, the red sequence
galaxies will be redder than any other galaxies in the cluster or at
lower redshifts. Hence, by only picking galaxies redder than, and
clearly separated from, the red sequence, a robust selection of
background galaxies with negligible contamination can be
performed \citep[e.g.,][]{Okabe10}.

The robustness of a red background galaxy selection comes at a
considerable prize in terms of signal-to-noise, since most lensed
background galaxies are bluer than the red sequence and will be
omitted by this method. While such a selection is still feasible using
e.g., deep $V$ and $i'$-band data for a low-redshift ($z=0.18$)
cluster such as Abell 1689 \citep{Broadhurst05}, the number of red
background galaxies decreases rapidly with increasing cluster redshift
(because the red-sequence shifts to redder colors).

The number of source galaxies can be substantially increased by
avoiding the red-sequence, but including galaxies that are
sufficiently bluer. Such a selection will inevitably include some
residual contamination from actively star-forming blue dwarf galaxies
in the cluster, which may have a significantly shallower radial
distribution than their brighter, redder
counterparts \citep[e.g.,][find a slope $\alpha = 0.63$ for the radial
distribution of the "intermediate" dwarf population $-18 < R_{F606W} <
-15$ in Abell 2218]{Pracy04}. This residual contamination may be
estimated and removed using the methods outlined above, based on
galaxy overdensity as a function of
radius \citep[e.g.,][]{Hoekstra07}.

\paragraph{Exclusion of cluster galaxies in a color-color diagram:} 
Adding observations in a third passband enables a refined exclusion of
cluster galaxies based on their position in the color-color space
represented by the three filters.  By convolving template spectral
energy distributions derived empirically and/or from galaxy evolution
models with the bandpasses of the filters, the locations of different
types of galaxies in color-color space as a function of redshift may
be identified. This was used by \cite{Medezinski10} to study the
cluster light distribution and to minimize contamination of the sample
of source galaxies.

Thus, regions of color-color space where galaxies at redshifts lower
than, or similar to, the cluster may reside can be excluded from
further analysis. This provides a robust method to exclude foreground
and cluster galaxies which may also be checked against photometric
redshift catalogues. Recently, \cite{High12} applied this approach
using $gri$ observations of clusters with $0.28<z<0.43$. The resulting
source sample did not show appreciable clustering, suggesting that
this selection works well for intermediate redshift
clusters. \cite{Umetsu12} present results for their study of
MACS~J1206.2-0847, a massive $z=0.439$ cluster. For the blue sources
they find a constant number density with radius, but the red source
counts decline towards the center, a consequence of magnification.
Finally, \cite{Applegate12} compared the performance of such
color-cuts to the case where photometric redshift information is
available.

\paragraph{Photometric redshifts:} Having redshift information
for individual sources naturally provides the best separation of
sources and cluster galaxies. Although targeted weak lensing studies
have limited the number of filters, focusing instead on larger numbers
of clusters, the next generation cosmic shear studies will observe
large areas of the sky in multiple filters, providing large samples of
clusters {\it and} sources with redshifts. For
instance, \cite{Wittman03} provide an example of a massive galaxy
cluster at $z=0.68$ discovered in the Deep Lens Survey data.

The Cluster Lensing And Supernova survey with Hubble
\cite[CLASH;][]{Postman12} will target 25 clusters and provide
{\it HST} photometry in 16 filters. This will provide excellent photometric
redshift information for the sources, and will enable them to examine
the redshift dependence of the lensing signal in unprecedented detail,
which can be used to constrain cosmological
parameters \citep[e.g.,][]{Golse02, Jain03}.  The first results have been
published recently, based on strong lensing of multiple sources with
known redshifts behind the cluster
\citep{Jullo10} and by weak lensing measurements of a
stack of lower-mass galaxy groups \citep{Taylor12}. Given the small
differences predicted for different cosmological models, it is
presently unclear how powerful this technique may prove to
be. Any deviations between the assumed and actual cluster mass
distribution would introduce biases which might be difficult to
disentangle from the cosmological signal \citep{Zieser12}.

The amplitude of the lensing signal as a function of source redshift
can also be used to gain information about the 3D distribution of the
density field. This ``tomographic'' approach has been used to map the
mass density field towards the A901/2 supercluster 
\citep{Taylor04,Simon12}. While such data can be used to simultaneously
 derive redshifts and masses directly for structures along the line of
sight, well-defined structures such as rich galaxy clusters would in
any case be readily identifiable from the photometric data alone, and
the derived cluster redshifts from weak lensing tomography would have
much larger uncertainties (typically $\Delta z \sim 0.2$ or larger)
than any photometric redshifts that could be derived from the same
data set. Hence, the utility of this technique is more as a powerful
method of statistically probing structure growth as a function of
redshift based on large data sets for cosmic shear measurements,
rather than as an efficient method for finding large samples of galaxy
clusters.

\section{Mass reconstruction}
\label{sec:massrec}

In \S\ref{sec:weaklens} we already discussed that it is possible
to reconstruct the projected matter distribution from the
observed shear field. Unfortunately Eqn.~\ref{eq:ks}, derived
by \cite{KS93}, is not very practical, as its evaluation requires data
out to infinity.  Furthermore a smoothing of the results/data is
required because the shear is only sampled at the locations of the
source galaxies.  Finally, the reconstruction assumes that we have
measured the shear, which can deviate appreciably from the observed
reduced shear in the central regions of clusters. The latter problem
can, however, be solved by iteration \citep{Seitz95}.

The use of Eqn~\ref{eq:ks} leads to biased results if data for a
relatively small area are available (small in comparison to the extent
of the cluster). This was a serious issue for early ground-based
studies, carried out in the '90s, before wide-field imagers became
available. To solve this problem a number of finite-field inversion
techniques were developed \citep[e.g.][]{Seitz95,Squires96,
Seitz96,Seitz01}. Another approach is the use of maximum-likehood
methods, which avoid this problem
altogether \citep[e.g.][]{Bartelmann96b,Seitz98}. Modern weak lensing
studies of clusters employ wide field imagers, which alleviates the
issue by having observations over a finite area of the sky.  It can,
however, still be a relevant problem for {\it HST} observations of galaxy
clusters, which cover only a small area.

\subsection{Combination of Strong and Weak Lensing Data}

Strong and weak lensing observations of a cluster lens
naturally probe complementary regimes, as strong lensing information
can accurately constrain the matter distribution in the cluster center
whereas weak lensing traces mass out to the cluster outskirts. The
idea of combining both lensing methods into a single lensing mass
profile is hence straightforward, in particular as a means of breaking
the mass-sheet-degeneracy. In practice, however, combined strong and
weak lensing mass reconstructions continue to be technically
challenging.

Since the first developments of algorithms to combine strong and weak
lensing information \citep{Abdelsalam98,Bartelmann96b,Seitz98},
combined lensing methods have been applied to a small number of
clusters. Current methods include
the algorithms disussed in \citet{Bradac05a}, \citet{Cacciato06},
\cite{Diego07} and \citet{Merten09}. One of the most important results, based 
in part on a joint strong and weak lensing reconstruction, is the
discovery of the dissociation between the dark matter distribution and
the ICM in the ``Bullet cluster'' 1ES\,0657$-$558 \citep[also see
below]{Bradac06}. Combining information from both regimes is
relatively straightforward when considering parametric methods, such
as the one developed by \cite{Jullo07}, which as used by \cite{Limousin10}
to study the massive $z=0.545$ cluster MACS~J1423.8+2404.

These state-of-the-art joint strong and weak lensing algorithms are
based on optimising a suitable function $\chi^{2}[\Psi({\bf
x})]=\chi^{2}_{\mathrm{SL}}[\Psi({\bf x})]+
\chi^{2}_{\mathrm{WL}}[\Psi({\bf x})]+R[\Psi({\bf x })]$ 
with the deflection potential $\Psi({\bf x})$ in every cell of a grid
as a free parameter. The minimisation problem can in general only be
solved iteratively and a regularisation term $R[\Psi({\bf x})]$ is
needed to ensure numerical stability and smoothness of the resulting
deflection potential. Complexity arises due to the different nature of
strong lensing and weak lensing constraints, reflected in the
respective $\chi^{2}$-functions, notably in the error estimation.
Strong lensing gives a small number (in terms of degrees of freedom)
of highly significant constraints, localised in a small area within
the cluster's Einstein radius. These have to be calibrated well
against the much larger number of weak lensing galaxies which are
distributed over a large field-of-view.  These issues can be addressed
by using a grid of adaptive mesh size, giving special care to finite
differencing interpolation.

\begin{figure*}
\begin{center}
\hbox{%
\includegraphics[width=0.5\textwidth]{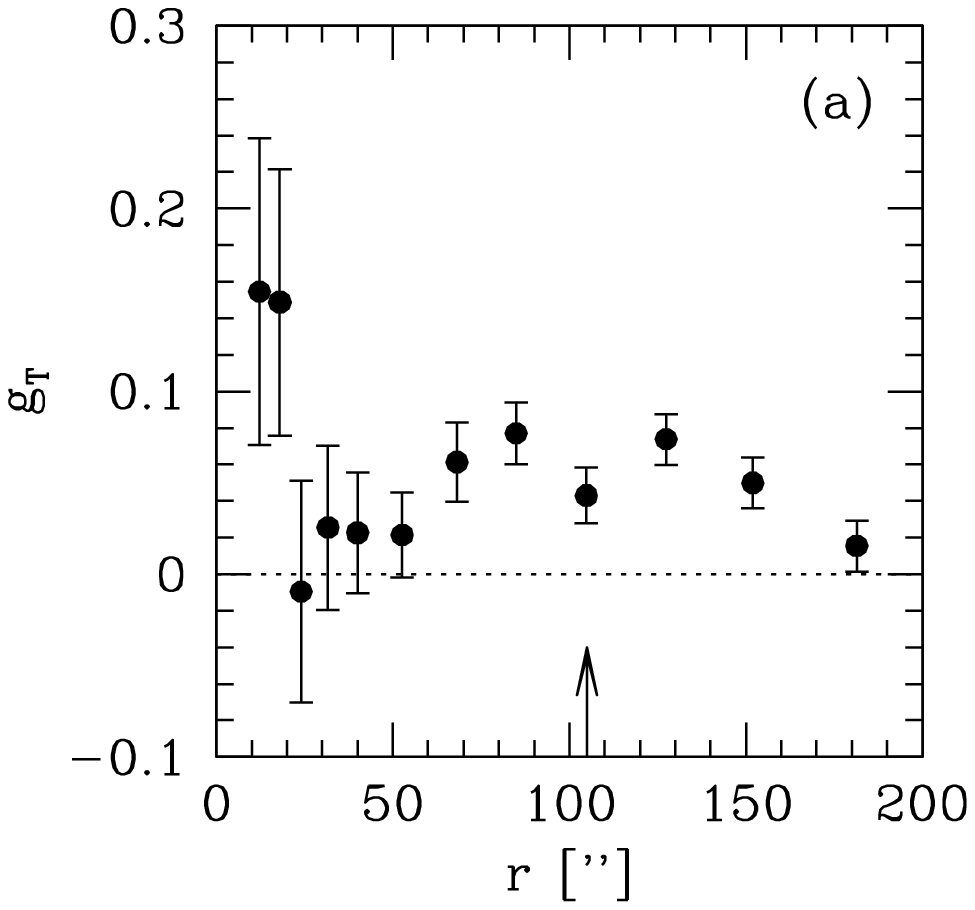}
\includegraphics[width=0.5\textwidth]{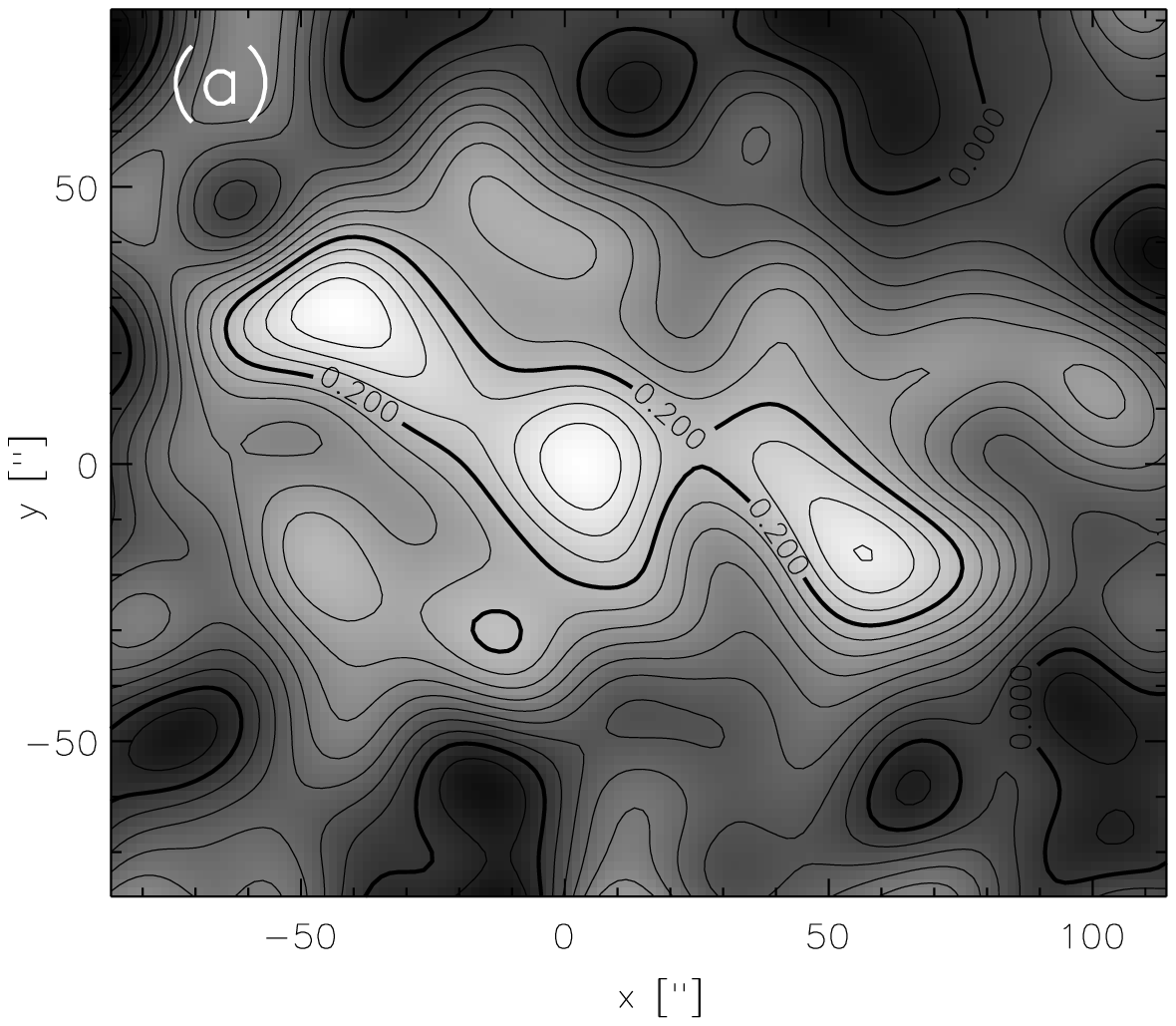}}
\end{center}
\caption{Figures from the weak lensing study of the $z=0.83$ cluster
MS1054-03 by \cite{Hoekstra00}. The left panel shows the azimuthally
averaged tangential distortion as a function of distance to the
cluster center. Note that the signal drops to zero at $r\sim30''$,
before rising again, which reflects the fact that the cluster consists
of multiple clumps, which is evident from the mass reconstruction
shown in the right panel. The mass distribution in the central regions
of this massive high redshift cluster is not well described by a simple
parametric model.}
\label{fig:ms1054}
\end{figure*}

\subsection{Results from Mass Reconstruction}

Non-parametric mass reconstructions are not particularly useful for
cluster mass determinations, because the observed shear field needs
to be smoothed to suppress noise in the mass map (especially when
considering mass estimates on small scales). Hence, mass measurements
are best carried out using parametric model fits to the data, or using
aperture mass statistics, as discussed in \S\ref{sec:weaklens}.
However, as many clusters are complex, especially at high redshifts
where they are still being formed, parametric model fits may not
always be appropriate.  This is highlighted in
Figure~\ref{fig:ms1054}. The left panel shows the lensing signal as a
function of radius for the massive $z=0.83$ cluster of galaxies
MS1054-03, studied using deep {\it HST} observations
by \cite{Hoekstra00}. The amplitude of the signal does not decrease
monotonically, and the mass reconstruction, shown in the right panel,
reveals the reason for this: the cluster consists of three distinct
clumps. The ability to identify significant substructures is an
important application of (non-parametric) cluster mass
reconstructions.

Hence, mass reconstructions of galaxy clusters can play an important
role in determining whether a cluster can be considered as a
``relaxed'' or a possibly ``merging'' system. Complex structure in the
mass map is, however, not necessarily related to merging activity, as
lensing probes the projected mass distribution. For
instance, \cite{Israel12} have presented a case of a $z\!=\!0.45$
cluster, likely seen through a filament at $z\!=\!0.22$, connecting
several structures to a massive foreground cluster. This mass
distribution results in an extended shear plateau and complicates the
mass determination of the more distant cluster. Using information from
the mass map in conjunction with other probes, in particular X-rays,
it is possible to account for such cases and retain them in the
analysis of a pre-defined X-ray sample. Another example was presented
in \cite{Israel10}, although in this case the foreground concentration
of galaxies does not seem to impede the weak lensing mass measurement.
Studying cluster maps thus helps the interpretation of anomalous
results in profile fits to cluster shear signals in the context of
cosmology from increasingly large and distant cluster samples.

However, arguably the most interesting class of objects for detailed
mass reconstruction studies are merging clusters. Obviously, the
description of the mass distribution by a smooth, highly symmetric
mass profile, such as Eqn.~\ref{eq:nfw}, breaks down. This is
particularly true in the case of a major merger. In these cases the
so-called ``non-parametric'' methods are clearly favoured.  What makes
mass reconstructions of merging clusters so important is the
possibility that during certain phases of a merger the
``collision-less'' dark matter and galaxy components can segregate
from the more dissipative ICM. Given a favourable alignment with
respect to the line-of-sight, this can lead to a measurable offset
between the lensing and X-ray surface brightness peaks. Over the last
years, a small number of such ``dissociative
mergers'' \citep{Dawson11} has been discovered and subsequently
studied using a range of methods and instruments.  The defining
feature of a dissociative merger is the offset between the ICM and
total mass peaks (as inferred by weak lensing).

The prototype of such systems is the ``Bullet cluster''
1ES\,0657$-$558. This system, an advanced stage of a merger in which
an infalling cluster has already passed through the core of the more
massive component almost in the plane of the sky, was the first in
which a significant separation between the X-ray and weak lensing mass
peaks of both subclusters was
discovered \citep{Clowe04b,Markevitch04}.  Using a finite field
inversion method and {\it HST} data, \citet{Clowe06} determined the
separation to be significant at the $8\sigma$ level, which was
confirmed in an analysis also incorporating strong lensing data
\citep{Bradac06}. The interpretation, also corroborated by follow-up studies
\citep[e.g.][]{Springel07}, is that while the ICM of the two clusters 
experienced deceleration due to the collision (as indicated best by
the shock visible with {\it Chandra}), the dark matter halos passed
through each other without significant interaction. An upper bound on
the self-interaction cross-section of dark matter particles can be
inferred from the displacement between the Dark Matter and gas
peaks. \citet{Markevitch04} constrain the cross-section to
$<\!1\,\mbox{cm}^{2}\,\mbox{g}^{-1}$, resulting in the most direct
piece of evidence for the existence of non-baryonic dark matter to
date\footnote{Even tighter constraints on the collision cross-section
were derived by \citet{Meneghetti01} who studied the impact of
collisonal dark matter on the formation of radial and giant arcs.}.

\begin{figure*}
\begin{center}
\includegraphics[width=\textwidth]{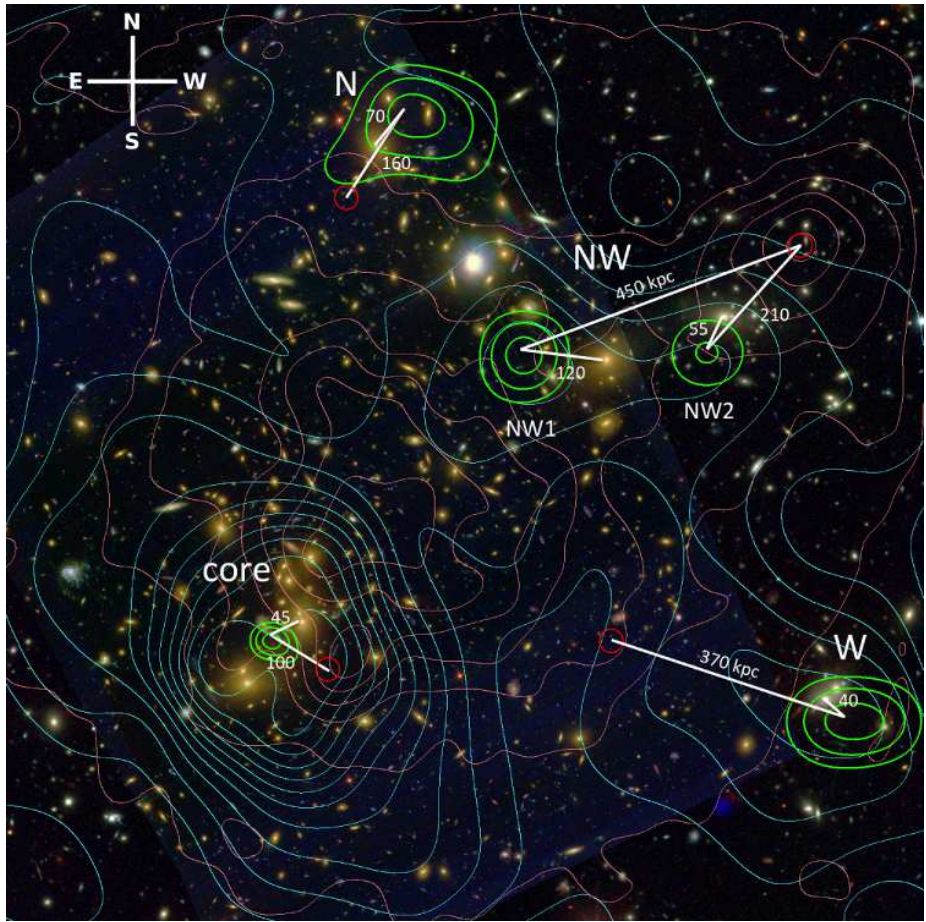}
\end{center}
\caption{Mass reconstruction of Abell~2744 (``Pandora's cluster''), overlaid
on a $4'\times 4'$ {\it HST}/VLT/Subaru false-color mosaic: The
reconstructed surface mass density is indicated by the cyan contours,
while X-ray surface brightness contours are given in magenta. Green
contours show mass peaks from bootstrap resampling.  Contour levels are
chosen at $86$\%, $61$\%, and $37$~\% of the respective maximum peak
likelihood. Relevant distances to bright galaxies and local gas peaks
are highlighted by rulers. Figure from \citet{Merten11}.}
\label{fig:A2744}
\end{figure*}

Another interesting system is Abell~2744, where at least four
different subclumps each with masses $>\!10^{14}\,\mathrm{M}_{\odot}$,
were discovered by applying the \citet{Merten09} technique to a
combination of strong and weak lensing data collected using
the \emph{Hubble Space Telescope}, VLT, and Subaru
telescopes. Using these data, as well as deep {\it Chandra} observations,
\cite{Merten11} find that Abell~2744 is a very complex merger. This
is evident from their results shown in Figure~\ref{fig:A2744}. Two of
the components strongly resemble the ``Bullet cluster'', with a
separation of $17''$ between lensing and X-ray peaks in the core
and a matching $30''$ offset for the smaller Northern
component. Following the approach of \cite{Markevitch04} a dark matter
self-interaction cross-section of
$\sigma/m\!<\!3\pm1\,\mbox{cm}^{2}\,\mbox{g}^{-1}$ can be inferred.

Abell~2744 features another interesting structure: towards the
North-West, two peaks are detected in the mass map, and one peak in
the {\it Chandra} map, but they are all separated from each other and
the nearest bright galaxies. In addition, \citet{Merten11} report a
further mass peak without ICM on the Western side. As an explanation
of the formation of the ``ghost'',``dark'', and ``stripped''
clumps, \cite{Merten11} suggest a complicated merger scenario. In this
picture, at least two merger events occurred almost simultaneously,
involving up to four different substructures colliding with the core
of Abell~2744.

Abell~520 is another example of a complicated merging system. Like
1ES\,0657$-$558 it shows clear evidence of a shock in the X-ray
observations. Using deep CFHT observations, \cite{Mahdavi07} found a
significant peak in the mass reconstruction that coincided with the
peak of the X-ray emission, but where almost no cluster galaxies were
present. This result was confirmed by \cite{Jee12}, using deep {\it
HST} WFPC2 observations, although even deeper ACS observations by
\cite{Clowe12} do not show evidence of a peak. These interesting 
clusters highlight the importance of mass reconstructions
as well as the combination of strong and weak lensing information, if
available.

\section{Cluster halo properties from weak lensing}
\label{sec:properties}

In \S\ref{sec:mass} we review applications of weak lensing to estimate
the masses of clusters, and the subsequent comparison to baryonic
tracers. In this section we discuss observational tests of two key
predictions from numerical simulations of structure formation in a
cold dark matter dominated universe.  We first review constraints on
the radial density profiles, which are predicted to be well described
by Eqn.~\ref{eq:nfw}. In \S\ref{sec:shapes} we review measurements of
the ellipticities of cluster halos, which can be compared to the CDM
predictions.

To examine the density profiles at large radii, weak gravitational
lensing is well suited, especially when combined with strong lensing
studies. This is because the weak lensing signal can be measured out
to large radii, thus probing the gravitational potential on large
scales. Even if they were measurable, the results from most dynamical
techniques would be difficult to interpret because the outskirts are
not fully virialized. We note that it is possible to study the
velocity field of the cluster infall region, which provides a direct
measure of the escape velocity and thus the cluster mass. In practice
this is done by identifying the caustics in redshift
space \citep{Diaferio97,Diaferio99}. This technique has been applied
to a number of local clusters of galaxies. \cite{Rines03}
and \cite{Diaferio05} compared weak lensing estimates to the masses
inferred from the identification of caustics in redshift space for
three clusters and found good agreement on scales of $\sim 2$Mpc.

\begin{figure*}
\begin{center}
\includegraphics[width=\textwidth]{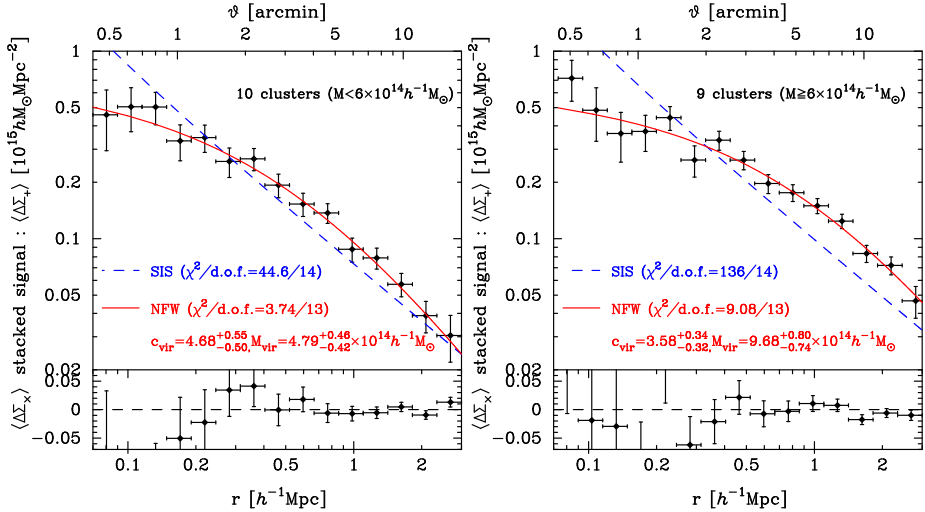}
\end{center}
\caption{Stacked surface mass density contrast $\langle\Delta\Sigma\rangle$
of $19$ LoCuSS clusters. Points with error bars show the measured tangential 
shear signals of clusters below (left panel) and above (right panel) a mass of 
$6\times10^{14}h^{-1}\mbox{M}_{\odot}$. 
The solid red lines gives the best NFW fit, while the dashed blue lines gives
the best SIS fit to the data. The lower sections of both panels show the 
corresponding stacked cross-components. 
Figure from \citet{Okabe10}.}
\label{fig:Ok10_f8}
\end{figure*}

\subsection{Density Profiles}
\label{sec:profiles}

One of the most important quantities that can be inferred from weak lensing
is the density profile of clusters. Although this can be measured for
individual clusters, the mention of a radial profile implicitly
assumes a high degree of spherical symmetry, which may not be
appropriate. Furthermore cosmic noise reduces the precision when
considering only weak lensing measurements \citep{Hoekstra03a}.  Hence,
a better question to ask is what the average density profile of
clusters looks like. Provided the selection of the cluster sample is
not (too) sensititive to the orientation of the cluster, the ensemble
averaged cluster should be rather spherical. These results can be
directly compared to the predictions of $N$-body dark matter
simulations.

Weak lensing studies of the density profile start with the tangential
shear profile $g_{\mathrm{t}}(r)$, which then can be converted into a
projected density profile $\kappa(r)$. Strong lensing and combined
methods typically yield the $\kappa$ profile via the lensing
potential. Alternatively one can fit a parametrized model to the
tangential shear. The latter approach was used by \cite{Okabe10} who
studied the shear profiles of $0.15\!<\!z\!<\!0.30$ clusters observed
by LoCuSS.

Out of the $22$ clusters that have color information (for background
selection and dilution correction; see \S\ref{sec:contam}) and do not
show a complex mass distribution, $19$ are modelled well by an NFW fit
(i.e.\ better than using an SIS fit).  Stacking the shear signals of
these $19$ clusters, the projected density contrast was measured over
a wide radial range ($70 h^{-1}\mbox{kpc}$ to $3 h^{-1}\mbox{Mpc}$;
Fig.~\ref{fig:Ok10_f8}).  Fitting to the stacked signals of the ten
lower-mass\footnote{\citet{Okabe10} use a virialisation overdensity
based on their adopted ``concordance'' $\Lambda$CDM cosmology,
$\Delta_{\mathrm{vir}}\!\approx\!112$ at $z\!=\!0.2$.}
($M_{\mathrm{vir}}\!<\!6\times10^{14}h^{-1}\mbox{M}_{\odot}$) and
higher-mass
($M_{\mathrm{vir}}\!\geq\!6\times10^{14}h^{-1}\mbox{M}_{\odot}$)
clusters, an NFW model is strongly favoured over an SIS model, given
the noticeable curvature in the measured profile.  Even at the largest
tested cluster-centric separations, determined by the field-of-view of
Subaru/SuprimeCam, the signal follows an NFW function. Matching
expectations, the stacked lower-mass cluster signal follows a slightly
more concentrated NFW profile than the higher-mass systems.

\subsubsection{Results from stacking}

To average out the effects of the cluster large-scale structure
stacking is clearly convenient. It also allows us to extend the range
to much lower cluster masses and study their \emph{statistical}
properties. The stacking process has to be performed in a sensible
way, meaning e.g.\ taking into account the redshift scaling of the
signals from clusters within a (redshift) bin. \citet{Johnston07a}
developed the extraction of 3D density and mass profiles from stacked
clusters and performed numerical simulations to test the performance.

Naturally, stacking methods are even more sensitive to miscentering
errors than weak lensing studies of individual clusters. But this can
be modeled (see \S\ref{sec:center}) and the downside is outweighed by
the cancellation of unrelated large-scale structure contaminants.
Furthermore, the analysis tends to reduce additive errors in the shape
analysis, because they do not contribute coherently, unless the
clusters are targeted and always appear on the same position in the
field-of-view.

The SDSS has made important contributions to the statistical studies
of lenses, ranging from galaxies to clusters of galaxies. For
example \citet{Mandelbaum06} used the luminous red galaxy (LRG) sample
to define a sample of galaxy groups and clusters of galaxies and
fitted NFW models to the lensing signal $\Delta\Sigma(r)$. They found
that the resulting concentrations agree well with $\Lambda$CDM
predictions. \cite{Mandelbaum06} also review extensively possible
systematic effects. The mass range was extended to lower masses
in \cite{Mandelbaum08} who thus provided constraints on the
concentration parameter over more than two orders of magnitude in mass
($10^{12}-5\times 10^{14}\msun$).

Combining both spatial and color-space clustering with the presence
of a BCG candidate \citep{Koester07}, the lensing signal for the
$0.1\!\leq\!z\!\leq\!0.3$. MaxBCG cluster catalog was studied
by \cite{Sheldon09}. The clusters were subdivided either by luminosity
(see Figure~\ref{fig:Johnston07}) or by richness $N_{200}$.
\citet{Sheldon09} report significant $\Delta\Sigma$ profiles in all
tested richness and (total $i$-band) luminosity bins, spanning a scale
between small galaxy groups and clusters. These measurements were
analysed further by \cite{Johnston07b} who fitted a halo model (as
discussed in \S\ref{sec:halomodel} and shown in Fig.~\ref{fig:Johnston07}).
The excellent precision of the measurements enabled \cite{Johnston07b}
to determine the concentration as a function of halo mass, finding 
good agreement with previous estimates.

\subsubsection{Results for individual clusters}

Although stacking analyses can provide excellent constraints on the
average density profiles over a wide range in mass, it is nonetheless
interesting to study individual systems, even if the results are
limited by the large scale structure. The study of mass density
profiles using gravitational lensing are discussed in detail
by \cite{Meneghetti12}, but here we discuss briefly some recent
results.

\cite{Zitrin12} used $16$ band {\it HST} imaging from the CLASH survey,
as well as VLT/VIMOS spectroscopy to construct a strong lensing model
for the massive $z\!=\!0.44$ cluster MACS~J1206$-$0847 based on $47$
images of $12$ background sources. The mass profile
resulting from the strong lensing solution exhibits a slope of
$\mathrm{d\,log}\,\Sigma/\mathrm{d\,log}\,r = -0.55\pm0.10$, measured
over a range of $5\!\lesssim\!r\!\lesssim\!300\,\mbox{kpc}$, or out to
about twice the Einstein radius. This is typical of a relaxed and
rather concentrated cluster, and corroborates the notion of
MACS~J1206$-$0847 being relaxed, despite its unusually high X-ray
luminosity and temperature. 

\begin{figure*}
\begin{center}
 \includegraphics[width=\textwidth]{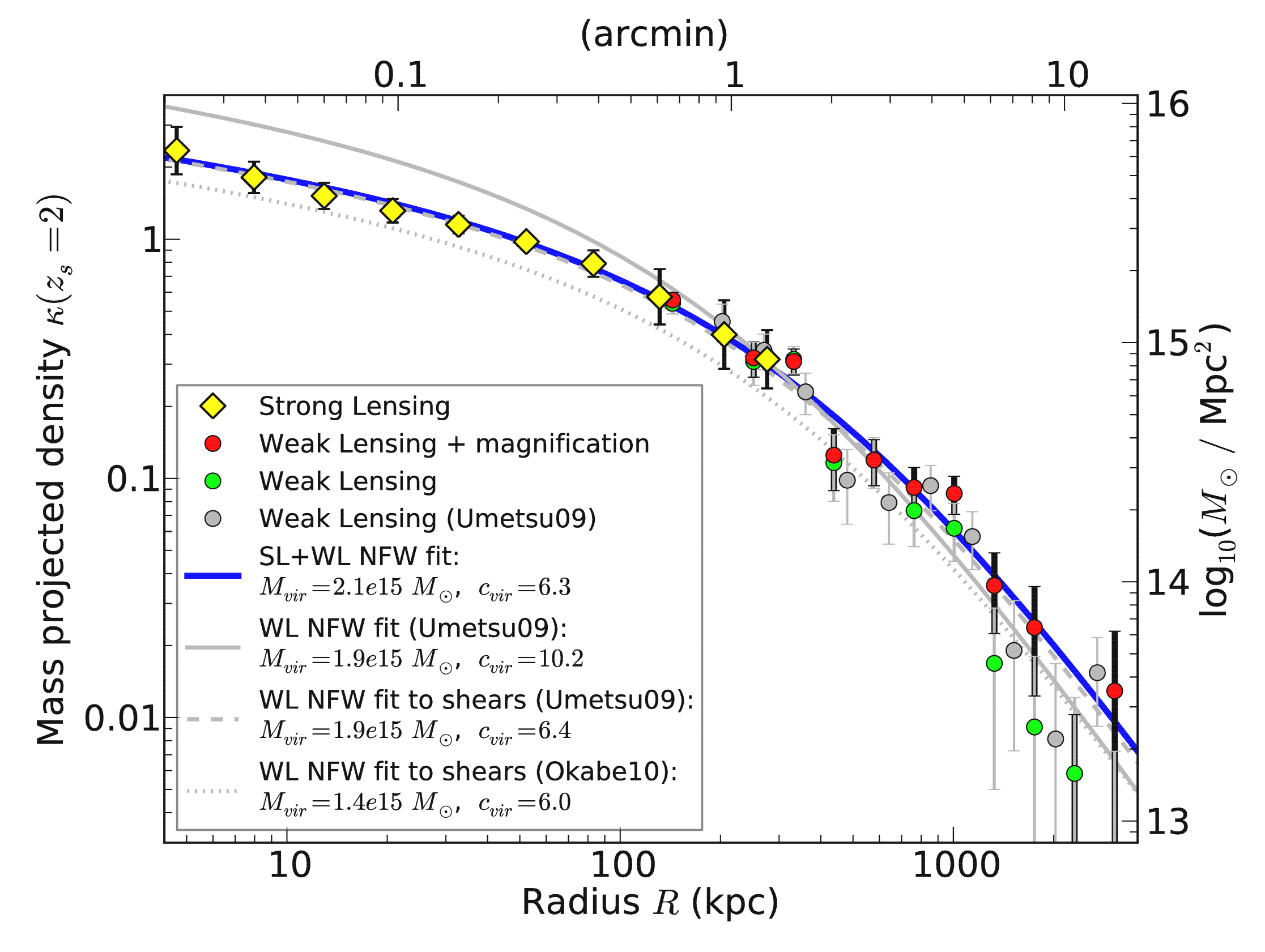}
 \includegraphics[width=\textwidth]{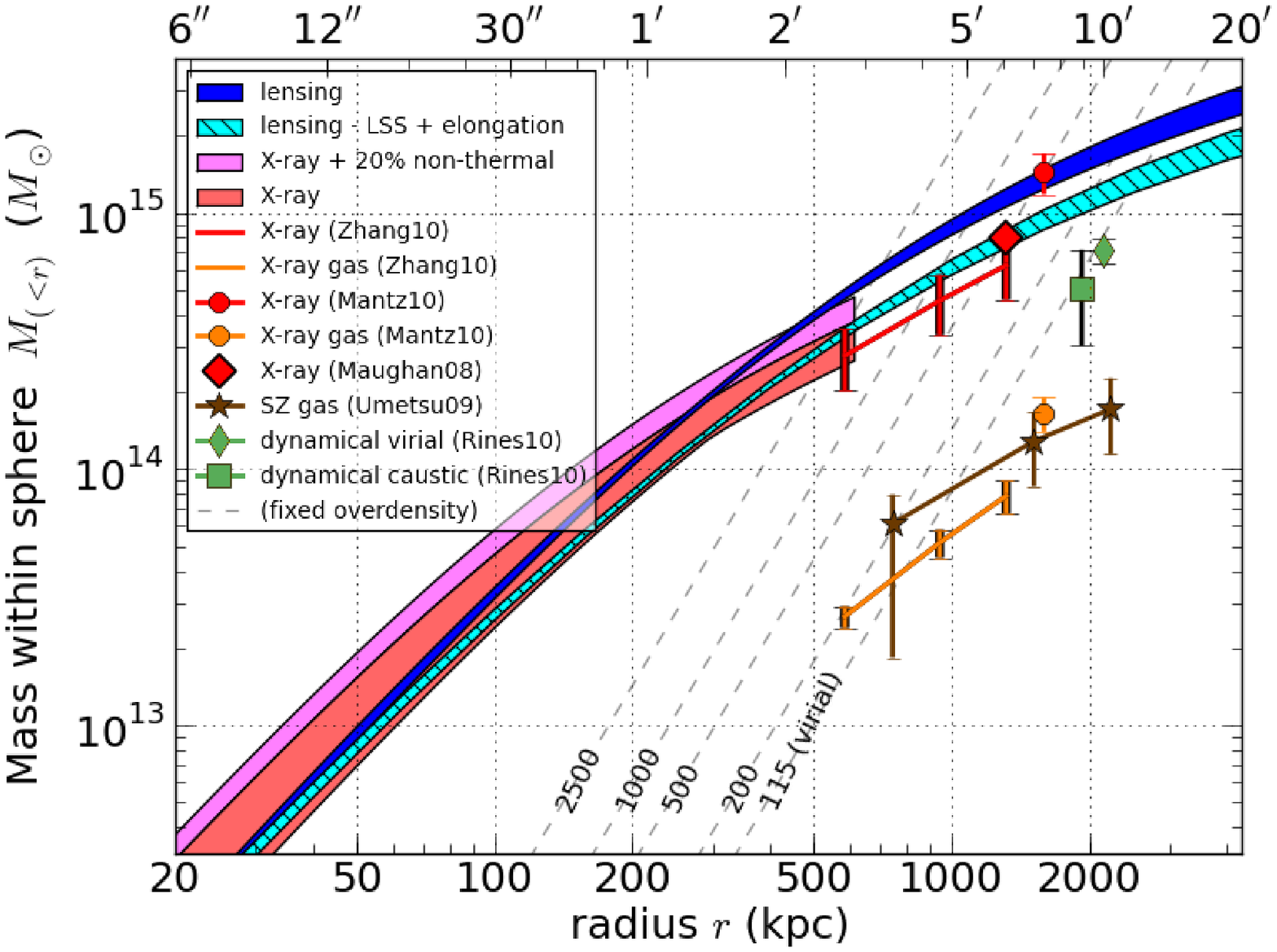}
\end{center}
\caption{\textit{Upper panel:} Projected mass density of Abell~2261.
Shown are various strong and weak lensing constraints to the $\kappa$ profile
and the respective NFW fits (see legend).
\textit{Lower panel:} Mass profiles of Abell~2261. Shown are cluster masses
within a sphere of radius $r$ from a variety of observables (see legend). 
For comparison, dashed lines show fixed overdensities $\Delta$. 
Figures from \citet{Coe12}.}
\label{fig:coe12}
\end{figure*}

\citet{Coe12} determined the inner mass profile of Abell~2261, also
as part of the CLASH survey. The {\it HST} imaging was combined with
wide-field Subaru and KPNO observations. Combining strong and weak
lensing, and also including gravitational magnification to break the
mass-sheet degeneracy, the authors find the profiles from strong and
weak lensing to agree well in the overlapping region (upper panel of
Fig.~\ref{fig:coe12}). Both the breaking of the mass-sheet degeneracy
and the careful selection of background galaxies using color
information are important in achieving this agreement.  The resulting
mass profile is well represented by an NFW fit of moderate
concentration ($c_{\mathrm{vir}}\!=\!6.2\pm0.3$), contrasting earlier
findings \citep[e.g.,][]{Umetsu09} of a concentration
$c_{\mathrm{vir}}\!>\!10$.  While \citet{Coe12} find that
the hydrostatic mass estimate based on their {\it Chandra} analysis
agrees with the lensing mass profile at the cluster center, at
$r_{2500}$, the outermost radius covered with {\it Chandra}, it is
$\sim\!35$~\% lower than the lensing mass (lower panel of
Fig.~\ref{fig:coe12}).

A 20\% deviation from hydrostatic equilibrium is necessary to
alleviate the discrepancy between the X-ray and lensing mass
profiles. Although non-thermal pressure sources can at least partly
explain the difference, halo elongation along the line of sigh can
reconcile the X-ray and lensing observations, as found
by \citet{Coe12}: the agreement between profiles improves when a 2:1
axis ratio, consistent with expectations from simulated clusters, is
assumed (cf.\ Fig.~\ref{fig:coe12}). We refer the reader
to \cite{Limousin12} in this volume for a review on this particlar
topic.

\subsection{Shapes of Dark Matter Halos}
\label{sec:shapes}

An important prediction of numerical simulations of structure
formation is that dark matter halos are triaxial, which complicates
the comparison of weak lensing masses to other methods. This
prediction can be studied using gravitational lensing. Strong lensing
studies can be used to study the inner regions, although the
interpretation of the results may be complicated by the presence of
baryons.

The shape of the dark matter halo at larger radii can be studied
using weak gravitational lensing. Although one could use mass
reconstructions for this purpose, it is better to examine the effect
of an elliptical mass distribution on the shear field. For
instance, \cite{Hoekstra98} looked at the azimuthal variation of the
tangential shear around the cluster MS1358+62. They found that the
cluster mass distribution was elongated with an axis ratio and
position angle that agreed well with strong lensing results.

\begin{figure*}
\begin{center}
\vspace{-0.0cm}
 \includegraphics[width=0.45\textwidth]{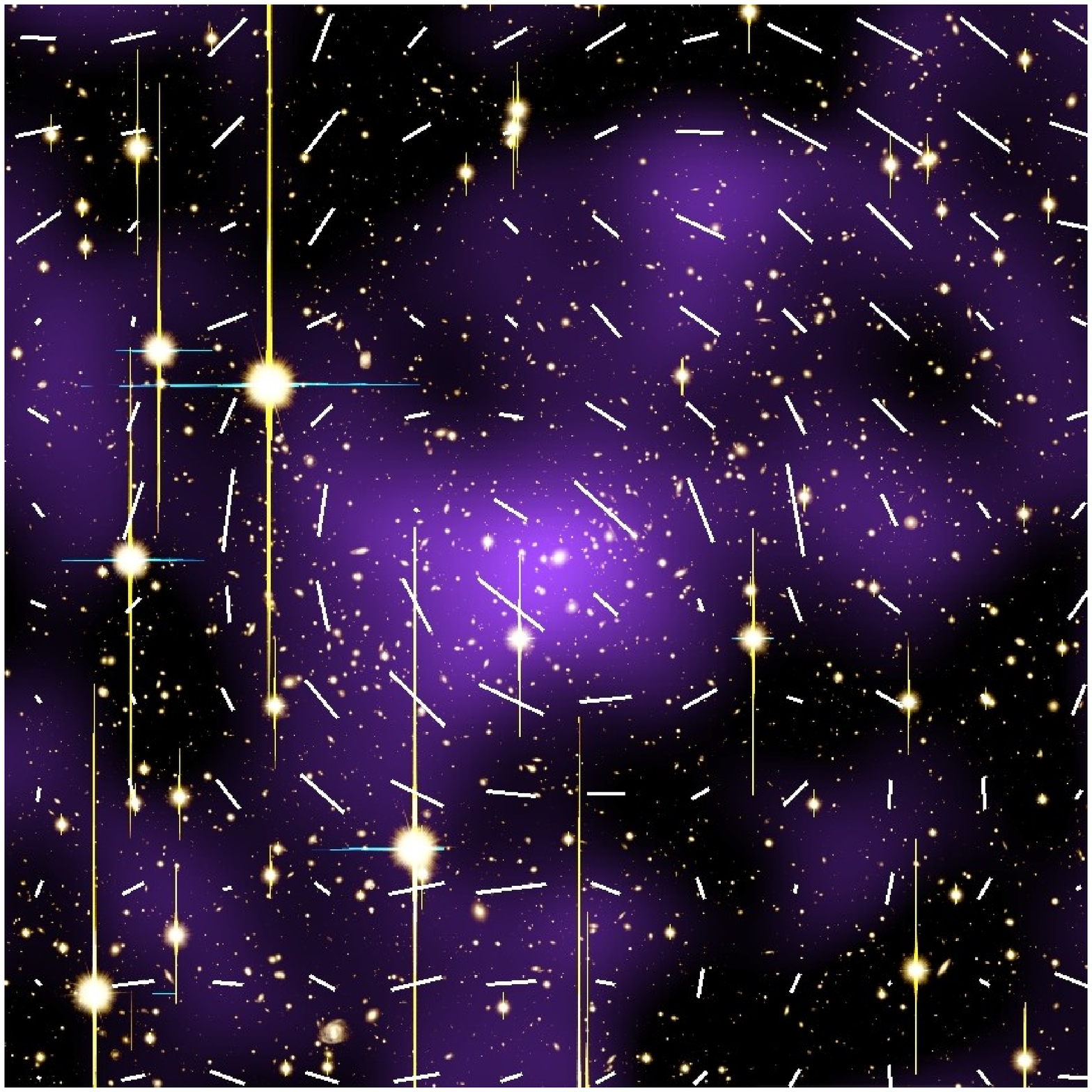}
 \includegraphics[width=0.45\textwidth]{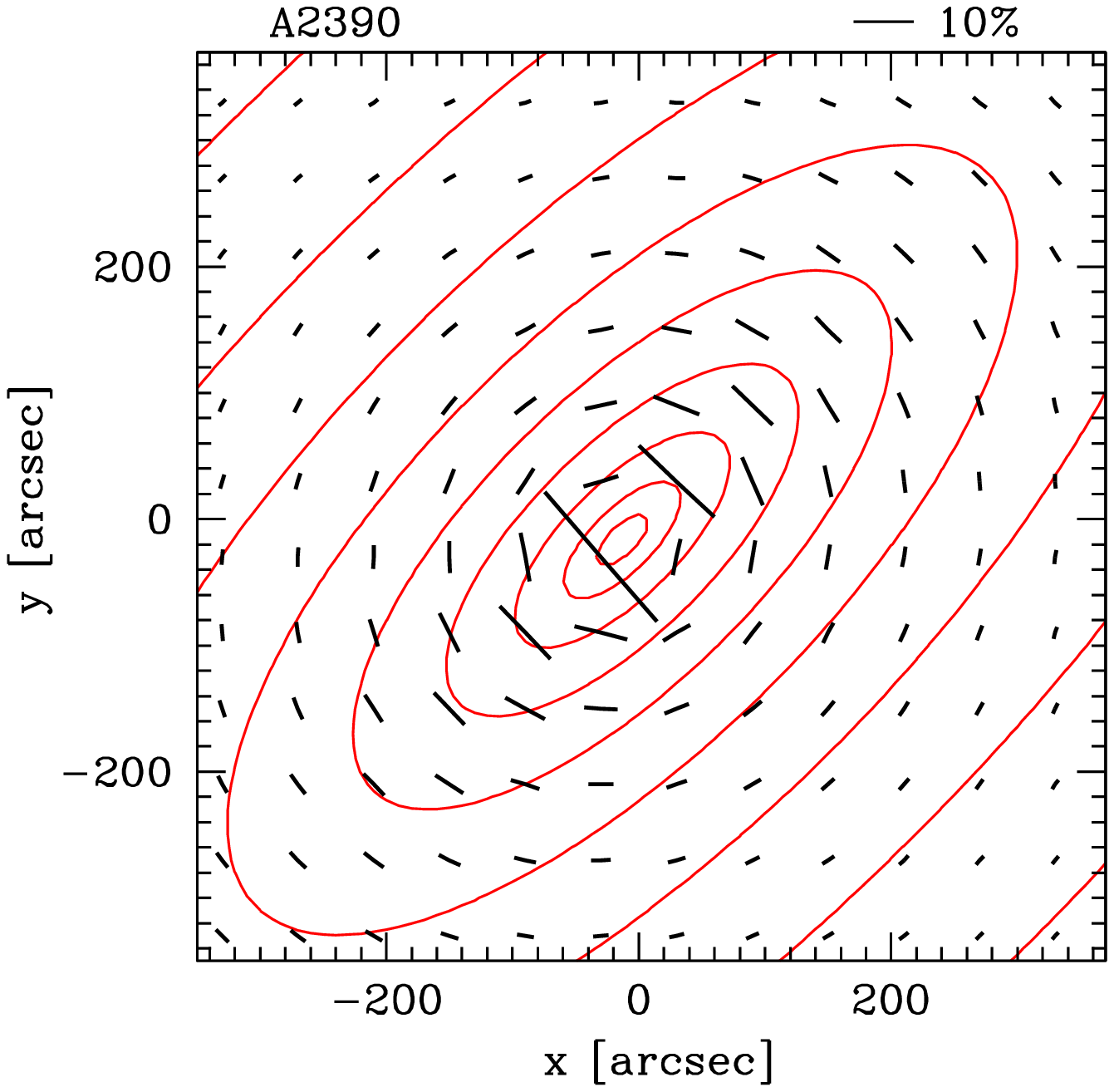}
\end{center}
\caption{Shear field of Abell~2390. Sticks in both panels show the average
shear measured in cells of $1'$ mesh size. The measured field in the left
panel has been smoothed by a Gaussian of $\approx\!1.6'$ FWHM. 
Also shown is the projected surface mass map, overlaid on the Subaru image. The
right panel presents the shear field by the best-fitting elliptical NFW model.
Figure from \citet{Oguri10}.}
\label{fig:shapes}
\end{figure*}

A more extensive study was carried out by \cite{Oguri10} who analysed
a subsample of the Local Cluster Substructure Survey (LoCuSS)
sample observed with Subaru's SuprimeCam. \cite{Oguri10} fit an
elliptical model (with a radial NFW model) to the 2D shear map of 25
clusters. Their fit accounts for covariance between the errors in the
grid cells caused by cosmic noise. An example of one of the clusters
in their sample is presented in Figure~\ref{fig:shapes}. For $18$ out
of their $25$ clusters, the elliptical shear model proves to be a good
fit. From these clusters, \cite{Oguri10} obtain a mean halo
ellipticity\footnote{Here, the ellipticity of an ellipse with major and minor
axes $a$ and $b$ is defined as $e\!=\!1-b/a$.} of 
$\langle e\rangle\!=\!0.46\pm0.04$ (where the uncertainty is obtained via a
Monte Carlo technique). This value is in good agreement with the prediction
of $\langle e\rangle\!=\!0.42$ for simulated triaxial dark matter halos
\citep[e.g.,][]{Jing02}. 

Recently, \citet{Oguri12} published a new measurement of the halo
ellipticity, based on a combined strong/weak lensing study of $28$
galaxy clusters selected from the Sloan Giant Arcs Survey.  The
combination of the strong and weak lensing data is performed using the
\citet{Oguri09} method. Using both an independent sample and a different
technique, the authors find a mean halo ellipticity of $\langle
e\rangle\!=\!0.47\pm0.06$, confirming the results of \citet{Oguri10},
and in good agreement with $\Lambda$CDM expectations.
\citet{Oguri12} also detect a correlation of the halo ellipticity measured 
by lensing with the distribution of luminous ($L\!\gtrsim\!L_{\ast})$
galaxies.

\section{Cluster masses from weak lensing}
\label{sec:mass}

There is an ongoing effort to use the observed number density of
clusters of galaxies as a precision probe of cosmological parameters
\citep[see e.g.,][for a recent review]{Allen11}. A correct interpretation
of the observations relies on the ability to relate the observables to
the mass. The calibration of such ``scaling relations'', reviewed in
detail by \cite{Giodini12} in this volume, is still an important limiting
factor for the precision of recent cluster surveys. Weak gravitational
lensing masses, however, are playing an increasingly important role
and we highlight a few results in this section.

One reason for the popularity of weak lensing is that the method is
considered sufficiently mature, thanks to improved shape measurements
as discussed in \S\ref{sec:shapemeasurement}. Also important is the fact
that the cluster lensing signal can be measured out to large radii.
Finally, there is increasing evidence that non-gravitational processes
such as AGN feedback play a role in cluster formation. Furthermore,
pressure support by turbulent motion of the ICM leads to biases in the
hydrostatic masses \citep[e.g.,][]{Nagai07,Mahdavi08,Mahdavi12}, which
need to be calibrated observationally.

As weak lensing masses are insensitive to the dynamical state of the
cluster, the results can be readily compared to cosmological numerical
simulations. Note that this does not imply that weak lensing masses
themselves are unbiased, only that the biases can be reliably
quantified using numerical simulations. The main complication to
compare lensing masses to other tracers is the fact that clusters are
complicated three-dimensional structures. Although the measurement of
the lensing signal itself does not require assumptions about the
detailed geometry of the cluster, comparison to other observations or
predictions does rely on the assumed geometry.

Several studies have examined the impact of this. For instance,
{\cite{Meneghetti10} focused specifically on the comparison of lensing
masses and X-ray estimates using hydrodynamical simulations. From this
study it is clear that the triaxility of the halo limits the
usefulness of comparing results for individual systems, but large
samples can provide key constraints on the gas physics in clusters.
Note that it is possible to account for the triaxiality in the
modeling by combining multi-wavelength data \citep[see][]{Limousin12,
Morandi12}. Alternatively, \cite{Corless08} and \cite{Corless09}
investigated the fitting of triaxial models to the shear field
instead. They showed that this approach, using priors obtained from
numerical simulations can indeed reduce the bias in the recovered weak
lensing mass. 

It has also become clear that the bias in the (deprojected) weak
lensing mass depends on the methodology and the radial range that is
used.  For instance \cite{Becker11} have shown that NFW model fits to
the signal at large radii ($>10'$) biases the mass low by $\sim 6\%$,
whereas the bias is smaller when the outer radius of the fit is
limited to $1-2r_{\rm vir}$. If left unaccounted for, this leads to a
significant contribution to the systematic error budget for studies
that contain 10 or more clusters. Large cluster surveys therefore need
to compare the algorithm used to infer the mass to such simulations, as
was done by \cite{High12}.

\subsection{Results from stacking analyses}

As discussed in \S\ref{sec:weaklens} the signal-to-noise is too low for low
mass systems, or clusters at high redshifts. By stacking of the
lensing signal from an ensemble of objects we can nonetheless provide
constraints on their (average) mass. This was first done for a sample
of 50 galaxy groups from the CNOC2 field galaxy redshift survey in
\cite{Hoekstra01b} who found fair agreement between the lensing and
dynamical mass. Although extended by \cite{Parker05} who studied
of 116 groups, these studies were limited by the relatively small
number of systems.

\paragraph{SDSS:} As in so many areas of astronomy, the SDSS also has had 
a great impact in the study of cluster masses, in particular the
low-mass end. The large survey area provides the opportunity to stack
the signals of many clusters and study their mass as a function of
binned baryonic properties. As discussed in \S\ref{sec:halomodel} the
resulting cluster-mass cross-correlation signal (shown in
Figure~\ref{fig:Johnston07}) can be modeled, which in turn yields mass
estimates for the various ensembles. The ensemble-averaged weak
lensing signal and subsequent mass modelling were presented
in \cite{Sheldon09} and \cite{Johnston07b}, respectively. The left
panel of Figure~\ref{fig:massrichness} shows the resulting $M_{200}$
as a function of the optical luminosity. The right panel of
Figure~\ref{fig:massrichness} compares the mass to the cluster
richness (quantified by $N_{200}$, the number of red-sequence galaxies
within $r_{200}$). \cite{Johnston07b} also compared to the dynamical
masses determined by \cite{Becker07} and found good agreement
\emph{on average} (see the red points in Fig.~\ref{fig:massrichness}.
This suggests that the difference between the dark matter and galaxy
velocity dispersions is fairly small. Finally, \cite{Rykoff08}
compared the SDSS lensing masses to stacked X-ray luminosity estimates
from {\it ROSAT} All-Sky Survey data \citep[also see][]{Giodini12}.

\begin{figure}[t!]
\begin{center}
\hbox{%
\includegraphics[width=0.55\textwidth]{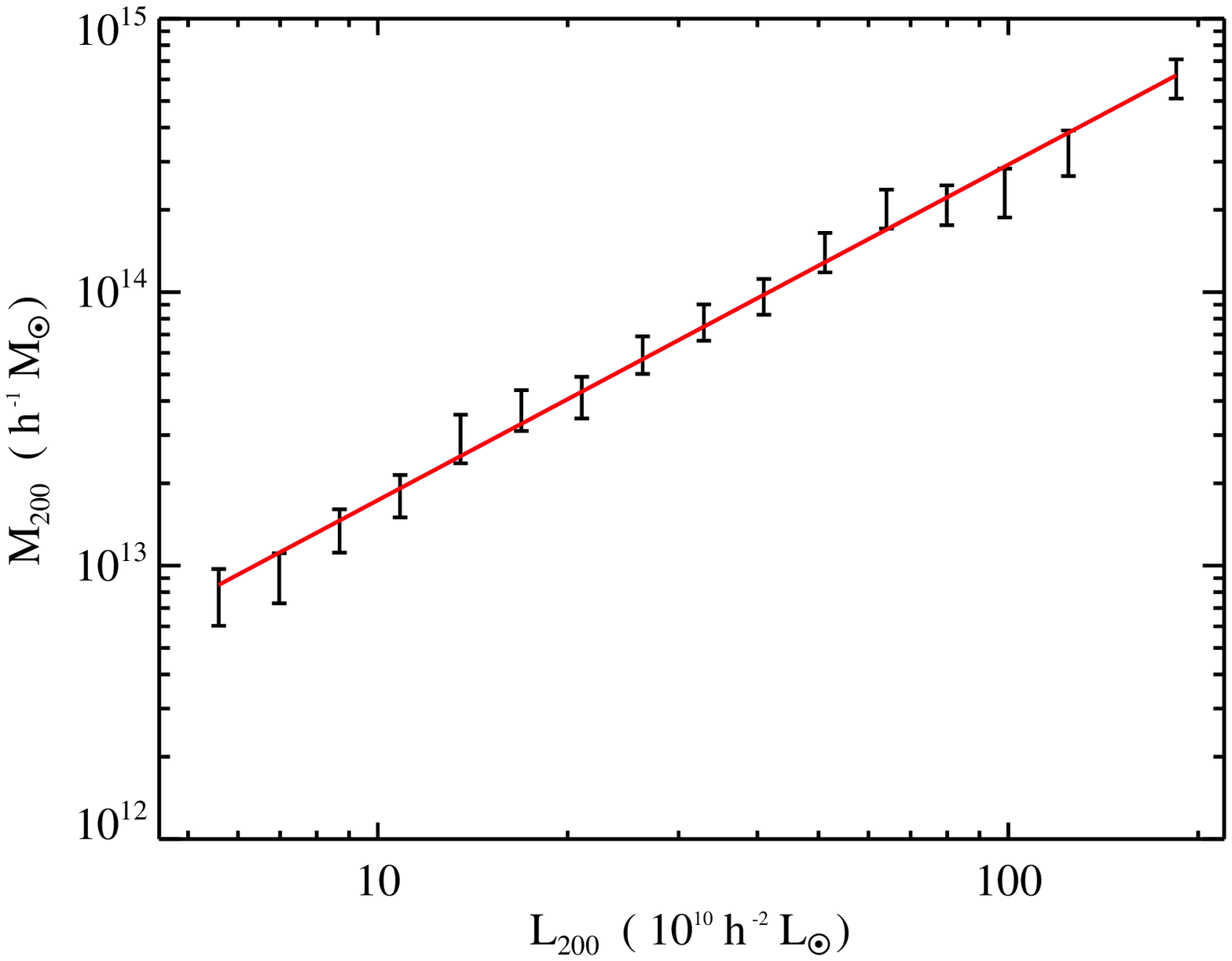}
\includegraphics[width=0.45\textwidth]{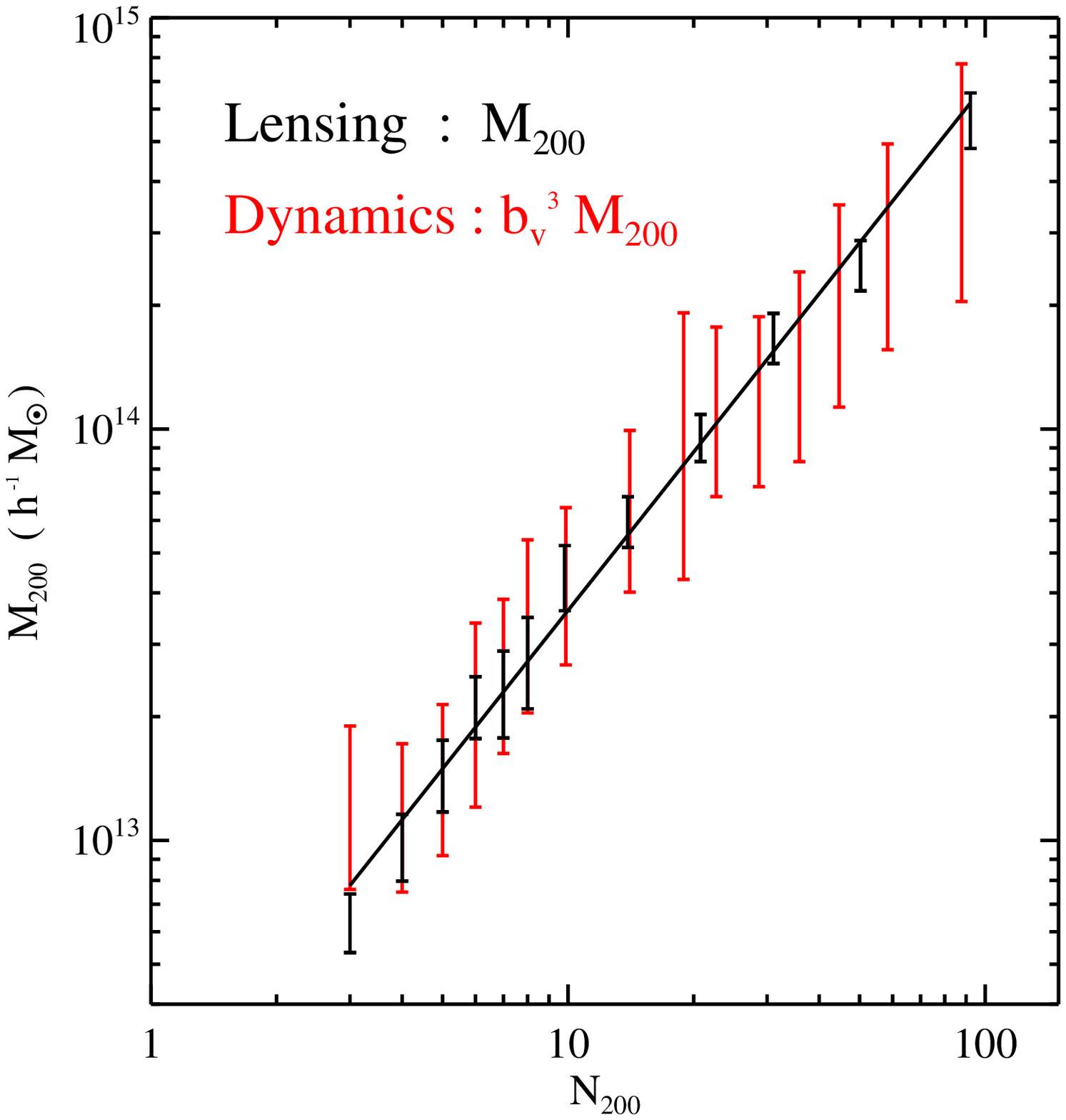}}
\caption{\footnotesize From \cite{Johnston07b}. {\it Left panel:}
Lensing mass $M_{200}$ as a function of the optical luminosity.  {\it
Right panel:} The black points show the best fit weak lensing mass
$M_{200}$ as a function of cluster richness $N_{200}$. The scaling
relation is well described by a single power law with exponent
$\alpha=1.28$.  The red points show the dynamical mass estimates
from \cite{Becker07}, which agree well with the lensing masses.}
\label{fig:massrichness}
\end{center}
\end{figure}

\paragraph{COSMOS:} Although the SDSS analyses have been able to
provide excellent results for the group regime, the shallowness of the
data limits the constraints to relatively low redshift. Furthermore,
the low signal-to-noise ratio per system prevents efficient
multi-wavelength studies. Such work is best done using deep
high-quality data, which is not (yet) available for large areas.  The
COSMOS field of $1.64$ square degrees has been studied extensively,
and a wide range of data, including optical ({\it HST} and VLT) and
X-ray ({\it XMM-Newton} and {\it Chandra}) observations, have been
obtained. \citet{Leauthaud10} used this data set to investigate the
weak lensing properties of a sample of $206$ X-ray--selected galaxy
groups with $0.2\!<\!z\!<\!0.9$ from \citet{Finoguenov07}. The source
catalogue is based on earlier COSMOS results \citep{Leauthaud07}.

\begin{figure*}
\begin{center}
 \includegraphics[width=0.7\textwidth]{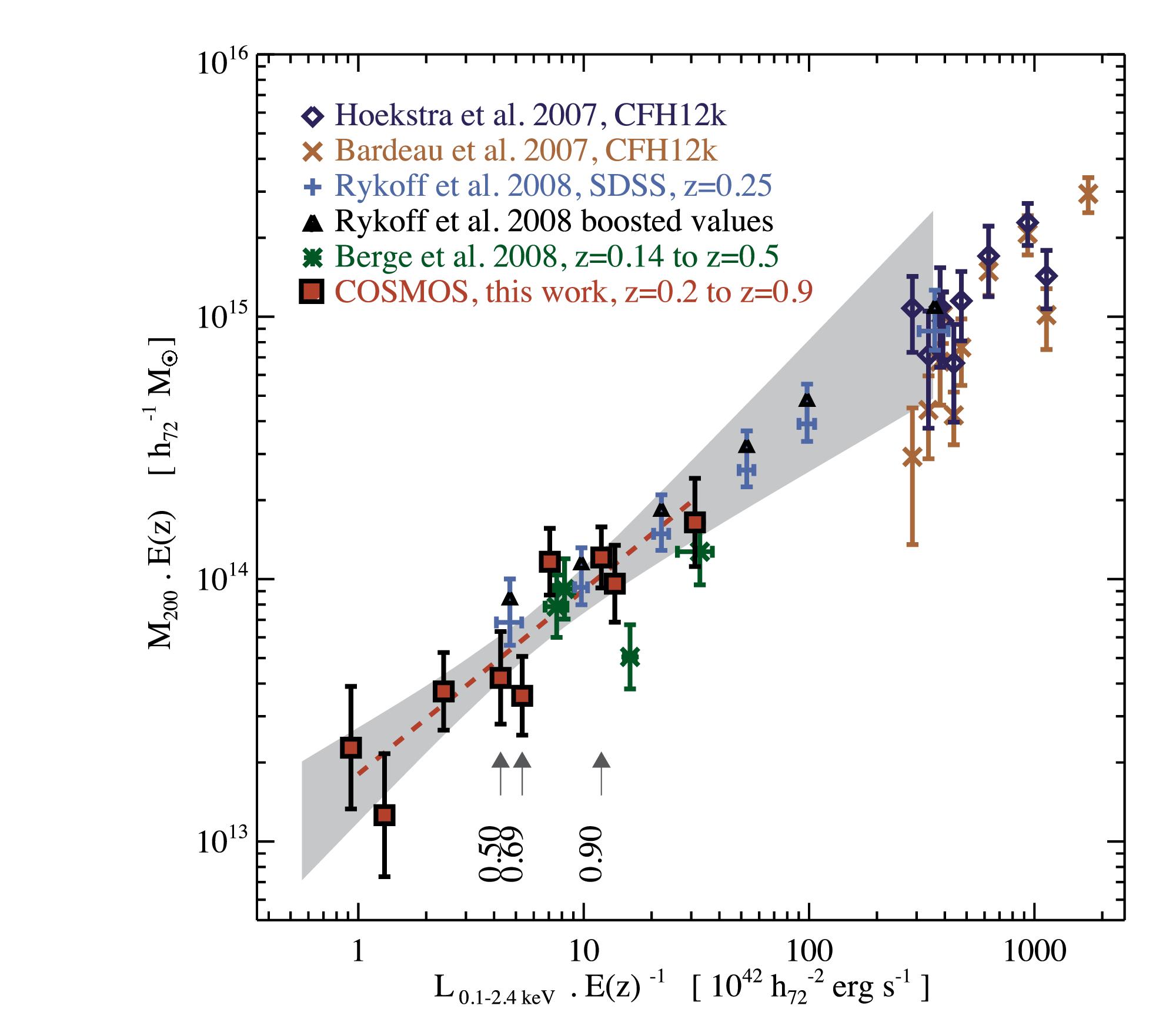}
\end{center}
\caption{Figure from \cite{Leauthaud10}; Scaling relation between 
$M^{\mathrm{wl}}$ and $L_{\mathrm{X}}$, comparing stacking results
from COSMOS galaxy groups \citep[squares]{Leauthaud10} with earlier
results. Arrows point at the three highest redshift COSMOS bins.  The
grey area gives the $68$\% confidence regions of the scaling relation
derived from the COSMOS data.}
\label{fig:le10}
\end{figure*}

\cite{Leauthaud10} stack the shear signal in nine $L_{\mathrm{X}}$ bins
and constrain the slope $\alpha$ of the
$M_{200}^{\mathrm{wl}}$--$L_{\mathrm{X}}$ relation. The results are
reproduced in Figure~\ref{fig:le10}. While their COSMOS data alone
yield $\alpha\!=\!0.66\pm0.14$, a combination with earlier results
tightens constraints to $\alpha\!=\!0.64\pm0.03$, inconsistent with
the self-similar prediction of $\alpha\!=\!0.75$.  Comparing to
earlier studies of the $M_{200}^{\mathrm{wl}}$--$L_{\mathrm{X}}$ relation for
clusters, \citet{Leauthaud10} find overall agreement within the error
bars of their measurement (Fig.~\ref{fig:le10}). The inclusion of
group data improves the knowledge on the scaling relation,
illustrating the usefulness of weak lensing as a calibration tool to
the next generation of X-ray cluster cosmology surveys
\citep[also see the comparison between results in][]{Hoekstra11a}.

\subsection{Recent studies of samples of individual clusters}

The first weak lensing studies targeted massive clusters of galaxies
in order to maximize the signal-to-noise. A number of clusters were
studied in the '90s with relatively small cameras. Although these
pioneering studies have proven critical to advance weak gravitational
lensing to its current state, these masses do not meet the standards
of current state-of-the-art analyses. Modern results are based on deep
wide-field imaging data, and analysed using well-tested shape
measurement algorithms (see e.g. \S\ref{sec:shapemeasurement}).
Consequently, we will limit the discussion in this section to only the
most recent results that study ever larger samples of clusters.

Most of the recent increase in the number of clusters with individual
weak lensing mass estimates has come from ground-based observations,
but larger samples are also being studied using {\it HST}, such as the
ongoing Cluster Lensing And Supernova survey with Hubble (CLASH).  The
most recent published large sample based on {\it HST} data, is the
study by \cite{Jee11} who presented weak lensing masses for a sample
of 22 high-redshift ($0.91<z<1.46$) clusters of
galaxies. Unfortunately the large number of targets with deep imaging
comes at the price of a relatively small field-of-view.  As a result
the masses have been derived from NFW fits to the observed tangential
shear profiles. Although one has little choice in this case, this may
lead to biases because \cite{Jee11} studied clusters that have only
recently formed, and thus may show large deviations from a simple NFW
model. Furthermore, the lack of photometric redshifts for the sources
leads to additional uncertainties as discussed
in \S\ref{sec:zsource}. For the 15 clusters with X-ray temperature
measurements \cite{Jee11} find a power-law slope of $1.54\pm0.23$ for
the $M-T_X$ relation, consistent with the self-similar value. However,
compared to measurements at low redshift, the normalization is lower
by $20-30\%$, suggesting some mild evolution.

The study of massive clusters using ground-based telescopes has also 
progressed tremendously in recent years, and here we will focus on the
three largest studies. We mention briefly a number of other ongoing
studies as well. \cite{Israel12} presented results for an initial
sample of 7 clusters drawn from the 400d galaxy cluster survey, with
the final study targeting 36 $0.35<z<0.9$ clusters. \cite{High12}
presented masses for 5 clusters from the South Pole Telescope (SPT)
survey, with a larger sample already observed. \cite{Foex12} presented
results for a sample of 11 X-ray bright $0.4<z<0.6$ clusters observed
with MegaCam at CFHT. The large ground-based projects that we will review in
more detail are the Local Cluster Substructure
Survey \citep[LoCUSS;][]{Okabe10,Zhang10,Marrone12}, the Canadian
Cluster Comparison
Project \cite[CCCP;][]{Hoekstra07,Mahdavi08,Hoekstra12,Mahdavi12} and
the Weighing the Giants
Project \cite[WtG;][]{VonderLinden12,Kelly12,Applegate12}.

\paragraph{LoCuSS:} \cite{Okabe10} presented weak lensing masses for
29 clusters with $0.15<z<0.3$ using deep Suprime-Cam data obtained
with the Subaru telescope. This sample presents a subset of the full
LoCuSS sample, which is a systematic multi-wavelength survey of X-ray
luminous clusters of galaxies with $L_X>2\times 10^{44}$erg/s.
Importantly, no further selection was made based on other physical
properties. Results on the density profiles of clusters were already
reviewed in \S\ref{sec:profiles}. 

\cite{Zhang10} derived hydrostatic masses for 12 clusters using 
{\it XMM-Newton} observations, and compared these to the weak lensing
masses. They find a good agreement, as opposed to \cite{Mahdavi08} and
\cite{Mahdavi12}. We note here that comparison to other lensing results 
suggest that the \cite{Okabe10} masses are biased low (see
discussion below). Although the sample is small, \cite{Zhang10} find
that the X-ray gas mass shows the least amount of scatter when
compared to the weak lensing mass. A subset of 18 clusters was used
by \cite{Marrone12} to study the scaling relation between the SZ Compton
parameter $Y$ and weak lensing mass. They found an intrinsic scatter
of 20\% in the mass at fixed $Y$, with a suggestion of a dependence on
morphology.

\paragraph{CCCP:} The main aim of CCCP is to compare different baryonic
tracers of cluster mass and to study the cluster-to-cluster variations
in these relations.  Weak lensing masses for an initial sample of 20
clusters, based on archival CFHT observations, were presented
in \cite{Hoekstra07}. This sample was augmented with masses for
another 32 clusters in \cite{Hoekstra12}, based on CFHT Megacam
observations that targeted 30 clusters. The final sample covers
$0.15<z<0.55$ with most of the clusters around $z\sim 0.2$, and thus
overlapping partly with the sample studied by \cite{Okabe10}.
\cite{Hoekstra12} compare their aperture masses to SZ results from
\cite{Bonamente08} and \cite{Planck11}, concluding that the SZ signal
correlates well with weak lensing mass. The intrinsic scatter that
they measure is smaller but consistent with the results from
\cite{Marrone12}. 

The comparison with X-ray measurements is presented
in \cite{Mahdavi12}.  This work confirms the earlier finding
by \cite{Mahdavi08} that hydrostatic masses underestimate the weak
lensing masses by $\sim 10\%$ on average. Splitting the sample by
central entropy suggests X-ray and lensing masses agree for
low-entropy systems, whereas less relaxed clusters show a bias of
$15-20\%$. In support of the finding by \cite{Zhang10} for a sample of
12 clusters, \cite{Mahdavi12} find very little scatter in the gas mass
fractions, with the gas mass in fact being the most robust indicator
of weak lensing mass. This is to be contrasted by claims based on
numerical simulations that suggest that $Y_X$, the product of gas mass
and X-ray temperture, is a better
indicator \citep{Kravtsov06}. \cite{Mahdavi12} do note that the scatter
in $Y_X$ appears to be independent of dynamical state, which may be
beneficical for cosmological studies with galaxy clusters.

\paragraph{WtG:} \cite{Applegate12} presented weak masses for 51
of the most X-ray luminous clusters currently known. Compared to
LoCuSS and CCCP, this work includes a more detailed study of the
contamination by cluster members (see \S\ref{sec:contam}) thanks to
more extensive wavelength coverage, including 5-band photometry for 27
of the clusters. \cite{Applegate12} also quantify the measurement bias using
dedicated image simulations and develop a Bayesian framework to
include the source redshift information. Together with the
accompanying papers \citep{VonderLinden12,Kelly12} this work provides
another step forward in weak lensing cluster lensing, which is
particularly important for the study of high redshift clusters.

There is substantial overlap between the various cluster samples and
this was used by \cite{Applegate12} to compare results. They find that
the mass estimates for the overlapping clusters correlate very well,
but that the normalizations show significant variations. In particular
the \cite{Okabe10} masses are lower by 23\%, whereas the CCCP masses
are 12\% lower than the \cite{Applegate12} measurements. This may
point to biases in the weak lensing signal itself, but differences in
the subsequent steps of converting the signal into a mass may
contribute as well. For instances differences in the assumed source
redshift distributions, mass-concentration relations, fitting ranges,
etc. all contribute at the several percent level. An important task
for the near future is to better understand the impact of the choices
one needs to make.

\section{Halos of cluster members}
\label{sec:galhalo}

Galaxy properties depend on their local environment. The
morphology-density relation \citep{Dressler80} represents a well-known
example of this. In recent years large surveys, such as the SDSS have
enabled much more detailed studies of correlations of galaxy
properties. The cluster environment is rather hostile to infalling
galaxies. An important actor in changing the galaxy properties is
tidal stripping.  For instance, \cite{Cayatte90} found that the extent
of the HI gas in the galaxies of the Virgo cluster decreases when the
galaxies are closer to the cluster center.

Current studies have mostly focused on the baryonic aspects of the
galaxies, but a less studied question is how the properties of the
dark matter halos around galaxies change with environment. Simulations
and observations \citep[e.g.][]{Hoekstra04} have shown that galaxies
in the field are surrounded by extendend dark matter halos. Tidal
stripping of these halos should occur when galaxies (both infalling
and cluster members) interact with the gravitational tidal field of
the cluster. The study of the lensing signal around cluster galaxies
can provide unique constraints, and in this section we briefly review
some of the (relatively few) results that have been obtained to date.

\begin{figure*}
\begin{center}
\includegraphics[width=0.6\textwidth]{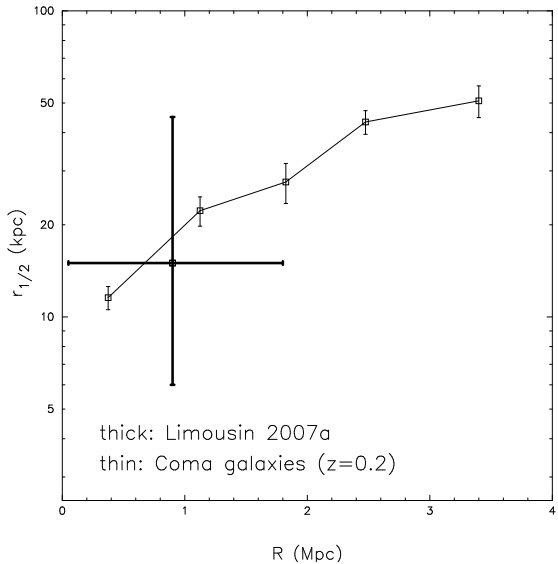}
\end{center}
\caption{\footnotesize From \citet{Limousin09}: simulated half-mass 
radius of dark matter halos as a function of cluster-centric radius
for a Coma-like cluster. The simulations show that halos are stripped
as they are located near the cluster center. The thick data point with
errors are derived from observations by \cite{Limousin07} and in
general agreement with the simulations.
\label{fig:sim_limousin}}
\end{figure*}

\subsection{Comparison with numerical simulations}

Cosmological numerical simulations suggests the presence of
significant substructure in dark matter halos. In the case of galaxy
scale halos these have not (yet) been detected; this has been referred
to as the missing satellite problem \citep[e.g.][]{Klypin99b}. In the
case of galaxy clusters these sub-halos correspond to the halos around
cluster galaxies. Whether or not numerical simulations capture the
formation of such halos is an important question which can be
addressed using cluster galaxy-galaxy lensing studies. For instance,
lack of numerical resolution can lead to premature destruction of simulated
halos, a process known as over-merging \citep[e.g.][]{Klypin99a}.

These considerations suggest that predictions of the galaxy-galaxy
lensing signal in clusters require high resolution simulations.  Such
a study was carried out by \citet{Limousin09} who analysed
hydrodynamical N-body simulations of galaxy clusters to probe tidal
stripping of the dark matter subhalos. They investigated how the
extent of the dark matter halos of cluster galaxies, quantified by
the half mass radius $r_{1/2}$, depends on projected cluster-centric
distance $R$ and how it evolves as a function of redshift. 

As shown in Figure~\ref{fig:sim_limousin}, \citet{Limousin09} find a
clear trend for $r_{1/2}$ with cluster-centric distance: the closer
the galaxies are to the center of the cluster, the smaller the half
mass radius.  This trend is present for all redshifts that were
studied (from $z=0$ to $z=0.7$). At the present day, galaxy halos in
the central regions of clusters are found to be highly truncated, with
the most compact having half mass radius of 10 kpc.  The corresponding total
mass of the cluster galaxies is also found to increase with projected
cluster-centric distance and luminosity, but with more scatter than
the trend of $r_{1/2}$ with radius $R$. The thick data point in
Figure~\ref{fig:sim_limousin} indicates the results of a comparison
with galaxy-galaxy lensing results from \cite{Limousin07} (also see
below), indicating good agreement between the simulations and
observations.

\subsection{Results from cluster galaxy-galaxy lensing}

The shear field generated by a galaxy cluster is dominated by the
central halo, which is relatively smooth. However, when
resolved with enough precision, local anisotropies due to the presence
of cluster members can be discerned. Although the signal caused by a
galaxy lens is typically small, the underlying cluster convergence
boosts the lensing effect. As a result, a detailed modeling of strong
lensing in galaxy clusters can provide interesting constraints, as
well as clear examples of galaxy-galaxy lensing in clusters.

The more subtle deformation in the shapes of background galaxies
produced by the cluster lenses can be measured as well. A
complication is that the halo signal decreases $\propto 1/r^2$ beyond
the radius where the density profile is truncated. Hence it is
important to be able to probe the lensing signal on very small scales.
This is why the first measurements used {\it HST} observations. The study
by \citet{Geiger99} for the galaxy cluster Cl\,0939+4713 led to a
detection of galaxy-galaxy lensing, but the field-of-view was too
small to allow strong conclusions to be drawn about the mass
distribution of the cluster galaxies. A series of studies
by \citet{Priya1, Priya2, Priya3, priya4} also used {\it HST} observations.  In
their analysis they included constraints from multiple images systems
which boosts the precision of the measurements. 

\begin{figure}
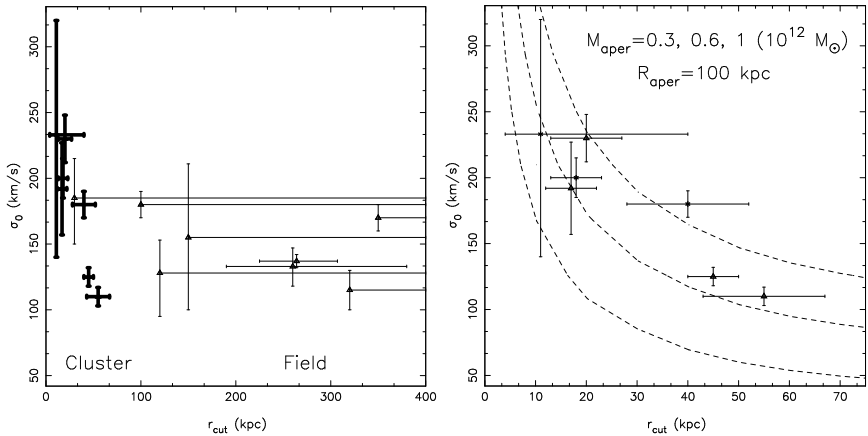

\begin{center}
\hbox{%
\includegraphics[width=0.47\textwidth]{limousin_trunc.eps}
\includegraphics[width=0.46\textwidth]{limousin_zoom.eps}}
\caption{\footnotesize From \cite{Limousin07}. {\it Left panel:}
Comparison of the cluster galaxy-galaxy lensing results (thick black
points) to measurements of field galaxies (thin ponints).  These
results indicate that the halos around cluster galaxies are
significantly smaller compared to galaxies in the field.  {\it Right
panel:} Zoom of the cluster results. Along the dotted lines the mass
within a projected radius of $R_{\rm aper}=100$~kpc is constant. The
three curves correspond to the values indicated in the plot.
\label{fig:trunc}}
\end{center}
\end{figure}

The {\it HST} studies are, however, typically confined to relatively small
cluster-centric radii, because of the small field-of-view. The
exception is the work by \citet{priya4} who used a large {\it HST} mosaic
covering up to 5 Mpc from the center of Cl\,0024+16. As a result
\citet{priya4} were able to probe the galaxy population in three radial 
bins and inferred a larger extent for the halos of galaxies located in
the outskirts of the cluster (i.e. at a cluster-centric distance
between 3 and 5 Mpc) compared to the galaxies in the core of
the cluster (between 0 and 3 Mpc), in agreement with the expectation
from numerical simulations.

Panoramic ground-based cameras, which have proven important for studies
of cosmic shear and cluster mass determinations have led to
ground-based constraints as well. \citet{Limousin07} studied a sample of
massive galaxy clusters observed using the \textsc{CFH12k} camera at
CFHT, which allowed them to probe the cluster galaxy population out to
a large radius ($\sim$ 2 Mpc). The results from \citet{Limousin07} are
reproduced in Figure~\ref{fig:trunc}. The measurements suggest small
truncation radii $r_{\rm cut}$ for galaxies located in massive
clusters, with half-mass radii typically less than 50~kpc. This is
significantly smaller compared to results from studies of field
galaxies \citep{Fischer00,Mckay01,Hoekstra03b,Hoekstra04}, which
suggest half-mass radii larger than 200~kpc for lenses of equivalent
luminosity.

The overall finding of these pioneering studies is that dark matter
halos in high density regions are stripped. The consequence is that
observational constraints are difficult to obtain, because they
require measurements very close to the lenses. Furthermore to study
the sizes as a function of cluster-centric distance requires clean
membership determination, because massive field galaxies can easily
bias the results. The increased statistics from next generation
surveys will improve the measurements, but in particular future
space based projects, such as {\it Euclid}, that combine wide-field
with high-resolution imaging, will provide a major step forward.

\section{Conclusions \& Outlook}

Gravitational lensing studies of clusters of galaxies have gained more
prominence in recent years. Strong lensing is now routinely observed
in deep HST observations, providing a unique (magnified) view of the
most distant galaxies, but also allowing a detailed modelling of the
mass distribution in the central regions of clusters \citep[see][for a
more in-depth discussion of this application]{Meneghetti12}. In this
review we focused on the determination of cluster masses, where
weak gravitational lensing has become an important tool. As shown
in \S\ref{sec:massrec}, it allows for unique studies of the mass
distribution in clusters of galaxies. The results reviewed
in \S\ref{sec:mass} demonstrate that weak lensing studies have become
an important observational link between the results from numerical
simulations and measurements from a number of baryonic tracers.

Cluster samples are increasing rapidly in size thanks to dedicated
optical, SZ and X-ray surveys. Similarly the number of clusters for
which weak lensing masses have been determined is increasing, with the
most recent results presenting masses for several tens of
clusters \citep{Okabe10,Hoekstra12, Applegate12}. Consequently, fewer
results will be dominated by the statistical uncertanties caused by
the intrinsic source ellipticity (although this cannot be avoided when
considering individual systems). Observational systematics, such as
the correction for the (anisotropic) smearing by the PSF which was
discussed in \S\ref{sec:shapemeasurement}, have largely been
considered a problem for cosmic shear studies, but when considering a
sample of $\sim 100$ massive clusters the statistical errors are $\sim
1\%$, and thus quite comparable to the level needed and tested for
cosmic shear. Cluster studies will, however, continue to benefit of
further development in shape measurement techniques. Nonetheless a
careful characterization of systematics is critical for current and
future analyses.

For cluster studies, the dominant source of uncertainty remains the
interpretation of the signal, which is complicated by the fact that
clusters are neither round nor described by simple parametric density
profiles. Comparison with numerical simulations indicate that weak
lensing masses can be biased, but that the level depends on the
details of the analysis. For instance, biases are small when NFW
models are fit to the lensing signal within the virial radius. Masses
derived from aperture masses may have smaller biases, which
nonetheless need to be quantified. Finally, as discussed
in \S\ref{sec:cosmicnoise}, gravitational lensing is senstive to all
matter along the line-of-sight, which introduces another source of
noise, which should be included when quoting the uncertainty in the
mass measurement. This is particularly important when considering
constraints on density profiles.

As reviewed in \S\ref{sec:redshift}, ever since the first weak lensing
studies of clusters has the limited knowledge of the source redshifts
been a dominant source of bias. Thanks to deep, multi-wavelength,
wide-angle surveys the average redshift distribution is now well
established for ground based studies of intermediate redshift
clusters. However, in the case of high-redshift ($z\sim 1$) clusters,
the lack of source redshift information is still a concern. This is a
key regime for weak lensing studies, because those clusters are still
dynamically young, and lensing masses are needed to calibrate other
observables. It is possible to use distributions inferred from other,
deep observations, but current data sets may not be sufficiently
representative. Furthermore, the field-to-field variation increases
the uncertainty in the mass measurement, which is already noisy due to
the small number of galaxies behind the cluster.

None of these complications pose insurmountable problems for an
increasingly more important role of weak gravitational lensing to
study a range of properties of galaxy clusters. Furthermore, new
questions can be addressed with more and better data. For instance the
studies of the properties of dark matter halos in dense environments
have to date been limited. These benefit from redshift information for
the lenses to establish membership and from future space based
observations that can probe the signal down to small radii. To
conclude, the role of weak gravitational lensing in the study of
galaxy clusters is increasing and thus the outlook for this area of
research is positive, especially with the selection of the {\it
Euclid} mission by ESA \citep{Euclid}. This exciting project will
provide a unique data set for the study the distribution of matter in
clusters, using both strong and weak gravitational lensing.

\begin{acknowledgements}
We would like to thank ISSI for their hospitality. HH acknowledges
support from NWO Vidi grant 639.042.814 and Marie Curie IRG Grant
230924. We also thank Elisabetta Semboloni for a careful reading
of this manuscript.
\end{acknowledgements}

\bibliography{mass}   

\end{document}



%% file: mass.bbl
\begin{thebibliography}{192}
\expandafter\ifx\csname natexlab\endcsname\relax\def\natexlab#1{#1}\fi

\bibitem[{{Abdelsalam} {et~al}\mbox{.}(1998){Abdelsalam}, {Saha}, \&
  {Williams}}]{Abdelsalam98}
{Abdelsalam} H.~M., {Saha} P., {Williams} L.~L.~R., 1998, \aj, 116, 1541

\bibitem[{{Allen} {et~al}\mbox{.}(2011){Allen}, {Evrard}, \& {Mantz}}]{Allen11}
{Allen} S.~W., {Evrard} A.~E., {Mantz} A.~B., 2011, \araa, 49, 409

\bibitem[{{Applegate} {et~al}\mbox{.}(2012){Applegate}, {von der Linden},
  {Kelly}, {Allen}, {Allen}, {Burchat}, {Burke}, {Ebeling}, {Mantz}, \&
  {Morris}}]{Applegate12}
{Applegate} D.~E. {et~al.}, 2012, ArXiv e-prints

\bibitem[{{Bacon} {et~al}\mbox{.}(2006){Bacon}, {Goldberg}, {Rowe}, \&
  {Taylor}}]{Bacon06}
{Bacon} D.~J., {Goldberg} D.~M., {Rowe} B.~T.~P., {Taylor} A.~N., 2006, \mnras,
  365, 414

\bibitem[{{Bah{\'e}} {et~al}\mbox{.}(2012){Bah{\'e}}, {McCarthy}, \&
  {King}}]{Bahe12}
{Bah{\'e}} Y.~M., {McCarthy} I.~G., {King} L.~J., 2012, \mnras, 421, 1073

\bibitem[{{Bartelmann}(1995)}]{Bartelmann95}
{Bartelmann} M., 1995, \aap, 299, 11

\bibitem[{{Bartelmann}(1996)}]{Bartelmann96}
{Bartelmann} M., 1996, \aap, 313, 697

\bibitem[{{Bartelmann}(2010)}]{Bartelmann10}
{Bartelmann} M., 2010, Classical and Quantum Gravity, 27, 233001

\bibitem[{{Bartelmann} {et~al}\mbox{.}(2012){Bartelmann}, {Meneghetti}, \& {et
  al.}}]{Bartelmann12}
{Bartelmann} M., {Meneghetti} M., {et al.}, 2012, this volume

\bibitem[{{Bartelmann} {et~al}\mbox{.}(1996){Bartelmann}, {Narayan}, {Seitz},
  \& {Schneider}}]{Bartelmann96b}
{Bartelmann} M., {Narayan} R., {Seitz} S., {Schneider} P., 1996, \apjl, 464,
  L115

\bibitem[{{Bartelmann} \& {Schneider}(2001)}]{Bartelmann01}
{Bartelmann} M., {Schneider} P., 2001, \physrep, 340, 291

\bibitem[{{Bauer} {et~al}\mbox{.}(2012){Bauer}, {Baltay}, {Ellman}, {Jerke},
  {Rabinowitz}, \& {Scalzo}}]{Bauer12}
{Bauer} A.~H., {Baltay} C., {Ellman} N., {Jerke} J., {Rabinowitz} D., {Scalzo}
  R., 2012, \apj, 749, 56

\bibitem[{{Becker} \& {Kravtsov}(2011)}]{Becker11}
{Becker} M.~R., {Kravtsov} A.~V., 2011, \apj, 740, 25

\bibitem[{{Becker} {et~al}\mbox{.}(2007){Becker}, {McKay}, {Koester},
  {Wechsler}, {Rozo}, {Evrard}, {Johnston}, {Sheldon}, {Annis}, {Lau},
  {Nichol}, \& {Miller}}]{Becker07}
{Becker} M.~R. {et~al.}, 2007, \apj, 669, 905

\bibitem[{{Bernstein}(2010)}]{Bernstein10}
{Bernstein} G.~M., 2010, \mnras, 406, 2793

\bibitem[{{Bildfell} {et~al}\mbox{.}(2008){Bildfell}, {Hoekstra}, {Babul}, \&
  {Mahdavi}}]{Bildfell08}
{Bildfell} C., {Hoekstra} H., {Babul} A., {Mahdavi} A., 2008, \mnras, 389, 1637

\bibitem[{{Bonamente} {et~al}\mbox{.}(2008){Bonamente}, {Joy}, {LaRoque},
  {Carlstrom}, {Nagai}, \& {Marrone}}]{Bonamente08}
{Bonamente} M., {Joy} M., {LaRoque} S.~J., {Carlstrom} J.~E., {Nagai} D.,
  {Marrone} D.~P., 2008, \apj, 675, 106

\bibitem[{{Brada{\v c}} {et~al}\mbox{.}(2006){Brada{\v c}}, {Clowe},
  {Gonzalez}, {Marshall}, {Forman}, {Jones}, {Markevitch}, {Randall},
  {Schrabback}, \& {Zaritsky}}]{Bradac06}
{Brada{\v c}} M. {et~al.}, 2006, \apj, 652, 937

\bibitem[{{Brada{\v c}} {et~al}\mbox{.}(2005){Brada{\v c}}, {Schneider},
  {Lombardi}, \& {Erben}}]{Bradac05a}
{Brada{\v c}} M., {Schneider} P., {Lombardi} M., {Erben} T., 2005, \aap, 437,
  39

\bibitem[{{Brainerd} {et~al}\mbox{.}(1996){Brainerd}, {Blandford}, \&
  {Smail}}]{Brainerd96}
{Brainerd} T.~G., {Blandford} R.~D., {Smail} I., 1996, \apj, 466, 623

\bibitem[{{Bridle} {et~al}\mbox{.}(2010){Bridle}, {Balan}, {Bethge}, {Gentile},
  {Harmeling}, {Heymans}, {Hirsch}, {Hosseini}, {Jarvis}, {Kirk}, {Kitching},
  {Kuijken}, {Lewis}, {Paulin-Henriksson}, {Sch{\"o}lkopf}, {Velander},
  {Voigt}, {Witherick}, {Amara}, {Bernstein}, {Courbin}, {Gill}, {Heavens},
  {Mandelbaum}, {Massey}, {Moghaddam}, {Rassat}, {R{\'e}fr{\'e}gier}, {Rhodes},
  {Schrabback}, {Shawe-Taylor}, {Shmakova}, {van Waerbeke}, \&
  {Wittman}}]{GREAT08}
{Bridle} S. {et~al.}, 2010, \mnras, 405, 2044

\bibitem[{{Broadhurst} {et~al}\mbox{.}(2005){Broadhurst}, {Takada}, {Umetsu},
  {Kong}, {Arimoto}, {Chiba}, \& {Futamase}}]{Broadhurst05}
{Broadhurst} T., {Takada} M., {Umetsu} K., {Kong} X., {Arimoto} N., {Chiba} M.,
  {Futamase} T., 2005, \apjl, 619, L143

\bibitem[{{Broadhurst} {et~al}\mbox{.}(1995){Broadhurst}, {Taylor}, \&
  {Peacock}}]{Broadhurst95}
{Broadhurst} T.~J., {Taylor} A.~N., {Peacock} J.~A., 1995, \apj, 438, 49

\bibitem[{{Bryan} \& {Norman}(1998)}]{Bryan98}
{Bryan} G.~L., {Norman} M.~L., 1998, \apj, 495, 80

\bibitem[{{Cacciato} {et~al}\mbox{.}(2006){Cacciato}, {Bartelmann},
  {Meneghetti}, \& {Moscardini}}]{Cacciato06}
{Cacciato} M., {Bartelmann} M., {Meneghetti} M., {Moscardini} L., 2006, \aap,
  458, 349

\bibitem[{{Cayatte} {et~al}\mbox{.}(1990){Cayatte}, {van Gorkom}, {Balkowski},
  \& {Kotanyi}}]{Cayatte90}
{Cayatte} V., {van Gorkom} J.~H., {Balkowski} C., {Kotanyi} C., 1990, \aj, 100,
  604

\bibitem[{{Clowe} {et~al}\mbox{.}(2006{\natexlab{a}}){Clowe}, {Brada{\v c}},
  {Gonzalez}, {Markevitch}, {Randall}, {Jones}, \& {Zaritsky}}]{Clowe96}
{Clowe} D., {Brada{\v c}} M., {Gonzalez} A.~H., {Markevitch} M., {Randall}
  S.~W., {Jones} C., {Zaritsky} D., 2006{\natexlab{a}}, \apjl, 648, L109

\bibitem[{{Clowe} {et~al}\mbox{.}(2006{\natexlab{b}}){Clowe}, {Brada{\v c}},
  {Gonzalez}, {Markevitch}, {Randall}, {Jones}, \& {Zaritsky}}]{Clowe06}
{Clowe} D., {Brada{\v c}} M., {Gonzalez} A.~H., {Markevitch} M., {Randall}
  S.~W., {Jones} C., {Zaritsky} D., 2006{\natexlab{b}}, \apjl, 648, L109

\bibitem[{{Clowe} {et~al}\mbox{.}(2004{\natexlab{a}}){Clowe}, {De Lucia}, \&
  {King}}]{Clowe04}
{Clowe} D., {De Lucia} G., {King} L., 2004{\natexlab{a}}, \mnras, 350, 1038

\bibitem[{{Clowe} {et~al}\mbox{.}(2004{\natexlab{b}}){Clowe}, {Gonzalez}, \&
  {Markevitch}}]{Clowe04b}
{Clowe} D., {Gonzalez} A., {Markevitch} M., 2004{\natexlab{b}}, \apj, 604, 596

\bibitem[{{Clowe} {et~al}\mbox{.}(1998){Clowe}, {Luppino}, {Kaiser}, {Henry},
  \& {Gioia}}]{Clowe98}
{Clowe} D., {Luppino} G.~A., {Kaiser} N., {Henry} J.~P., {Gioia} I.~M., 1998,
  \apjl, 497, L61

\bibitem[{{Clowe} {et~al}\mbox{.}(2012){Clowe}, {Markevitch}, {Brada{\v c}},
  {Gonzalez}, {Chung}, {Massey}, \& {Zaritsky}}]{Clowe12}
{Clowe} D., {Markevitch} M., {Brada{\v c}} M., {Gonzalez} A.~H., {Chung} S.~M.,
  {Massey} R., {Zaritsky} D., 2012, \apj, 758, 128

\bibitem[{{Coe} {et~al}\mbox{.}(2006){Coe}, {Ben{\'{\i}}tez}, {S{\'a}nchez},
  {Jee}, {Bouwens}, \& {Ford}}]{Coe06}
{Coe} D., {Ben{\'{\i}}tez} N., {S{\'a}nchez} S.~F., {Jee} M., {Bouwens} R.,
  {Ford} H., 2006, \aj, 132, 926

\bibitem[{{Coe} {et~al}\mbox{.}(2012){Coe}, {Umetsu}, {Zitrin}, {Donahue},
  {Medezinski}, {Postman}, {Carrasco}, {Anguita}, {Geller}, {Rines},
  {Diaferio}, {Kurtz}, {Bradley}, {Koekemoer}, {Zheng}, {Nonino}, {Molino},
  {Mahdavi}, {Lemze}, {Infante}, {Ogaz}, {Melchior}, {Host}, {Ford}, {Grillo},
  {Rosati}, {Jim{\'e}nez-Teja}, {Moustakas}, {Broadhurst}, {Ascaso}, {Lahav},
  {Bartelmann}, {Ben{\'{\i}}tez}, {Bouwens}, {Graur}, {Graves}, {Jha},
  {Jouvel}, {Kelson}, {Moustakas}, {Maoz}, {Meneghetti}, {Merten}, {Riess},
  {Rodney}, \& {Seitz}}]{Coe12}
{Coe} D. {et~al.}, 2012, ArXiv e-prints

\bibitem[{{Corless} \& {King}(2007)}]{Corless07}
{Corless} V.~L., {King} L.~J., 2007, \mnras, 380, 149

\bibitem[{{Corless} \& {King}(2008)}]{Corless08}
{Corless} V.~L., {King} L.~J., 2008, \mnras, 390, 997

\bibitem[{{Corless} {et~al}\mbox{.}(2009){Corless}, {King}, \&
  {Clowe}}]{Corless09}
{Corless} V.~L., {King} L.~J., {Clowe} D., 2009, \mnras, 393, 1235

\bibitem[{{Dawson} {et~al}\mbox{.}(2011){Dawson}, {Wittman}, {Jee}, {Gee},
  {Hughes}, {Tyson}, {Schmidt}, {Thorman}, {Bradac}, {Miyazaki}, {Lemaux}, \&
  {Utsumi}}]{Dawson11}
{Dawson} W.~A. {et~al.}, 2011, ArXiv:1110.4391

\bibitem[{{de Putter} \& {White}(2005)}]{dePutter05}
{de Putter} R., {White} M., 2005, New Astronomy, 10, 676

\bibitem[{{Diaferio}(1999)}]{Diaferio99}
{Diaferio} A., 1999, \mnras, 309, 610

\bibitem[{{Diaferio} \& {Geller}(1997)}]{Diaferio97}
{Diaferio} A., {Geller} M.~J., 1997, \apj, 481, 633

\bibitem[{{Diaferio} {et~al}\mbox{.}(2005){Diaferio}, {Geller}, \&
  {Rines}}]{Diaferio05}
{Diaferio} A., {Geller} M.~J., {Rines} K.~J., 2005, \apjl, 628, L97

\bibitem[{{Diego} {et~al}\mbox{.}(2007){Diego}, {Tegmark}, {Protopapas}, \&
  {Sandvik}}]{Diego07}
{Diego} J.~M., {Tegmark} M., {Protopapas} P., {Sandvik} H.~B., 2007, \mnras,
  375, 958

\bibitem[{{Dodelson}(2004)}]{Dodelson04}
{Dodelson} S., 2004, \prd, 70, 023008

\bibitem[{{Dressler}(1980)}]{Dressler80}
{Dressler} A., 1980, \apj, 236, 351

\bibitem[{{Duffy} {et~al}\mbox{.}(2010){Duffy}, {Schaye}, {Kay}, {Dalla
  Vecchia}, {Battye}, \& {Booth}}]{Duffy10}
{Duffy} A.~R., {Schaye} J., {Kay} S.~T., {Dalla Vecchia} C., {Battye} R.~A.,
  {Booth} C.~M., 2010, \mnras, 405, 2161

\bibitem[{{Erben} {et~al}\mbox{.}(2001){Erben}, {Van Waerbeke}, {Bertin},
  {Mellier}, \& {Schneider}}]{Erben01}
{Erben} T., {Van Waerbeke} L., {Bertin} E., {Mellier} Y., {Schneider} P., 2001,
  \aap, 366, 717

\bibitem[{{Finoguenov} {et~al}\mbox{.}(2007){Finoguenov}, {Guzzo}, {Hasinger},
  {Scoville}, {Aussel}, {B{\"o}hringer}, {Brusa}, {Capak}, {Cappelluti},
  {Comastri}, {Giodini}, {Griffiths}, {Impey}, {Koekemoer}, {Kneib},
  {Leauthaud}, {Le F{\`e}vre}, {Lilly}, {Mainieri}, {Massey}, {McCracken},
  {Mobasher}, {Murayama}, {Peacock}, {Sakelliou}, {Schinnerer}, {Silverman},
  {Smol{\v c}i{\'c}}, {Taniguchi}, {Tasca}, {Taylor}, {Trump}, \&
  {Zamorani}}]{Finoguenov07}
{Finoguenov} A. {et~al.}, 2007, \apjs, 172, 182

\bibitem[{{Fischer} {et~al}\mbox{.}(2000){Fischer}, {McKay}, {Sheldon},
  {Connolly}, {Stebbins}, {Frieman}, {Jain}, {Joffre}, {Johnston}, {Bernstein},
  {Annis}, {Bahcall}, {Brinkmann}, {Carr}, {Csabai}, {Gunn}, {Hennessy},
  {Hindsley}, {Hull}, {Ivezi{\'c}}, {Knapp}, {Limmongkol}, {Lupton}, {Munn},
  {Nash}, {Newberg}, {Owen}, {Pier}, {Rockosi}, {Schneider}, {Smith},
  {Stoughton}, {Szalay}, {Szokoly}, {Thakar}, {Vogeley}, {Waddell}, {Weinberg},
  {York}, \& {SDSS Collaboration}}]{Fischer00}
{Fischer} P. {et~al.}, 2000, \aj, 120, 1198

\bibitem[{{Fo{\"e}x} {et~al}\mbox{.}(2012){Fo{\"e}x}, {Soucail},
  {Pointecouteau}, {Arnaud}, {Limousin}, \& {Pratt}}]{Foex12}
{Fo{\"e}x} G., {Soucail} G., {Pointecouteau} E., {Arnaud} M., {Limousin} M.,
  {Pratt} G.~W., 2012, \aap, 546, A106

\bibitem[{{Ford} {et~al}\mbox{.}(2012){Ford}, {Hildebrandt}, {Van Waerbeke},
  {Leauthaud}, {Capak}, {Finoguenov}, {Tanaka}, {George}, \& {Rhodes}}]{Ford12}
{Ford} J. {et~al.}, 2012, \apj, 754, 143

\bibitem[{{Geiger} \& {Schneider}(1999)}]{Geiger99}
{Geiger} B., {Schneider} P., 1999, \mnras, 302, 118

\bibitem[{{Giodini} {et~al}\mbox{.}(2012){Giodini}, {Lovisari}, \& {et
  al.}}]{Giodini12}
{Giodini} S., {Lovisari} L., {et al.}, 2012, this volume

\bibitem[{{Goldberg} \& {Bacon}(2005)}]{Goldberg05}
{Goldberg} D.~M., {Bacon} D.~J., 2005, \apj, 619, 741

\bibitem[{{Goldberg} \& {Natarajan}(2002)}]{Golberg02}
{Goldberg} D.~M., {Natarajan} P., 2002, \apj, 564, 65

\bibitem[{{Golse} {et~al}\mbox{.}(2002){Golse}, {Kneib}, \&
  {Soucail}}]{Golse02}
{Golse} G., {Kneib} J.-P., {Soucail} G., 2002, \aap, 387, 788

\bibitem[{{Gorenstein} {et~al}\mbox{.}(1988){Gorenstein}, {Shapiro}, \&
  {Falco}}]{Gorenstein88}
{Gorenstein} M.~V., {Shapiro} I.~I., {Falco} E.~E., 1988, \apj, 327, 693

\bibitem[{{Guzik} \& {Seljak}(2002)}]{Guzik02}
{Guzik} J., {Seljak} U., 2002, \mnras, 335, 311

\bibitem[{{Heymans} {et~al}\mbox{.}(2006){Heymans}, {Van Waerbeke}, {Bacon},
  {Berge}, {Bernstein}, {Bertin}, {Bridle}, {Brown}, {Clowe}, {Dahle}, {Erben},
  {Gray}, {Hetterscheidt}, {Hoekstra}, {Hudelot}, {Jarvis}, {Kuijken},
  {Margoniner}, {Massey}, {Mellier}, {Nakajima}, {Refregier}, {Rhodes},
  {Schrabback}, \& {Wittman}}]{STEP1}
{Heymans} C. {et~al.}, 2006, \mnras, 368, 1323

\bibitem[{{Heymans} {et~al}\mbox{.}(2012){Heymans}, {Van Waerbeke}, {Miller},
  {Erben}, {Hildebrandt}, {Hoekstra}, {Kitching}, {Mellier}, {Simon},
  {Bonnett}, {Coupon}, {Fu}, {Harnois-D'eraps}, {Hudson}, {Kilbinger},
  {Kuijken}, {Rowe}, {Schrabback}, {Semboloni}, {van Uitert}, {Vafaei}, \&
  {Velander}}]{Heymans12}
{Heymans} C. {et~al.}, 2012, ArXiv e-prints

\bibitem[{{High} {et~al}\mbox{.}(2012){High}, {Hoekstra}, {Leethochawalit}, {de
  Haan}, {Abramson}, {Aird}, {Armstrong}, {Ashby}, {Bautz}, {Bayliss}, {Bazin},
  {Benson}, {Bleem}, {Brodwin}, {Carlstrom}, {Chang}, {Cho}, {Clocchiatti},
  {Conroy}, {Crawford}, {Crites}, {Desai}, {Dobbs}, {Dudley}, {Foley},
  {Forman}, {George}, {Gladders}, {Gonzalez}, {Halverson}, {Harrington},
  {Holder}, {Holzapfel}, {Hoover}, {Hrubes}, {Jones}, {Joy}, {Keisler}, {Knox},
  {Lee}, {Leitch}, {Liu}, {Lueker}, {Luong-Van}, {Mantz}, {Marrone},
  {McDonald}, {McMahon}, {Mehl}, {Meyer}, {Mocanu}, {Mohr}, {Montroy},
  {Murray}, {Natoli}, {Nurgaliev}, {Padin}, {Plagge}, {Pryke}, {Reichardt},
  {Rest}, {Ruel}, {Ruhl}, {Saliwanchik}, {Saro}, {Sayre}, {Schaffer}, {Shaw},
  {Schrabback}, {Shirokoff}, {Song}, {Spieler}, {Stalder}, {Staniszewski},
  {Stark}, {Story}, {Stubbs}, {Suhada}, {Tokarz}, {van Engelen}, {Vanderlinde},
  {Vieira}, {Vikhlinin}, {Williamson}, {Zahn}, \& {Zenteno}}]{High12}
{High} F.~W. {et~al.}, 2012, ArXiv e-prints

\bibitem[{{Hildebrandt} {et~al}\mbox{.}(2012){Hildebrandt}, {Erben}, {Kuijken},
  {van Waerbeke}, {Heymans}, {Coupon}, {Benjamin}, {Bonnett}, {Fu}, {Hoekstra},
  {Kitching}, {Mellier}, {Miller}, {Velander}, {Hudson}, {Rowe}, {Schrabback},
  {Semboloni}, \& {Ben{\'{\i}}tez}}]{Hildebrandt12}
{Hildebrandt} H. {et~al.}, 2012, \mnras, 421, 2355

\bibitem[{{Hildebrandt} {et~al}\mbox{.}(2011){Hildebrandt}, {Muzzin}, {Erben},
  {Hoekstra}, {Kuijken}, {Surace}, {van Waerbeke}, {Wilson}, \&
  {Yee}}]{Hildebrandt11}
{Hildebrandt} H. {et~al.}, 2011, \apjl, 733, L30

\bibitem[{{Hildebrandt} {et~al}\mbox{.}(2009){Hildebrandt}, {van Waerbeke}, \&
  {Erben}}]{Hildebrandt09}
{Hildebrandt} H., {van Waerbeke} L., {Erben} T., 2009, \aap, 507, 683

\bibitem[{{Hirata} \& {Seljak}(2004)}]{Hirata04}
{Hirata} C.~M., {Seljak} U., 2004, \prd, 70, 063526

\bibitem[{{Hoekstra}(2001)}]{Hoekstra01a}
{Hoekstra} H., 2001, \aap, 370, 743

\bibitem[{{Hoekstra}(2003)}]{Hoekstra03a}
{Hoekstra} H., 2003, \mnras, 339, 1155

\bibitem[{{Hoekstra}(2004)}]{Hoekstra04}
{Hoekstra} H., 2004, \mnras, 347, 1337

\bibitem[{{Hoekstra}(2007)}]{Hoekstra07}
{Hoekstra} H., 2007, \mnras, 379, 317

\bibitem[{{Hoekstra} {et~al}\mbox{.}(2011{\natexlab{a}}){Hoekstra}, {Donahue},
  {Conselice}, {McNamara}, \& {Voit}}]{Hoekstra11a}
{Hoekstra} H., {Donahue} M., {Conselice} C.~J., {McNamara} B.~R., {Voit} G.~M.,
  2011{\natexlab{a}}, \apj, 726, 48

\bibitem[{{Hoekstra} {et~al}\mbox{.}(2000){Hoekstra}, {Franx}, \&
  {Kuijken}}]{Hoekstra00}
{Hoekstra} H., {Franx} M., {Kuijken} K., 2000, \apj, 532, 88

\bibitem[{{Hoekstra} {et~al}\mbox{.}(2003){Hoekstra}, {Franx}, {Kuijken},
  {Carlberg}, \& {Yee}}]{Hoekstra03b}
{Hoekstra} H., {Franx} M., {Kuijken} K., {Carlberg} R.~G., {Yee} H.~K.~C.,
  2003, \mnras, 340, 609

\bibitem[{{Hoekstra} {et~al}\mbox{.}(2001){Hoekstra}, {Franx}, {Kuijken},
  {Carlberg}, {Yee}, {Lin}, {Morris}, {Hall}, {Patton}, {Sawicki}, \&
  {Wirth}}]{Hoekstra01b}
{Hoekstra} H. {et~al.}, 2001, \apjl, 548, L5

\bibitem[{{Hoekstra} {et~al}\mbox{.}(1998){Hoekstra}, {Franx}, {Kuijken}, \&
  {Squires}}]{Hoekstra98}
{Hoekstra} H., {Franx} M., {Kuijken} K., {Squires} G., 1998, \apj, 504, 636

\bibitem[{{Hoekstra} {et~al}\mbox{.}(2002{\natexlab{a}}){Hoekstra}, {Franx},
  {Kuijken}, \& {van Dokkum}}]{Hoekstra02a}
{Hoekstra} H., {Franx} M., {Kuijken} K., {van Dokkum} P.~G.,
  2002{\natexlab{a}}, \mnras, 333, 911

\bibitem[{{Hoekstra} {et~al}\mbox{.}(2011{\natexlab{b}}){Hoekstra}, {Hartlap},
  {Hilbert}, \& {van Uitert}}]{Hoekstra11b}
{Hoekstra} H., {Hartlap} J., {Hilbert} S., {van Uitert} E., 2011{\natexlab{b}},
  \mnras, 412, 2095

\bibitem[{{Hoekstra} \& {Jain}(2008)}]{Hoekstra08}
{Hoekstra} H., {Jain} B., 2008, Annual Review of Nuclear and Particle Science,
  58, 99

\bibitem[{{Hoekstra} {et~al}\mbox{.}(2012){Hoekstra}, {Mahdavi}, {Babul}, \&
  {Bildfell}}]{Hoekstra12}
{Hoekstra} H., {Mahdavi} A., {Babul} A., {Bildfell} C., 2012, ArXiv e-prints

\bibitem[{{Hoekstra} {et~al}\mbox{.}(2002{\natexlab{b}}){Hoekstra}, {Yee},
  {Gladders}, {Barrientos}, {Hall}, \& {Infante}}]{Hoekstra02b}
{Hoekstra} H., {Yee} H.~K.~C., {Gladders} M.~D., {Barrientos} L.~F., {Hall}
  P.~B., {Infante} L., 2002{\natexlab{b}}, \apj, 572, 55

\bibitem[{{Ilbert} {et~al}\mbox{.}(2006){Ilbert}, {Arnouts}, {McCracken},
  {Bolzonella}, {Bertin}, {Le F{\`e}vre}, {Mellier}, {Zamorani}, {Pell{\`o}},
  {Iovino}, {Tresse}, {Le Brun}, {Bottini}, {Garilli}, {Maccagni}, {Picat},
  {Scaramella}, {Scodeggio}, {Vettolani}, {Zanichelli}, {Adami}, {Bardelli},
  {Cappi}, {Charlot}, {Ciliegi}, {Contini}, {Cucciati}, {Foucaud}, {Franzetti},
  {Gavignaud}, {Guzzo}, {Marano}, {Marinoni}, {Mazure}, {Meneux}, {Merighi},
  {Paltani}, {Pollo}, {Pozzetti}, {Radovich}, {Zucca}, {Bondi}, {Bongiorno},
  {Busarello}, {de La Torre}, {Gregorini}, {Lamareille}, {Mathez}, {Merluzzi},
  {Ripepi}, {Rizzo}, \& {Vergani}}]{Ilbert06}
{Ilbert} O. {et~al.}, 2006, \aap, 457, 841

\bibitem[{{Ilbert} {et~al}\mbox{.}(2009){Ilbert}, {Capak}, {Salvato}, {Aussel},
  {McCracken}, {Sanders}, {Scoville}, {Kartaltepe}, {Arnouts}, {Le Floc'h},
  {Mobasher}, {Taniguchi}, {Lamareille}, {Leauthaud}, {Sasaki}, {Thompson},
  {Zamojski}, {Zamorani}, {Bardelli}, {Bolzonella}, {Bongiorno}, {Brusa},
  {Caputi}, {Carollo}, {Contini}, {Cook}, {Coppa}, {Cucciati}, {de la Torre},
  {de Ravel}, {Franzetti}, {Garilli}, {Hasinger}, {Iovino}, {Kampczyk},
  {Kneib}, {Knobel}, {Kovac}, {Le Borgne}, {Le Brun}, {F{\`e}vre}, {Lilly},
  {Looper}, {Maier}, {Mainieri}, {Mellier}, {Mignoli}, {Murayama}, {Pell{\`o}},
  {Peng}, {P{\'e}rez-Montero}, {Renzini}, {Ricciardelli}, {Schiminovich},
  {Scodeggio}, {Shioya}, {Silverman}, {Surace}, {Tanaka}, {Tasca}, {Tresse},
  {Vergani}, \& {Zucca}}]{Ilbert09}
{Ilbert} O. {et~al.}, 2009, \apj, 690, 1236

\bibitem[{{Irwin} \& {Shmakova}(2005)}]{Irwin05}
{Irwin} J., {Shmakova} M., 2005, New Astronomy Reviews, 49, 83

\bibitem[{{Israel} {et~al}\mbox{.}(2010){Israel}, {Erben}, {Reiprich},
  {Vikhlinin}, {Hildebrandt}, {Hudson}, {McLeod}, {Sarazin}, {Schneider}, \&
  {Zhang}}]{Israel10}
{Israel} H. {et~al.}, 2010, \aap, 520, A58

\bibitem[{{Israel} {et~al}\mbox{.}(2012){Israel}, {Erben}, {Reiprich},
  {Vikhlinin}, {Sarazin}, \& {Schneider}}]{Israel12}
{Israel} H., {Erben} T., {Reiprich} T.~H., {Vikhlinin} A., {Sarazin} C.~L.,
  {Schneider} P., 2012, \aap, 546, A79

\bibitem[{{Jain} \& {Taylor}(2003)}]{Jain03}
{Jain} B., {Taylor} A., 2003, Physical Review Letters, 91, 141302

\bibitem[{{Jee} {et~al}\mbox{.}(2011){Jee}, {Dawson}, {Hoekstra}, {Perlmutter},
  {Rosati}, {Brodwin}, {Suzuki}, {Koester}, {Postman}, {Lubin}, {Meyers},
  {Stanford}, {Barbary}, {Barrientos}, {Eisenhardt}, {Ford}, {Gilbank},
  {Gladders}, {Gonzalez}, {Harris}, {Huang}, {Lidman}, {Rykoff}, {Rubin}, \&
  {Spadafora}}]{Jee11}
{Jee} M.~J. {et~al.}, 2011, \apj, 737, 59

\bibitem[{{Jee} {et~al}\mbox{.}(2012){Jee}, {Mahdavi}, {Hoekstra}, {Babul},
  {Dalcanton}, {Carroll}, \& {Capak}}]{Jee12}
{Jee} M.~J., {Mahdavi} A., {Hoekstra} H., {Babul} A., {Dalcanton} J.~J.,
  {Carroll} P., {Capak} P., 2012, \apj, 747, 96

\bibitem[{{Jing} \& {Suto}(2002)}]{Jing02}
{Jing} Y.~P., {Suto} Y., 2002, \apj, 574, 538

\bibitem[{{Joachimi} {et~al}\mbox{.}(2011){Joachimi}, {Mandelbaum}, {Abdalla},
  \& {Bridle}}]{Joachimi11}
{Joachimi} B., {Mandelbaum} R., {Abdalla} F.~B., {Bridle} S.~L., 2011, \aap,
  527, A26

\bibitem[{{Johnston} {et~al}\mbox{.}(2007{\natexlab{a}}){Johnston}, {Sheldon},
  {Tasitsiomi}, {Frieman}, {Wechsler}, \& {McKay}}]{Johnston07a}
{Johnston} D.~E., {Sheldon} E.~S., {Tasitsiomi} A., {Frieman} J.~A., {Wechsler}
  R.~H., {McKay} T.~A., 2007{\natexlab{a}}, \apj, 656, 27

\bibitem[{{Johnston} {et~al}\mbox{.}(2007{\natexlab{b}}){Johnston}, {Sheldon},
  {Wechsler}, {Rozo}, {Koester}, {Frieman}, {McKay}, {Evrard}, {Becker}, \&
  {Annis}}]{Johnston07b}
{Johnston} D.~E. {et~al.}, 2007{\natexlab{b}}, ArXiv e-prints

\bibitem[{{Jullo} {et~al}\mbox{.}(2007){Jullo}, {Kneib}, {Limousin},
  {El{\'{\i}}asd{\'o}ttir}, {Marshall}, \& {Verdugo}}]{Jullo07}
{Jullo} E., {Kneib} J.-P., {Limousin} M., {El{\'{\i}}asd{\'o}ttir} {\'A}.,
  {Marshall} P.~J., {Verdugo} T., 2007, New Journal of Physics, 9, 447

\bibitem[{{Jullo} {et~al}\mbox{.}(2010){Jullo}, {Natarajan}, {Kneib},
  {D'Aloisio}, {Limousin}, {Richard}, \& {Schimd}}]{Jullo10}
{Jullo} E., {Natarajan} P., {Kneib} J.-P., {D'Aloisio} A., {Limousin} M.,
  {Richard} J., {Schimd} C., 2010, Science, 329, 924

\bibitem[{{Kaiser} \& {Squires}(1993)}]{KS93}
{Kaiser} N., {Squires} G., 1993, \apj, 404, 441

\bibitem[{{Kaiser} {et~al}\mbox{.}(1995){Kaiser}, {Squires}, \&
  {Broadhurst}}]{KSB}
{Kaiser} N., {Squires} G., {Broadhurst} T., 1995, \apj, 449, 460

\bibitem[{{Kelly} {et~al}\mbox{.}(2012){Kelly}, {von der Linden}, {Applegate},
  {Allen}, {Allen}, {Burchat}, {Burke}, {Ebeling}, {Capak}, {Czoske},
  {Donovan}, {Mantz}, \& {Morris}}]{Kelly12}
{Kelly} P.~L. {et~al.}, 2012, ArXiv e-prints

\bibitem[{{Kitching} {et~al}\mbox{.}(2012){Kitching}, {Balan}, {Bridle},
  {Cantale}, {Courbin}, {Eifler}, {Gentile}, {Gill}, {Harmeling}, {Heymans},
  {Hirsch}, {Honscheid}, {Kacprzak}, {Kirkby}, {Margala}, {Massey}, {Melchior},
  {Nurbaeva}, {Patton}, {Rhodes}, {Rowe}, {Taylor}, {Tewes}, {Viola},
  {Witherick}, {Voigt}, {Young}, \& {Zuntz}}]{GREAT10}
{Kitching} T.~D. {et~al.}, 2012, \mnras, 423, 3163

\bibitem[{{Klypin} {et~al}\mbox{.}(1999{\natexlab{a}}){Klypin},
  {Gottl{\"o}ber}, {Kravtsov}, \& {Khokhlov}}]{Klypin99a}
{Klypin} A., {Gottl{\"o}ber} S., {Kravtsov} A.~V., {Khokhlov} A.~M.,
  1999{\natexlab{a}}, \apj, 516, 530

\bibitem[{{Klypin} {et~al}\mbox{.}(1999{\natexlab{b}}){Klypin}, {Kravtsov},
  {Valenzuela}, \& {Prada}}]{Klypin99b}
{Klypin} A., {Kravtsov} A.~V., {Valenzuela} O., {Prada} F., 1999{\natexlab{b}},
  \apj, 522, 82

\bibitem[{{Kneib} {et~al}\mbox{.}(1996){Kneib}, {Ellis}, {Smail}, {Couch}, \&
  {Sharples}}]{Kneib96}
{Kneib} J.-P., {Ellis} R.~S., {Smail} I., {Couch} W.~J., {Sharples} R.~M.,
  1996, \apj, 471, 643

\bibitem[{{Koester} {et~al}\mbox{.}(2007){Koester}, {McKay}, {Annis},
  {Wechsler}, {Evrard}, {Bleem}, {Becker}, {Johnston}, {Sheldon}, {Nichol},
  {Miller}, {Scranton}, {Bahcall}, {Barentine}, {Brewington}, {Brinkmann},
  {Harvanek}, {Kleinman}, {Krzesinski}, {Long}, {Nitta}, {Schneider},
  {Sneddin}, {Voges}, \& {York}}]{Koester07}
{Koester} B.~P. {et~al.}, 2007, \apj, 660, 239

\bibitem[{{Kravtsov} {et~al}\mbox{.}(2006){Kravtsov}, {Vikhlinin}, \&
  {Nagai}}]{Kravtsov06}
{Kravtsov} A.~V., {Vikhlinin} A., {Nagai} D., 2006, \apj, 650, 128

\bibitem[{{Kuijken}(1999)}]{Kuijken99}
{Kuijken} K., 1999, \aap, 352, 355

\bibitem[{{Kuijken}(2008)}]{Kuijken08}
{Kuijken} K., 2008, \aap, 482, 1053

\bibitem[{{Laureijs} {et~al}\mbox{.}(2011){Laureijs}, {Amiaux}, {Arduini},
  {Augu{\`e}res}, {Brinchmann}, {Cole}, {Cropper}, {Dabin}, {Duvet}, {Ealet},
  \& et~al.}]{Euclid}
{Laureijs} R. {et~al.}, 2011, ArXiv e-prints

\bibitem[{{Leauthaud} {et~al}\mbox{.}(2010){Leauthaud}, {Finoguenov}, {Kneib},
  {Taylor}, {Massey}, {Rhodes}, {Ilbert}, {Bundy}, {Tinker}, {George}, {Capak},
  {Koekemoer}, {Johnston}, {Zhang}, {Cappelluti}, {Ellis}, {Elvis}, {Giodini},
  {Heymans}, {Le F{\`e}vre}, {Lilly}, {McCracken}, {Mellier},
  {R{\'e}fr{\'e}gier}, {Salvato}, {Scoville}, {Smoot}, {Tanaka}, {Van
  Waerbeke}, \& {Wolk}}]{Leauthaud10}
{Leauthaud} A. {et~al.}, 2010, \apj, 709, 97

\bibitem[{{Leauthaud} {et~al}\mbox{.}(2007){Leauthaud}, {Massey}, {Kneib},
  {Rhodes}, {Johnston}, {Capak}, {Heymans}, {Ellis}, {Koekemoer}, {Le
  F{\`e}vre}, {Mellier}, {R{\'e}fr{\'e}gier}, {Robin}, {Scoville}, {Tasca},
  {Taylor}, \& {Van Waerbeke}}]{Leauthaud07}
{Leauthaud} A. {et~al.}, 2007, \apjs, 172, 219

\bibitem[{{Leonard} {et~al}\mbox{.}(2007){Leonard}, {Goldberg}, {Haaga}, \&
  {Massey}}]{Leonard07}
{Leonard} A., {Goldberg} D.~M., {Haaga} J.~L., {Massey} R., 2007, \apj, 666, 51

\bibitem[{{Leonard} \& {King}(2010)}]{Leonard10}
{Leonard} A., {King} L.~J., 2010, \mnras, 405, 1854

\bibitem[{{Limousin} {et~al}\mbox{.}(2010){Limousin}, {Ebeling}, {Ma},
  {Swinbank}, {Smith}, {Richard}, {Edge}, {Jauzac}, {Kneib}, {Marshall}, \&
  {Schrabback}}]{Limousin10}
{Limousin} M. {et~al.}, 2010, \mnras, 405, 777

\bibitem[{{Limousin} {et~al}\mbox{.}(2007){Limousin}, {Kneib}, {Bardeau},
  {Natarajan}, {Czoske}, {Smail}, {Ebeling}, \& {Smith}}]{Limousin07}
{Limousin} M., {Kneib} J.~P., {Bardeau} S., {Natarajan} P., {Czoske} O.,
  {Smail} I., {Ebeling} H., {Smith} G.~P., 2007, \aap, 461, 881

\bibitem[{{Limousin} {et~al}\mbox{.}(2012){Limousin}, {Morandi}, {Sereno},
  {Meneghetti}, {Ettori}, {Bartelmann}, \& {Verdugo}}]{Limousin12}
{Limousin} M., {Morandi} A., {Sereno} M., {Meneghetti} M., {Ettori} S.,
  {Bartelmann} M., {Verdugo} T., 2012, ArXiv e-prints

\bibitem[{{Limousin} {et~al}\mbox{.}(2009){Limousin}, {Sommer-Larsen},
  {Natarajan}, \& {Milvang-Jensen}}]{Limousin09}
{Limousin} M., {Sommer-Larsen} J., {Natarajan} P., {Milvang-Jensen} B., 2009,
  \apj, 696, 1771

\bibitem[{{Luppino} \& {Kaiser}(1997)}]{LK97}
{Luppino} G.~A., {Kaiser} N., 1997, \apj, 475, 20

\bibitem[{{Mahdavi} {et~al}\mbox{.}(2007){Mahdavi}, {Hoekstra}, {Babul},
  {Balam}, \& {Capak}}]{Mahdavi07}
{Mahdavi} A., {Hoekstra} H., {Babul} A., {Balam} D.~D., {Capak} P.~L., 2007,
  \apj, 668, 806

\bibitem[{{Mahdavi} {et~al}\mbox{.}(2012){Mahdavi}, {Hoekstra}, {Babul},
  {Bildfell}, {Jeltema}, \& {Henry}}]{Mahdavi12}
{Mahdavi} A., {Hoekstra} H., {Babul} A., {Bildfell} C., {Jeltema} T., {Henry}
  J.~P., 2012, ArXiv e-prints

\bibitem[{{Mahdavi} {et~al}\mbox{.}(2008){Mahdavi}, {Hoekstra}, {Babul}, \&
  {Henry}}]{Mahdavi08}
{Mahdavi} A., {Hoekstra} H., {Babul} A., {Henry} J.~P., 2008, \mnras, 384, 1567

\bibitem[{{Mandelbaum} {et~al}\mbox{.}(2006){Mandelbaum}, {Hirata}, {Ishak},
  {Seljak}, \& {Brinkmann}}]{Mandelbaum06}
{Mandelbaum} R., {Hirata} C.~M., {Ishak} M., {Seljak} U., {Brinkmann} J., 2006,
  \mnras, 367, 611

\bibitem[{{Mandelbaum} {et~al}\mbox{.}(2005){Mandelbaum}, {Hirata}, {Seljak},
  {Guzik}, {Padmanabhan}, {Blake}, {Blanton}, {Lupton}, \&
  {Brinkmann}}]{Mandelbaum05}
{Mandelbaum} R. {et~al.}, 2005, \mnras, 361, 1287

\bibitem[{{Mandelbaum} {et~al}\mbox{.}(2008){Mandelbaum}, {Seljak}, \&
  {Hirata}}]{Mandelbaum08}
{Mandelbaum} R., {Seljak} U., {Hirata} C.~M., 2008, JCAP, 8, 6

\bibitem[{{Marian} {et~al}\mbox{.}(2010){Marian}, {Smith}, \&
  {Bernstein}}]{Marian10}
{Marian} L., {Smith} R.~E., {Bernstein} G.~M., 2010, \apj, 709, 286

\bibitem[{{Markevitch} {et~al}\mbox{.}(2004){Markevitch}, {Gonzalez}, {Clowe},
  {Vikhlinin}, {Forman}, {Jones}, {Murray}, \& {Tucker}}]{Markevitch04}
{Markevitch} M., {Gonzalez} A.~H., {Clowe} D., {Vikhlinin} A., {Forman} W.,
  {Jones} C., {Murray} S., {Tucker} W., 2004, \apj, 606, 819

\bibitem[{{Marrone} {et~al}\mbox{.}(2012){Marrone}, {Smith}, {Okabe},
  {Bonamente}, {Carlstrom}, {Culverhouse}, {Gralla}, {Greer}, {Hasler},
  {Hawkins}, {Hennessy}, {Joy}, {Lamb}, {Leitch}, {Martino}, {Mazzotta},
  {Miller}, {Mroczkowski}, {Muchovej}, {Plagge}, {Pryke}, {Sanderson},
  {Takada}, {Woody}, \& {Zhang}}]{Marrone12}
{Marrone} D.~P. {et~al.}, 2012, \apj, 754, 119

\bibitem[{{Massey} {et~al}\mbox{.}(2007){Massey}, {Heymans}, {Berg{\'e}},
  {Bernstein}, {Bridle}, {Clowe}, {Dahle}, {Ellis}, {Erben}, {Hetterscheidt},
  {High}, {Hirata}, {Hoekstra}, {Hudelot}, {Jarvis}, {Johnston}, {Kuijken},
  {Margoniner}, {Mandelbaum}, {Mellier}, {Nakajima}, {Paulin-Henriksson},
  {Peeples}, {Roat}, {Refregier}, {Rhodes}, {Schrabback}, {Schirmer}, {Seljak},
  {Semboloni}, \& {van Waerbeke}}]{STEP2}
{Massey} R. {et~al.}, 2007, \mnras, 376, 13

\bibitem[{{Massey} {et~al}\mbox{.}(2012){Massey}, {Hoekstra}, {Kitching},
  {Rhodes}, {Cropper}, {Amiaux}, {Harvey}, {Mellier}, {Meneghetti}, {Miller},
  {Paulin-Henriksson}, {Pires}, {Scaramella}, \& {Schrabback}}]{Massey12}
{Massey} R. {et~al.}, 2012, ArXiv e-prints

\bibitem[{{McKay} {et~al}\mbox{.}(2001){McKay}, {Sheldon}, {Racusin},
  {Fischer}, {Seljak}, {Stebbins}, {Johnston}, {Frieman}, {Bahcall},
  {Brinkmann}, {Csabai}, {Fukugita}, {Hennessy}, {Ivezic}, {Lamb}, {Loveday},
  {Lupton}, {Munn}, {Nichol}, {Pier}, \& {York}}]{Mckay01}
{McKay} T.~A. {et~al.}, 2001, ArXiv Astrophysics e-prints

\bibitem[{{Medezinski} {et~al}\mbox{.}(2010){Medezinski}, {Broadhurst},
  {Umetsu}, {Oguri}, {Rephaeli}, \& {Ben{\'{\i}}tez}}]{Medezinski10}
{Medezinski} E., {Broadhurst} T., {Umetsu} K., {Oguri} M., {Rephaeli} Y.,
  {Ben{\'{\i}}tez} N., 2010, \mnras, 405, 257

\bibitem[{{Melchior} \& {Viola}(2012)}]{Melchior12}
{Melchior} P., {Viola} M., 2012, \mnras, 424, 2757

\bibitem[{{Melchior} {et~al}\mbox{.}(2011){Melchior}, {Viola}, {Sch{\"a}fer},
  \& {Bartelmann}}]{Melchior11}
{Melchior} P., {Viola} M., {Sch{\"a}fer} B.~M., {Bartelmann} M., 2011, \mnras,
  412, 1552

\bibitem[{{Meneghetti} {et~al}\mbox{.}(2012){Meneghetti}, {Bartelmann}, \& {et
  al.}}]{Meneghetti12}
{Meneghetti} M., {Bartelmann} M., {et al.}, 2012, this volume

\bibitem[{{Meneghetti} {et~al}\mbox{.}(2010){Meneghetti}, {Rasia}, {Merten},
  {Bellagamba}, {Ettori}, {Mazzotta}, {Dolag}, \& {Marri}}]{Meneghetti10}
{Meneghetti} M., {Rasia} E., {Merten} J., {Bellagamba} F., {Ettori} S.,
  {Mazzotta} P., {Dolag} K., {Marri} S., 2010, \aap, 514, A93

\bibitem[{{Meneghetti} {et~al}\mbox{.}(2001){Meneghetti}, {Yoshida},
  {Bartelmann}, {Moscardini}, {Springel}, {Tormen}, \& {White}}]{Meneghetti01}
{Meneghetti} M., {Yoshida} N., {Bartelmann} M., {Moscardini} L., {Springel} V.,
  {Tormen} G., {White} S.~D.~M., 2001, \mnras, 325, 435

\bibitem[{{Merten} {et~al}\mbox{.}(2009){Merten}, {Cacciato}, {Meneghetti},
  {Mignone}, \& {Bartelmann}}]{Merten09}
{Merten} J., {Cacciato} M., {Meneghetti} M., {Mignone} C., {Bartelmann} M.,
  2009, \aap, 500, 681

\bibitem[{{Merten} {et~al}\mbox{.}(2011){Merten}, {Coe}, {Dupke}, {Massey},
  {Zitrin}, {Cypriano}, {Okabe}, {Frye}, {Braglia}, {Jim{\'e}nez-Teja},
  {Ben{\'{\i}}tez}, {Broadhurst}, {Rhodes}, {Meneghetti}, {Moustakas},
  {Sodr{\'e}}, {Krick}, \& {Bregman}}]{Merten11}
{Merten} J. {et~al.}, 2011, \mnras, 417, 333

\bibitem[{{Metzler} {et~al}\mbox{.}(2001){Metzler}, {White}, \&
  {Loken}}]{Metzler01}
{Metzler} C.~A., {White} M., {Loken} C., 2001, \apj, 547, 560

\bibitem[{{Miller} {et~al}\mbox{.}(2012){Miller}, {Heymans}, {Kitching}, {Van
  Waerbeke}, {Erben}, {Hildebrandt}, {Hoekstra}, {Mellier}, {Rowe}, {Coupon},
  {Dietrich}, {Fu}, {Harnois-Deraps}, {Hudson}, {Kilbinger}, {Kuijken},
  {Schrabback}, {Semboloni}, {Vafaei}, \& {Velander}}]{Miller12}
{Miller} L. {et~al.}, 2012, ArXiv e-prints

\bibitem[{{Miller} {et~al}\mbox{.}(2007){Miller}, {Kitching}, {Heymans},
  {Heavens}, \& {van Waerbeke}}]{Miller07}
{Miller} L., {Kitching} T.~D., {Heymans} C., {Heavens} A.~F., {van Waerbeke}
  L., 2007, \mnras, 382, 315

\bibitem[{{Miralda-Escude}(1991)}]{MiraldaEscude91}
{Miralda-Escude} J., 1991, \apj, 370, 1

\bibitem[{{Morandi} \& {Limousin}(2012)}]{Morandi12}
{Morandi} A., {Limousin} M., 2012, \mnras, 421, 3147

\bibitem[{{Nagai} {et~al}\mbox{.}(2007){Nagai}, {Vikhlinin}, \&
  {Kravtsov}}]{Nagai07}
{Nagai} D., {Vikhlinin} A., {Kravtsov} A.~V., 2007, \apj, 655, 98

\bibitem[{{Natarajan} {et~al}\mbox{.}(2002{\natexlab{a}}){Natarajan}, {Kneib},
  \& {Smail}}]{Priya2}
{Natarajan} P., {Kneib} J.-P., {Smail} I., 2002{\natexlab{a}}, \apjl, 580, L11

\bibitem[{{Natarajan} {et~al}\mbox{.}(1998){Natarajan}, {Kneib}, {Smail}, \&
  {Ellis}}]{Priya1}
{Natarajan} P., {Kneib} J.-P., {Smail} I., {Ellis} R.~S., 1998, \apj, 499, 600

\bibitem[{{Natarajan} {et~al}\mbox{.}(2009){Natarajan}, {Kneib}, {Smail},
  {Treu}, {Ellis}, {Moran}, {Limousin}, \& {Czoske}}]{priya4}
{Natarajan} P., {Kneib} J.-P., {Smail} I., {Treu} T., {Ellis} R., {Moran} S.,
  {Limousin} M., {Czoske} O., 2009, \apj, 693, 970

\bibitem[{{Natarajan} {et~al}\mbox{.}(2002{\natexlab{b}}){Natarajan}, {Loeb},
  {Kneib}, \& {Smail}}]{Priya3}
{Natarajan} P., {Loeb} A., {Kneib} J.-P., {Smail} I., 2002{\natexlab{b}},
  \apjl, 580, L17

\bibitem[{{Navarro} {et~al}\mbox{.}(1997){Navarro}, {Frenk}, \& {White}}]{NFW}
{Navarro} J.~F., {Frenk} C.~S., {White} S.~D.~M., 1997, \apj, 490, 493

\bibitem[{{Oguri} {et~al}\mbox{.}(2012){Oguri}, {Bayliss}, {Dahle}, {Sharon},
  {Gladders}, {Natarajan}, {Hennawi}, \& {Koester}}]{Oguri12}
{Oguri} M., {Bayliss} M.~B., {Dahle} H., {Sharon} K., {Gladders} M.~D.,
  {Natarajan} P., {Hennawi} J.~F., {Koester} B.~P., 2012, \mnras, 420, 3213

\bibitem[{{Oguri} {et~al}\mbox{.}(2009){Oguri}, {Hennawi}, {Gladders}, {Dahle},
  {Natarajan}, {Dalal}, {Koester}, {Sharon}, \& {Bayliss}}]{Oguri09}
{Oguri} M. {et~al.}, 2009, \apj, 699, 1038

\bibitem[{{Oguri} {et~al}\mbox{.}(2010){Oguri}, {Takada}, {Okabe}, \&
  {Smith}}]{Oguri10}
{Oguri} M., {Takada} M., {Okabe} N., {Smith} G.~P., 2010, \mnras, 405, 2215

\bibitem[{{Okabe} {et~al}\mbox{.}(2010){Okabe}, {Takada}, {Umetsu}, {Futamase},
  \& {Smith}}]{Okabe10}
{Okabe} N., {Takada} M., {Umetsu} K., {Futamase} T., {Smith} G.~P., 2010,
  \pasj, 62, 811

\bibitem[{{Okura} {et~al}\mbox{.}(2008){Okura}, {Umetsu}, \&
  {Futamase}}]{Okura08}
{Okura} Y., {Umetsu} K., {Futamase} T., 2008, \apj, 680, 1

\bibitem[{{Parker} {et~al}\mbox{.}(2005){Parker}, {Hudson}, {Carlberg}, \&
  {Hoekstra}}]{Parker05}
{Parker} L.~C., {Hudson} M.~J., {Carlberg} R.~G., {Hoekstra} H., 2005, \apj,
  634, 806

\bibitem[{{Pedersen} \& {Dahle}(2007)}]{Pedersen07}
{Pedersen} K., {Dahle} H., 2007, \apj, 667, 26

\bibitem[{{Planck Collaboration} {et~al}\mbox{.}(2011){Planck Collaboration},
  {Ade}, {Aghanim}, {Arnaud}, {Ashdown}, {Aumont}, {Baccigalupi}, {Balbi},
  {Banday}, {Barreiro}, \& et~al.}]{Planck11}
{Planck Collaboration} {et~al.}, 2011, \aap, 536, A8

\bibitem[{{Postman} {et~al}\mbox{.}(2012){Postman}, {Coe}, {Ben{\'{\i}}tez},
  {Bradley}, {Broadhurst}, {Donahue}, {Ford}, {Graur}, {Graves}, {Jouvel},
  {Koekemoer}, {Lemze}, {Medezinski}, {Molino}, {Moustakas}, {Ogaz}, {Riess},
  {Rodney}, {Rosati}, {Umetsu}, {Zheng}, {Zitrin}, {Bartelmann}, {Bouwens},
  {Czakon}, {Golwala}, {Host}, {Infante}, {Jha}, {Jimenez-Teja}, {Kelson},
  {Lahav}, {Lazkoz}, {Maoz}, {McCully}, {Melchior}, {Meneghetti}, {Merten},
  {Moustakas}, {Nonino}, {Patel}, {Reg{\"o}s}, {Sayers}, {Seitz}, \& {Van der
  Wel}}]{Postman12}
{Postman} M. {et~al.}, 2012, \apjs, 199, 25

\bibitem[{{Pracy} {et~al}\mbox{.}(2004){Pracy}, {De Propris}, {Driver},
  {Couch}, \& {Nulsen}}]{Pracy04}
{Pracy} M.~B., {De Propris} R., {Driver} S.~P., {Couch} W.~J., {Nulsen}
  P.~E.~J., 2004, \mnras, 352, 1135

\bibitem[{{Rasia} {et~al}\mbox{.}(2012){Rasia}, {Meneghetti}, {Martino},
  {Borgani}, {Bonafede}, {Dolag}, {Ettori}, {Fabjan}, {Giocoli}, {Mazzotta},
  {Merten}, {Radovich}, \& {Tornatore}}]{Rasia12}
{Rasia} E. {et~al.}, 2012, New Journal of Physics, 14, 055018

\bibitem[{{Refregier} \& {Bacon}(2003)}]{Refregier03}
{Refregier} A., {Bacon} D., 2003, \mnras, 338, 48

\bibitem[{{Refregier} {et~al}\mbox{.}(2012){Refregier}, {Kacprzak}, {Amara},
  {Bridle}, \& {Rowe}}]{Refregier12}
{Refregier} A., {Kacprzak} T., {Amara} A., {Bridle} S., {Rowe} B., 2012,
  \mnras, 425, 1951

\bibitem[{{Rines} {et~al}\mbox{.}(2003){Rines}, {Geller}, {Kurtz}, \&
  {Diaferio}}]{Rines03}
{Rines} K., {Geller} M.~J., {Kurtz} M.~J., {Diaferio} A., 2003, \aj, 126, 2152

\bibitem[{{Rykoff} {et~al}\mbox{.}(2008){Rykoff}, {Evrard}, {McKay}, {Becker},
  {Johnston}, {Koester}, {Nord}, {Rozo}, {Sheldon}, {Stanek}, \&
  {Wechsler}}]{Rykoff08}
{Rykoff} E.~S. {et~al.}, 2008, \mnras, 387, L28

\bibitem[{{Schmidt} {et~al}\mbox{.}(2012){Schmidt}, {Leauthaud}, {Massey},
  {Rhodes}, {George}, {Koekemoer}, {Finoguenov}, \& {Tanaka}}]{Schmidt12}
{Schmidt} F., {Leauthaud} A., {Massey} R., {Rhodes} J., {George} M.~R.,
  {Koekemoer} A.~M., {Finoguenov} A., {Tanaka} M., 2012, \apjl, 744, L22

\bibitem[{{Schneider} \& {Bridle}(2010)}]{Schneider10}
{Schneider} M.~D., {Bridle} S., 2010, \mnras, 402, 2127

\bibitem[{{Schneider}(2006)}]{Schneider06}
{Schneider} P., 2006, in Saas-Fee Advanced Course 33: Gravitational Lensing:
  Strong, Weak and Micro, {G.~Meylan, P.~Jetzer, P.~North, P.~Schneider,
  C.~S.~Kochanek, \& J.~Wambsganss}, ed., pp. 269--451

\bibitem[{{Schrabback} {et~al}\mbox{.}(2010){Schrabback}, {Hartlap},
  {Joachimi}, {Kilbinger}, {Simon}, {Benabed}, {Brada{\v c}}, {Eifler},
  {Erben}, {Fassnacht}, {High}, {Hilbert}, {Hildebrandt}, {Hoekstra},
  {Kuijken}, {Marshall}, {Mellier}, {Morganson}, {Schneider}, {Semboloni}, {van
  Waerbeke}, \& {Velander}}]{Schrabback10}
{Schrabback} T. {et~al.}, 2010, \aap, 516, A63

\bibitem[{{Seitz} \& {Schneider}(1995)}]{Seitz95}
{Seitz} C., {Schneider} P., 1995, \aap, 297, 287

\bibitem[{{Seitz} \& {Schneider}(1997)}]{Seitz97}
{Seitz} C., {Schneider} P., 1997, \aap, 318, 687

\bibitem[{{Seitz} \& {Schneider}(1996)}]{Seitz96}
{Seitz} S., {Schneider} P., 1996, \aap, 305, 383

\bibitem[{{Seitz} \& {Schneider}(2001)}]{Seitz01}
{Seitz} S., {Schneider} P., 2001, \aap, 374, 740

\bibitem[{{Seitz} {et~al}\mbox{.}(1998){Seitz}, {Schneider}, \&
  {Bartelmann}}]{Seitz98}
{Seitz} S., {Schneider} P., {Bartelmann} M., 1998, \aap, 337, 325

\bibitem[{{Seljak}(2000)}]{Seljak00}
{Seljak} U., 2000, \mnras, 318, 203

\bibitem[{{Semboloni} {et~al}\mbox{.}(2011){Semboloni}, {Hoekstra}, {Schaye},
  {van Daalen}, \& {McCarthy}}]{Semboloni11}
{Semboloni} E., {Hoekstra} H., {Schaye} J., {van Daalen} M.~P., {McCarthy}
  I.~G., 2011, \mnras, 417, 2020

\bibitem[{{Sheldon} {et~al}\mbox{.}(2009){Sheldon}, {Johnston}, {Scranton},
  {Koester}, {McKay}, {Oyaizu}, {Cunha}, {Lima}, {Lin}, {Frieman}, {Wechsler},
  {Annis}, {Mandelbaum}, {Bahcall}, \& {Fukugita}}]{Sheldon09}
{Sheldon} E.~S. {et~al.}, 2009, \apj, 703, 2217

\bibitem[{{Simon} {et~al}\mbox{.}(2012){Simon}, {Heymans}, {Schrabback},
  {Taylor}, {Gray}, {van Waerbeke}, {Wolf}, {Bacon}, {Barden}, {B{\"o}hm},
  {H{\"a}u{\ss}ler}, {Jahnke}, {Jogee}, {van Kampen}, {Meisenheimer}, \&
  {Peng}}]{Simon12}
{Simon} P. {et~al.}, 2012, \mnras, 419, 998

\bibitem[{{Soucail} {et~al}\mbox{.}(1987){Soucail}, {Fort}, {Mellier}, \&
  {Picat}}]{Soucail87}
{Soucail} G., {Fort} B., {Mellier} Y., {Picat} J.~P., 1987, \aap, 172, L14

\bibitem[{{Springel} \& {Farrar}(2007)}]{Springel07}
{Springel} V., {Farrar} G.~R., 2007, \mnras, 380, 911

\bibitem[{{Squires} \& {Kaiser}(1996)}]{Squires96}
{Squires} G., {Kaiser} N., 1996, \apj, 473, 65

\bibitem[{{Taylor} {et~al}\mbox{.}(2004){Taylor}, {Bacon}, {Gray}, {Wolf},
  {Meisenheimer}, {Dye}, {Borch}, {Kleinheinrich}, {Kovacs}, \&
  {Wisotzki}}]{Taylor04}
{Taylor} A.~N. {et~al.}, 2004, \mnras, 353, 1176

\bibitem[{{Taylor} {et~al}\mbox{.}(2012){Taylor}, {Massey}, {Leauthaud},
  {George}, {Rhodes}, {Kitching}, {Capak}, {Ellis}, {Finoguenov}, {Ilbert},
  {Jullo}, {Kneib}, {Koekemoer}, {Scoville}, \& {Tanaka}}]{Taylor12}
{Taylor} J.~E. {et~al.}, 2012, \apj, 749, 127

\bibitem[{{Tyson} {et~al}\mbox{.}(1990){Tyson}, {Wenk}, \& {Valdes}}]{Tyson90}
{Tyson} J.~A., {Wenk} R.~A., {Valdes} F., 1990, \apjl, 349, L1

\bibitem[{{Umetsu} {et~al}\mbox{.}(2009){Umetsu}, {Birkinshaw}, {Liu}, {Wu},
  {Medezinski}, {Broadhurst}, {Lemze}, {Zitrin}, {Ho}, {Huang}, {Koch}, {Liao},
  {Lin}, {Molnar}, {Nishioka}, {Wang}, {Altamirano}, {Chang}, {Chang}, {Chang},
  {Chen}, {Han}, {Huang}, {Hwang}, {Jiang}, {Kesteven}, {Kubo}, {Li},
  {Martin-Cocher}, {Oshiro}, {Raffin}, {Wei}, \& {Wilson}}]{Umetsu09}
{Umetsu} K. {et~al.}, 2009, \apj, 694, 1643

\bibitem[{{Umetsu} {et~al}\mbox{.}(2011){Umetsu}, {Broadhurst}, {Zitrin},
  {Medezinski}, \& {Hsu}}]{Umetsu11}
{Umetsu} K., {Broadhurst} T., {Zitrin} A., {Medezinski} E., {Hsu} L.-Y., 2011,
  \apj, 729, 127

\bibitem[{{Umetsu} {et~al}\mbox{.}(2012){Umetsu}, {Medezinski}, {Nonino},
  {Merten}, {Zitrin}, {Molino}, {Grillo}, {Carrasco}, {Donahue}, {Mahdavi},
  {Coe}, {Postman}, {Koekemoer}, {Czakon}, {Sayers}, {Mroczkowski}, {Golwala},
  {Koch}, {Lin}, {Molnar}, {Rosati}, {Balestra}, {Mercurio}, {Scodeggio},
  {Biviano}, {Anguita}, {Infante}, {Seidel}, {Sendra}, {Jouvel}, {Host},
  {Lemze}, {Broadhurst}, {Meneghetti}, {Moustakas}, {Bartelmann},
  {Ben{\'{\i}}tez}, {Bouwens}, {Bradley}, {Ford}, {Jim{\'e}nez-Teja}, {Kelson},
  {Lahav}, {Melchior}, {Moustakas}, {Ogaz}, {Seitz}, \& {Zheng}}]{Umetsu12}
{Umetsu} K. {et~al.}, 2012, \apj, 755, 56

\bibitem[{{van Uitert} {et~al}\mbox{.}(2012){van Uitert}, {Hoekstra},
  {Schrabback}, {Gilbank}, {Gladders}, \& {Yee}}]{vanUitert12}
{van Uitert} E., {Hoekstra} H., {Schrabback} T., {Gilbank} D.~G., {Gladders}
  M.~D., {Yee} H.~K.~C., 2012, \aap, 545, A71

\bibitem[{{van Uitert} {et~al}\mbox{.}(2011){van Uitert}, {Hoekstra},
  {Velander}, {Gilbank}, {Gladders}, \& {Yee}}]{vanUitert11}
{van Uitert} E., {Hoekstra} H., {Velander} M., {Gilbank} D.~G., {Gladders}
  M.~D., {Yee} H.~K.~C., 2011, \aap, 534, A14

\bibitem[{{von der Linden} {et~al}\mbox{.}(2012){von der Linden}, {Allen},
  {Applegate}, {Kelly}, {Allen}, {Ebeling}, {Burchat}, {Burke}, {Donovan},
  {Morris}, {Blandford}, {Erben}, \& {Mantz}}]{VonderLinden12}
{von der Linden} A. {et~al.}, 2012, ArXiv e-prints

\bibitem[{{Walsh} {et~al}\mbox{.}(1979){Walsh}, {Carswell}, \&
  {Weymann}}]{Walsh79}
{Walsh} D., {Carswell} R.~F., {Weymann} R.~J., 1979, \nat, 279, 381

\bibitem[{{Wittman} {et~al}\mbox{.}(2003){Wittman}, {Margoniner}, {Tyson},
  {Cohen}, {Becker}, \& {Dell'Antonio}}]{Wittman03}
{Wittman} D., {Margoniner} V.~E., {Tyson} J.~A., {Cohen} J.~G., {Becker} A.~C.,
  {Dell'Antonio} I.~P., 2003, \apj, 597, 218

\bibitem[{{Wright} \& {Brainerd}(2000)}]{Wright00}
{Wright} C.~O., {Brainerd} T.~G., 2000, \apj, 534, 34

\bibitem[{{Zhang} {et~al}\mbox{.}(2010){Zhang}, {Okabe}, {Finoguenov}, {Smith},
  {Piffaretti}, {Valdarnini}, {Babul}, {Evrard}, {Mazzotta}, {Sanderson}, \&
  {Marrone}}]{Zhang10}
{Zhang} Y.-Y. {et~al.}, 2010, \apj, 711, 1033

\bibitem[{{Zieser} \& {Bartelmann}(2012)}]{Zieser12}
{Zieser} B., {Bartelmann} M., 2012, ArXiv e-prints

\bibitem[{{Zitrin} {et~al}\mbox{.}(2012){Zitrin}, {Rosati}, {Nonino}, {Grillo},
  {Postman}, {Coe}, {Seitz}, {Eichner}, {Broadhurst}, {Jouvel}, {Balestra},
  {Mercurio}, {Scodeggio}, {Ben{\'{\i}}tez}, {Bradley}, {Ford}, {Host},
  {Jimenez-Teja}, {Koekemoer}, {Zheng}, {Bartelmann}, {Bouwens}, {Czoske},
  {Donahue}, {Graur}, {Graves}, {Infante}, {Jha}, {Kelson}, {Lahav}, {Lazkoz},
  {Lemze}, {Lombardi}, {Maoz}, {McCully}, {Medezinski}, {Melchior},
  {Meneghetti}, {Merten}, {Molino}, {Moustakas}, {Ogaz}, {Patel}, {Regoes},
  {Riess}, {Rodney}, {Umetsu}, \& {Van der Wel}}]{Zitrin12}
{Zitrin} A. {et~al.}, 2012, \apj, 749, 97

\bibitem[{{Zwicky}(1937)}]{Zwicky37}
{Zwicky} F., 1937, \apj, 86, 217

\end{thebibliography}
